\documentclass[manuscript]{aastex62}

\submitjournal{ApJ}
\shorttitle{Shocks in Sunspot Umbra}
\shortauthors{Kayshap, Tripathi and Jelinek}
\begin{document}
\title{Dynamics of Sunspot Shock Waves in the Chromosphere and Transition Region}
%%--------------------------------------------
\correspondingauthor{Pradeep Kayshap}
\email{pradeep.kayshap@vitbhopal.ac.in}
\author[0000-0002-0786-7307]{Pradeep Kayshap}
\affil{Vellore Institute of Technology (VIT) Bhopal, 466114, India} 
\affil{Inter University Centre for Astronomy $\&$ Astrophysics, Pune 411007, India}
\author[0000-0003-1689-6254]{Durgesh Tripathi}
\affil{Inter University Centre for Astronomy $\&$ Astrophysics, Pune 411007, India}
\author{P. Jel\'\i nek}
\affil{University of South Bohemia, Faculty of Science, Institute of Physics, Brani\v sovsk\'a 1760, CZ -- 370 05 \v{C}esk\'e Bud\v{e}jovice, Czech Republic}
%%--------------------------------------------
\begin{abstract}
%%--------------------------------------------

We study the dynamics of shock waves observed in the umbra of a sunspot using the spectroscopic observations from the Interface Region Imaging Spectrometer (IRIS). The presence of the shock significantly deforms the shape of the spectral lines of \ion{Mg}{2}, \ion{C}{2}, and \ion{Si}{4}. We found that \ion{C}{2}~1335.66~{\AA} and \ion{Si}{4}~1393.75~{\AA} show double-peaked profiles that change to a single peak later on. However, the \ion{Mg}{2}~h~2803.53~{\AA} line first shows flat-top profiles that change into double-peaked followed by the single peak. To study the shock dynamics, we isolate the shock component from the spectra by fitting two Gaussians. We find that the lifetime of the shock is largest in \ion{Mg}{2}~h~2803.53~{\AA} line. Moreover, the plasma motion shows both acceleration and deceleration phase of the shock. Yet, in \ion{C}{2}~1335.66~{\AA} and \ion{Si}{4}~1393.75~{\AA}, only deceleration phase is observed. We observe a strong correlation between the largest blueshift of the shock and deceleration for all three spectral lines. We find a positive (negative) correlation between intensities contributed due to the shocks in \ion{Mg}{2} and \ion{C}{2} (\ion{Si}{4}). This is suggestive that the shocks are first amplified in \ion{C}{2}, followed by a decline in the height range corresponding to \ion{Si}{4}. These results \textbf{may} indicate the dissipation of shocks above the formation height of \ion{C}{2}, and the shocks may have important roles in the dynamics of the upper chromosphere and transition region above sunspots.

%%--------------------------------------------
\end{abstract}
%%---------------------------------------------------
\keywords{line: profiles --- Sun: chromosphere --- Sun: transition region --- Sun:sunspots}
%%---------------------------------------------------
\section{Introduction} \label{sec:intro}
%%---------------------------------------------------
Waves and oscillations are ubiquitous in the sunspots. Due to their vital role in channelizing the energy in the solar atmosphere, they are one of the most studied features in the solar atmosphere
\citep[see, e.g.][]{Bogdan2006, Kho2015, Zhao2016, Sharma2017, Kayshap2018b, Kayshap2020, Sharma2020} 

\cite{Lites1979} showed that the low-frequency waves ($<$4 mHz; 4 minutes) are evanescent and those at high frequency propagate upward. \cite{Carlsson1997} explained the formation of chromospheric bright grains due to the steepening of 3-mins waves into shocks. \cite{Wikstol2000, Bloomfield2004} reported the propagation of 3-mins oscillation in the quiet-Sun networks. \cite{DePon2003} studied the propagation of 5-min oscillations in plage regions, while \cite{Zhao2016} also reported the propagation of p-modes into solar corona in the active-region and discussed their importance in the dynamics of the upper solar atmosphere.

\citet{Centeno2006} investigated the propagation of waves in sunspot umbrae, tiny pores, and facular regions. They used \ion{Si}{1}~10827~{\AA} and \ion{He}{1} 10830~{\AA} spectral lines formed in the photosphere and chromosphere, respectively and observed the propagation of 3 or 5 minutes oscillation from the photosphere into the chromosphere depending on the region. Moreover, they demonstrated that wave propagation depends on in-situ physical conditions. \citet{Kayshap2018b, Kayshap2020} showed the presence of 3 minutes oscillation in the inter-network and 5 minutes in plage regions.

During the propagation, waves very often exhibit their non-linear characteristic. For example, the appearance of a saw-tooth pattern in wavelength-time ($\lambda${--}t) for \ion{Ca}{2} observed in inter-network. \citet{Carlsson1994, Carlsson1997} suggested that such patters are due to shocks \citep[see also][]{voort2003}. \citet{Tian2014} reported the presence of saw-tooth patterns in the time series of intensities, Doppler velocities, and full width at half maximum (FWHM) of \ion{Si}{4}, \ion{C}{2}, and \ion{Mg}{2}~k spectral lines observed by the Interface Region Imaging Spectrometer \citep[IRIS;][]{DePon2009}. Using the observations recorded by IRIS and New Solar Telescope (NST; \citealt{Goode2010}), \cite{Yur2015} studied the propagation of shock in the solar atmosphere from \ion{Mg}{2}~k to \ion{Si}{4} and reported a time-lag of about 0~s to 40~s.

As described earlier, there are several observations of shocks' formation in the lower and middle solar atmosphere. Nevertheless, their role in plasma heating is not fully understood \citep[see, e.g.,][]{Kho2015}. \citet{Grant2015} presented the first evidence of plasma heating due to shock in the chromosphere of a sunspot umbra. 

The spectral lines formed in the chromosphere and transition region show double-peaked profiles during the passes of shocks \citep[see, e.g.,][]{voort2003, Centeno2006, Tian2014}. One peak corresponds to the shock and the other to the downflows. However, authors have fitted a single Gaussian to determine the properties of the shocks. This does not allow the shock to be resolved. The excellent spectral resolution of IRIS allows such a study. This is the main aim of this paper.

Here we have utilized a sit-n-stare observation of a sunspot recorded by IRIS. We resolve the two components of the line profile and study the dynamics of the shocks in detail. We present the data and observation in \S\ref{sec:obs_res} and data analysis and results in \S\ref{res}. We summarize our results and conclude in \S\ref{sec:sum_con}.

%%---------------------------------------------------
\section{Observations and Data} \label{sec:obs_res}
%%---------------------------------------------------

To identify and study the behavior of the shocks in a sunspot, we have used IRIS observations taken in three spectral lines, viz. \ion{Mg}{2}~2803.52~{\AA} ($\log\,T[K]=4.0$), \ion{C}{2}~1335.70~{\AA} ($\log\,T[K]=4.4$), and \ion{Si}{4}~1393.77~{\AA} ($\log\,T[K]=4.9$). These three lines cover chromosphere as well as lower and middle transition region. The sunspot studied here is associated with the \textsl{AR~12546}, located near the disk center on  May 20, 2016. IRIS observed this sunspot in sit-n-stare mode between 13:17:58~UT and 16:25:07~UT, with an exposure time of $\sim$15.0~s. We have used level~2 data that is calibrated for dark current, flat field, thermal orbital variation \citep[see][for details]{DePon2009}. For context, we have also utilized observations from Helioseismic and Magnetic Imager \citep[HMI,][]{hmi} and the Atmospheric Imaging Assembly \citep[AIA,][]{aia} onboard the Solar Dynamics Observatory.

%%---------------------------------------------------
\section{Data Analysis and Results}\label{res}
%%---------------------------------------------------

Panel A in Fig.~\ref{fig:ref_fig} displays the HMI line-of-sight (LOS) magnetogram. Panels B \& C display the corresponding slit-jaw-image (SJI) taken in 2832.0~{\AA} continuum and 1400~{\AA}.  The vertical blue (panel A) and white lines (panels B \& C) locate the slit position. The two blue horizontal marks on the slit locate the region considered for further study.

We chose a location at $Y=84.31\arcsec$ on the slit to derive wavelength-time ($\lambda${--}t) plots for \ion{Si}{4}, \ion{C}{2} and \ion{Mg}{2}~h lines. We show these plots in the 2nd row of Fig.~\ref{fig:ref_fig}. To convert x-scale into velocity we use the reference wavelengths from CHIANTI viz. 1393.75 for \ion{Si}{4}, 1335.66 for \ion{C}{2}, and 2803.53 for \ion{Mg}{2}~h.

The $\lambda${--}t plots all the three spectral lines reveal a sudden transition from redshift to blueshift at the beginning of the observation. The blueshift changes to redshift again, albeit gentle. We note that within a duration of $\sim$5 minutes, there are about 2 to 3 sudden jumps from red to blue to red. The jump is most pronounced in \ion{Si}{4}, followed by \ion{C}{2} and \ion{Mg}{2}~h. The presence of such patterns in the $\lambda${--}t plots suggests the presence of shocks \citep[][]{Tian2014}. 

In the bottom row of Fig.~\ref{fig:ref_fig}, we plot the spectral profiles at the time of presence (absence) of the shock in black (red). We define the presence (absence) of shock as the time when we observe the strongest blueshifts (redshift) in the $\lambda${--}t diagram. The black (red) plus symbols mark the presence (absence) of shock in panels D, E \& F of Fig.~\ref{fig:ref_fig}. 

We observe that at the time when the shock is present, all three spectral lines exhibit double-peaked profiles with larger width (black curves in panels G, H and I). Thus we conclude that there are two different components in the line profiles. One blueshifted due to shock and the other redshifted as a result of downflows. 

Next, we study the time evolution of the shock. We plot the spectral line profiles for \ion{Si}{4} (top row), \ion{C}{2} (middle row) and \ion{Mg}{2} (bottom row) in Fig.~\ref{fig:all_profile}. We show the line profiles in black. Note that these profiles are also obtained at $Y=84.31\arcsec$, albeit at different times as labelled. Based on the number of peaks observed, we fit either a double-Gaussian (red) or a single Gaussian (blue) to the spectral lines. Note that we have performed free double Gaussian fits to the profiles. We have shown the fit parameters in the respective panels. We note that the spectral lines of \ion{Si}{4} and \ion{C}{2} at first show double-peaked profiles (first two columns). At later times they change into a single-peaked profile (third column). 

We observed that the double-peaked profile for \ion{Mg}{2} appeared about a few seconds earlier than \ion{C}{2} and \ion{Si}{4}. Thus, we plot the line profile for \ion{Mg}{2} (bottom) from an earlier time. We note that at first, the \ion{Mg}{2} line shows a flat-top profile.  These then change into double-peaked followed by a single peak profile. We find that the flat-top profiles are well represented by two Gaussians with the same amplitude. Thus we conclude that these profiles consist of two components. The first Gaussian is due to shock and the second is due to downflows.

%%----------------------------------------------------------------
\subsection{Statistical Study of Shock and its characteristics} \label{stats}
%%----------------------------------------------------------------
In total, we identify 71 such shocks in $\sim$97 mins of observation. We have derived their velocities, acceleration, and contribution to the total line intensity. We plot the time evolution of the Doppler shift of the shock identified at $Y = -84.31\arcsec$ in Fig.~\ref{fig:shock_acel_main} A. 

For the plots shown in Fig.~\ref{fig:shock_acel_main}, black color is for \ion{Mg}{2}~h line, blue for \ion{C}{2} and red is for \ion{Si}{4} line. The vertical black dashed line corresponds to the time of the first occurrence of the shock in \ion{Si}{4} and \ion{C}{2}. Note that the shocks appeared in \ion{Mg}{2}~h line about 50~s earlier than in \ion{Si}{4} and \ion{C}{2}. Moreover, the lifetime of shocks was longer in \ion{Mg}{2} (about 110s) than that in \ion{C}{2} and \ion{Si}{4} ($\sim$80 s). Following \citealt{voort2003, Tian2014}, we define lifetime as the duration within which the spectral lines move from the strongest blueshift to the strongest redshift. In principle, this is the time for which the local plasma parcel is affected due to shock. 

Figure~\ref{fig:shock_acel_main}.A, in \ion{Mg}{2}, we observe the acceleration of the shock for the first 30~s. It attained a constant speed for the next 30~s and decelerated thereafter. But, in \ion{C}{2} and \ion{Si}{4}, we only observe the deceleration phase. We further note that, the deceleration is slower in \ion{Mg}{2} than those in \ion{C}{2} and \ion{Si}{4}. The plots further reveal that the shock appeared in \ion{C}{2} and \ion{Si}{4} exactly at the time when it attained a constant velocity in \ion{Mg}{2}. 

We estimate the deceleration profile of the shock in the three spectral lines. For this, we have fitted straight lines to the velocity-time plots. Note that for such fittings, we have only considered the last four data points of \ion{Mg}{2}. We find that the shocks show similar deceleration in \ion{C}{2} and \ion{Si}{4} ($\sim$220~m~s$^{-2}$). Moreover, the shock decelerated with $\sim$143~m~s$^{-2}$ in \ion{Mg}{2}, which is slower than \ion{C}{2} and \ion{Si}{4}

We have followed the above-described procedure for all the 71 shocks and derive their deceleration profiles. We show a scatter plot between the derived deceleration and the largest blueshifts of shocks in Fig.~\ref{fig:shock_acel_main}.B. The over-plotted solid lines are linear fit to the data in corresponding colors. We have also shown the Pearson's coefficients. The plots show a strong correlation between deceleration and the largest blueshift for all the three spectral lines, suggesting that faster shocks decelerate faster.

We plot the distribution of the largest blueshifts in \ion{Mg}{2}, \ion{C}{2} and \ion{Si}{4} in the bottom row of Fig.~\ref{fig:shock_acel_main}. The mean velocities are 7.95 km s$^{-1}$, 9.34 km s$^{-1}$, and 8.34 km s$^{-1}$ for \ion{Mg}{2} (black), \ion{C}{2} (blue) and \ion{Si}{4}, respectively. We note that shocks have highest (lowest) mean velocity in \ion{C}{2} (\ion{Si}{4})

%%---------------------------------------------
\subsection{Contribution of shock in the line intensity}\label{inte}
%%---------------------------------------------

Depending on the lifetime, we may have about 6 to 10 spectral profiles for any given shock. We have used the spectral profile with the strongest blue shift to derive the intensity contribution due to these shocks. Such profiles are very well resolved with two peaks. The contribution due to shock is essentially the intensity in the blue-shifted component. We derive the intensities in all three spectral lines. Finally, we perform a correlation study between the intensities due to shock in different lines.

In Fig.~\ref{fig:shock_correl}, we show the scatter plots between intensity contributions due to shock in \ion{C}{2} \& \ion{Mg}{2} (panel A) and \ion{Si}{4} \& \ion{Mg}{2} (panel B). We observe a positive correlation between \ion{C}{2} and \ion{Mg}{2} and a negative correlation between \ion{Si}{4} and \ion{Mg}{2}. Such correlations suggest that the shocks which are seen in \ion{Mg}{2} may have been amplified in \ion{C}{2} and decline at heights corresponding to \ion{Si}{4} line. It is important to note that \ion{Mg}{2} and \ion{C}{2} lines are optically thick. Their formation and intensities depend on several factors viz. temperature, opacity, density etc. \citep[see e.g.,][]{Leen2013,Rathore2015}. Thus, caution must be taken in the interpretation of these results.

%%-------------------------------------------------
\subsection{Footpoints of Fan-loops $\&$ shocks}
%%-------------------------------------------------

Fan loops are often rooted inside the umbra of sunspots \citep[e.g.][]{Sch1999, Wine2002, Young2012, Chitta2016,Ghosh2017}. In Fig.~\ref{fig:fan_loop} (top panel), we compare the IRIS SJI images studied here with the corresponding AIA~171~{\AA} images. The over-plotted blue contours in the left panel outline the umbra's and penumbra's boundaries. We derive the intensity levels contours from SJI taken in 2832~{\AA} channel. We observe that fan loops are rooted within the sunspot umbra (left panel in Fig~\ref{fig:fan_loop}). We locate the footprints of fan loops with white arrows in the middle and right panels. It is interesting to note that the locations of the footpoints of fan-loops match with those for shocks. Thus, it provides an opportunity to study the possible role of shocks in the fan loops' heating.

In the bottom panel of Fig.~\ref{fig:fan_loop}, we plot the intensity light curve for 171~{\AA} (black) and \ion{Si}{4} (blue), derived at the location indicated by the arrow. The time series corresponds to the first 40 minutes of the observation. During this time at that location, we identified four shocks at that location. We mark these by vertical dashed lines. The complete time series is shown in Fig.~\ref{fig:iris_aia}. Note that we have detrended both the light curves, as shown in Fig.~\ref{fig:iris_aia}. The plot reveals that in \ion{Si}{4}, there is a clear time lag between the peak intensity and identification time of shocks. We attribute this to the fact that we identify shocks based on the Doppler velocities, based on the conclusions made by \citealt{Centeno2006} and \citealt{Tian2014}, who demonstrated that phase-lag exists between velocity and intensity in the case of magneto-acoustic waves. Moreover, we observe a strong correlation between the light curves of 171~{\AA} and \ion{Si}{4}. Such correlation indicates that these shocks propagate further into the lower corona.

%%------------------------------------------------------
\section{Summary $\&$ Conclusions} \label{sec:sum_con}
%%------------------------------------------------------

In this work, we have identified shocks in the umbra of a sunspot and studied their properties. To do this, we have used IRIS observation of a sunspot in sit-n-stare mode. For this work we have used three spectral lines viz. \ion{Mg}{2}~k~2803.52~{\AA}, \ion{C}{2}~1335.55~{\AA} and \ion{Si}{4}~1393.77~{\AA} lines. Using a 97~minutes long observation sequence, we have identified 71 shocks. To study their counterpart in the lower corona, we have used the corresponding observations taken from AIA 171~{\AA}.

\begin{figure}
\centering
\includegraphics[trim=3.0cm 1.0cm 2.0cm 0.5cm, scale=0.90]{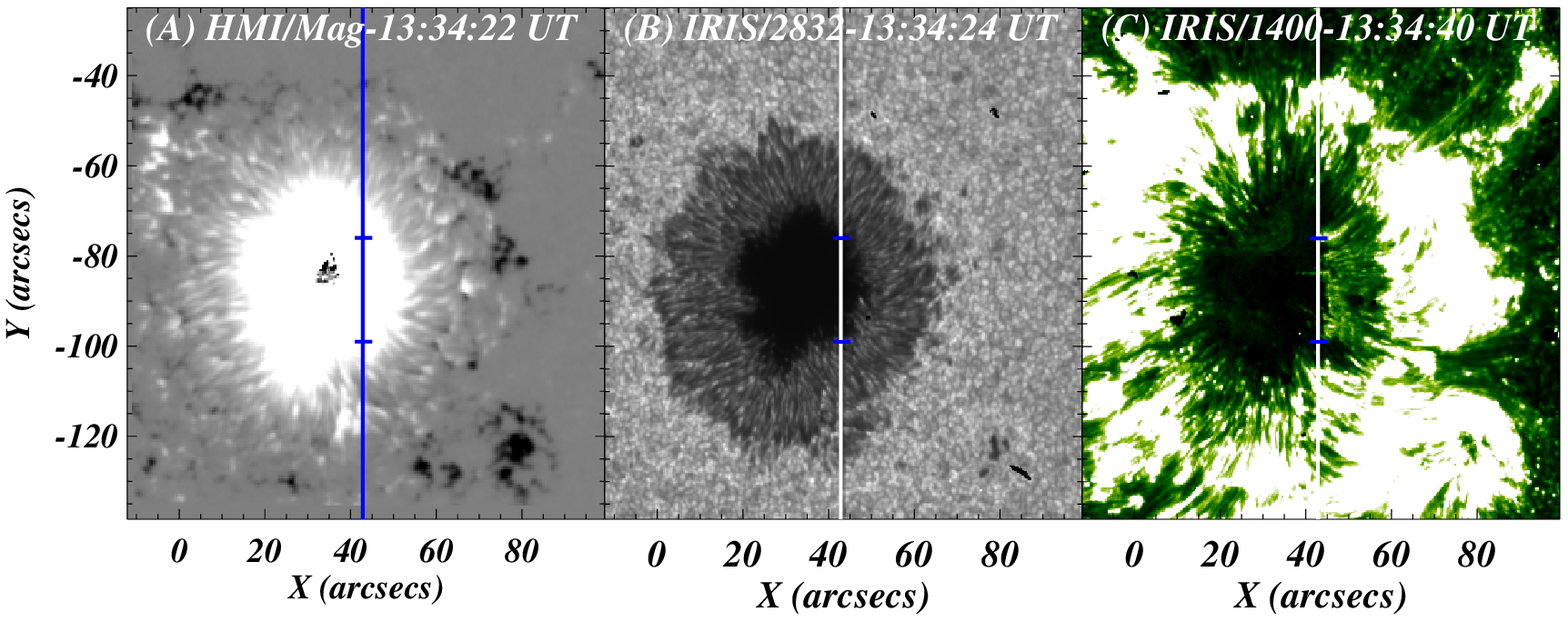}
\includegraphics[trim=4.0cm 1.0cm 4.0cm 4.5cm, scale=0.95]{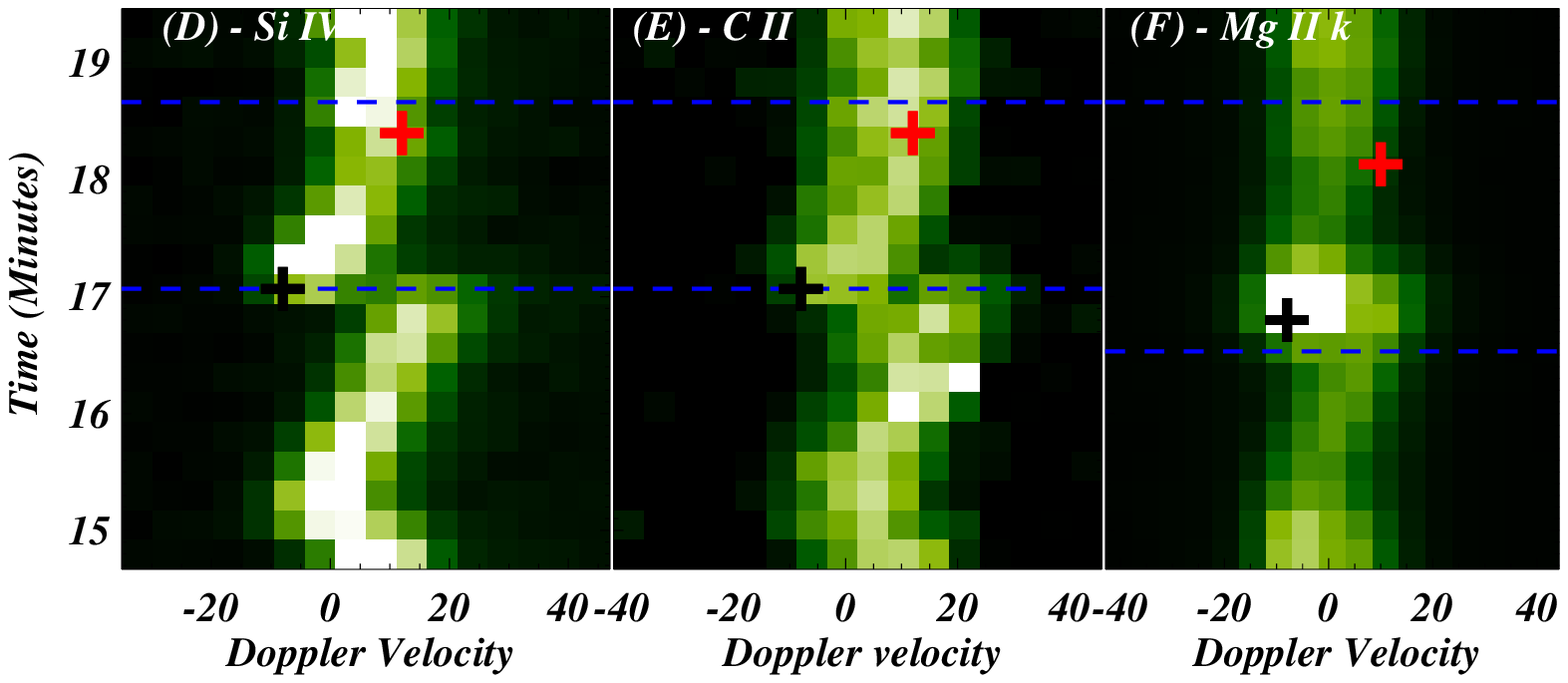}
\includegraphics[trim=2.0cm 5.0cm 2.0cm 4.5cm, scale=0.90]{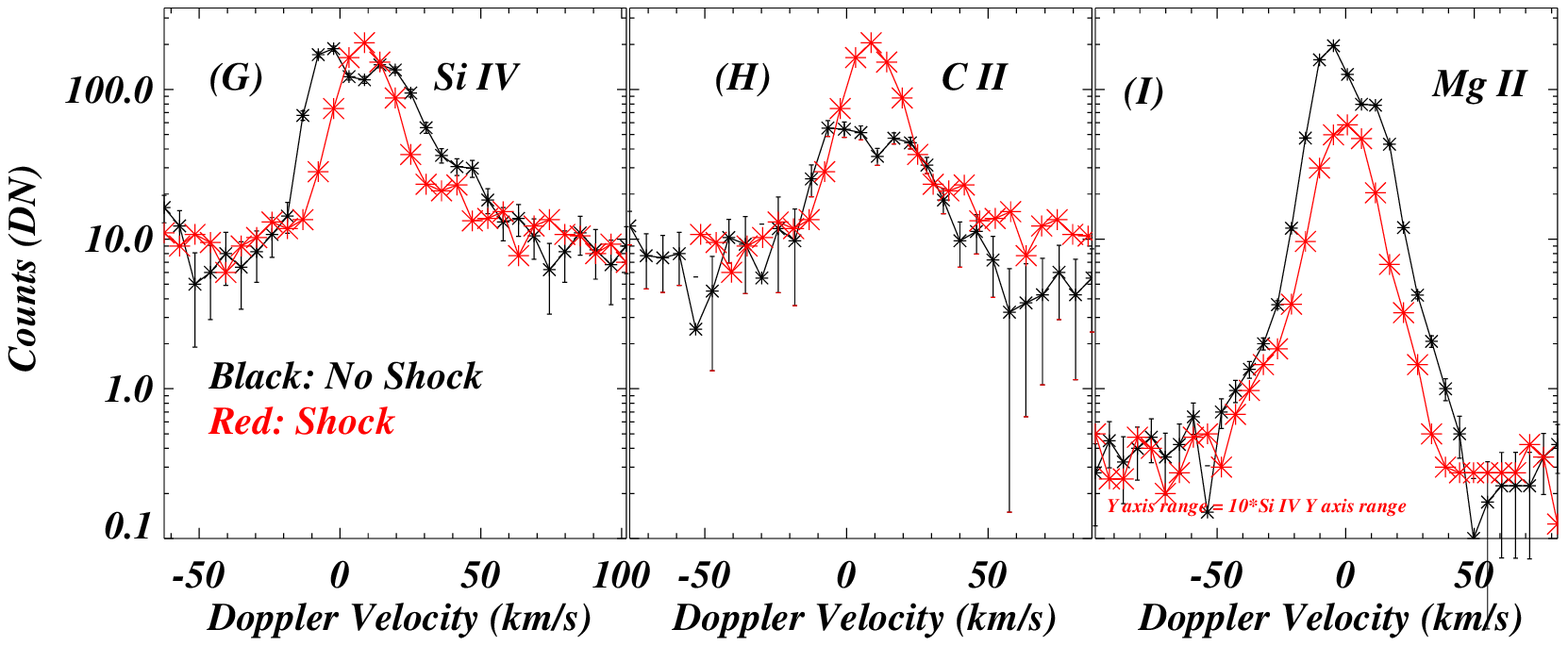}
\caption{HMI Line-of-sight (LOS) magnetogram (panel A) corresponding to the sunspot region observed with IRIS SJI 2382~{\AA} \& 1400~{\AA} channels (panel B \& C). The vertical blue (panel A) and white (panels B \& C) lines locate the IRIS slit. The blue horizontal marks locate the region which is used for further study. Wavelength-time ($\lambda${--}t) diagram and spectral line profiles obtained at Y = -84.31$"$ is shown middle and bottom rows as labeled. The spectral line profiles are obtained at two different times marked by plus symbols in middle row.}\label{fig:ref_fig}
\end{figure}

\begin{figure}
\centering
\mbox{
\includegraphics[trim=0.0cm 0.5cm 1.5cm 0.0cm,scale=0.37]{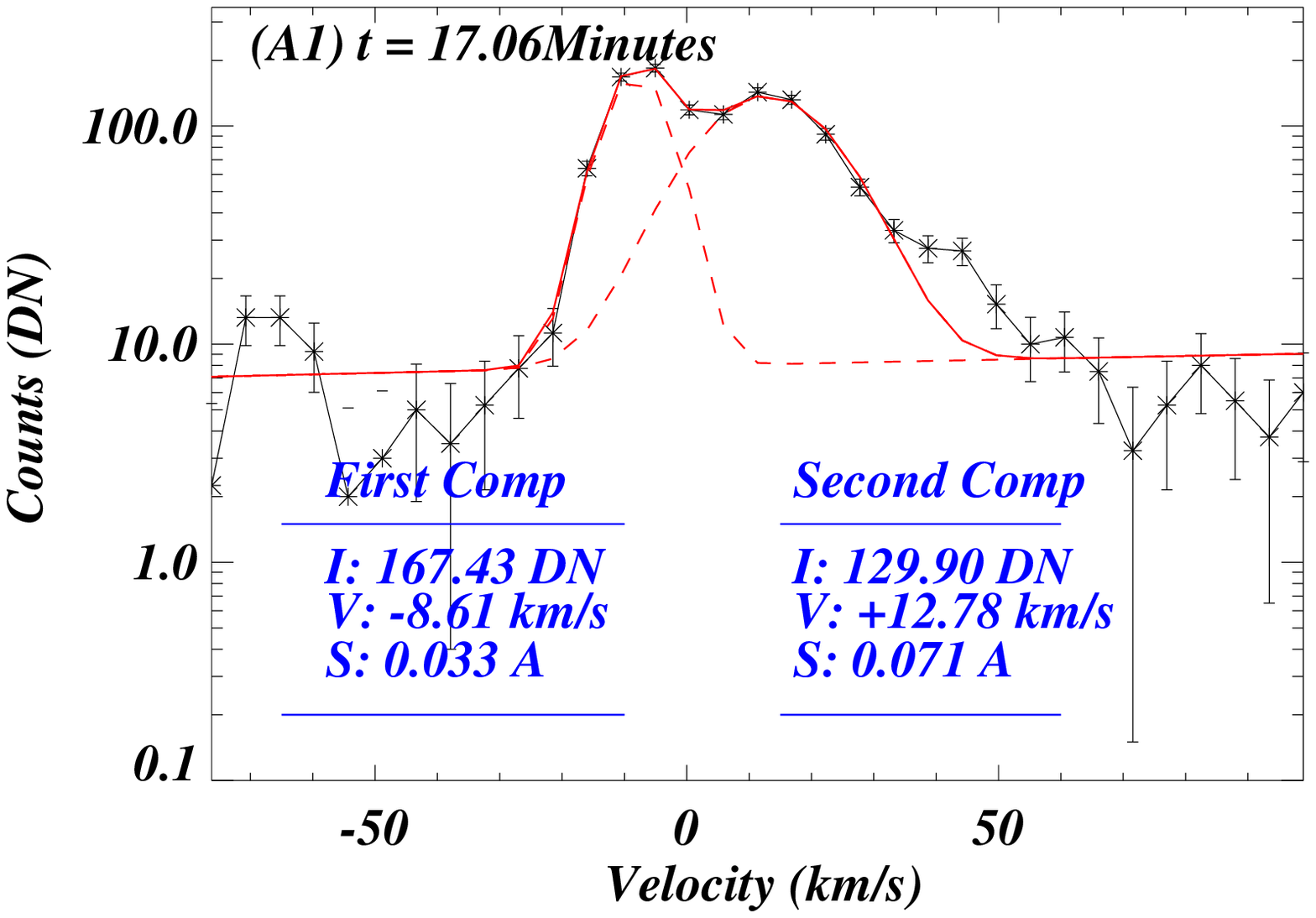}
\includegraphics[trim=3.0cm 0.5cm 1.5cm 0.0cm,scale=0.37]{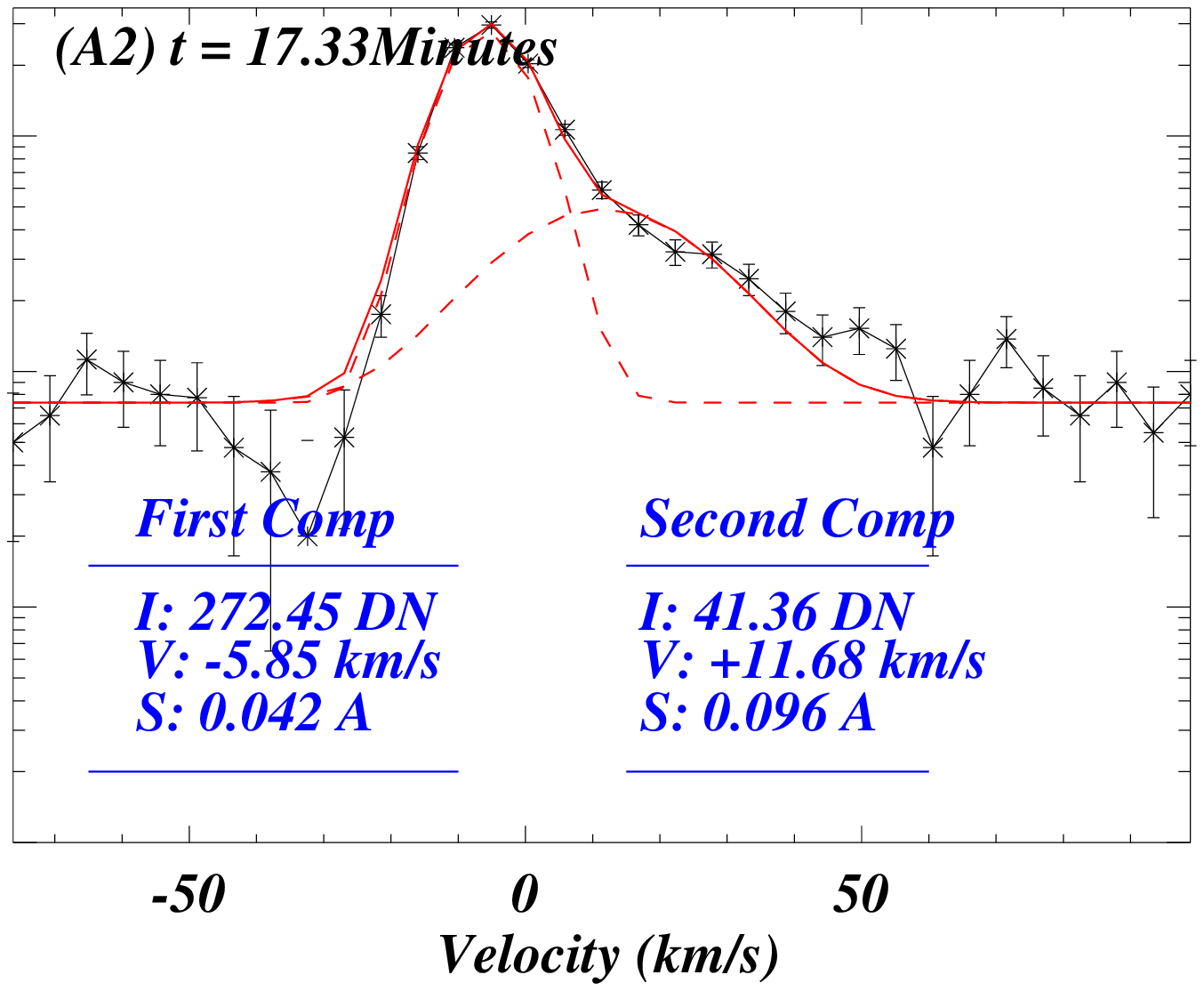}
\includegraphics[trim=3.0cm 0.5cm 1.5cm 0.0cm,scale=0.37]{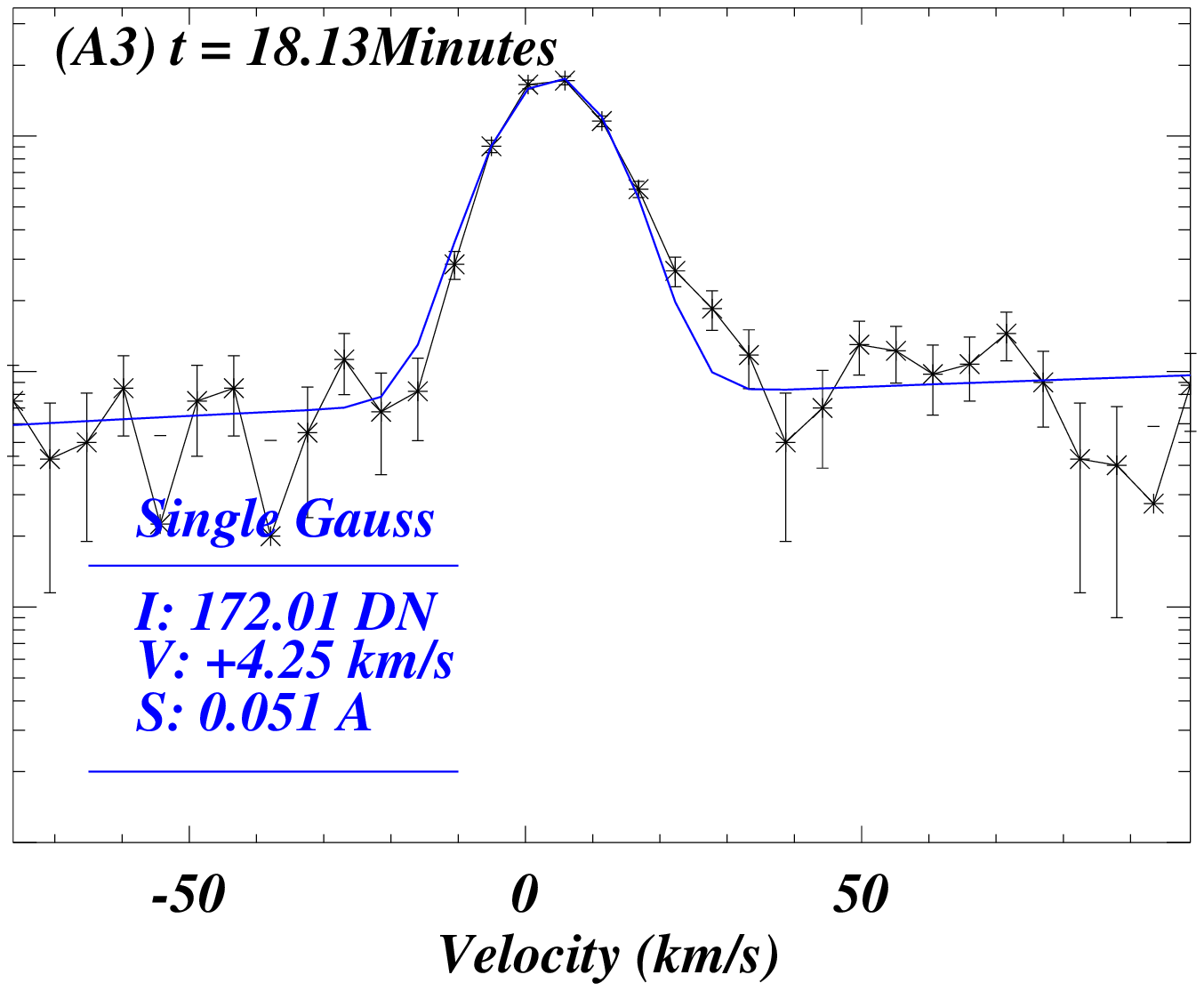}
}
\mbox{
\includegraphics[trim=0.0cm 0.5cm 1.5cm 0.0cm,scale=0.37]{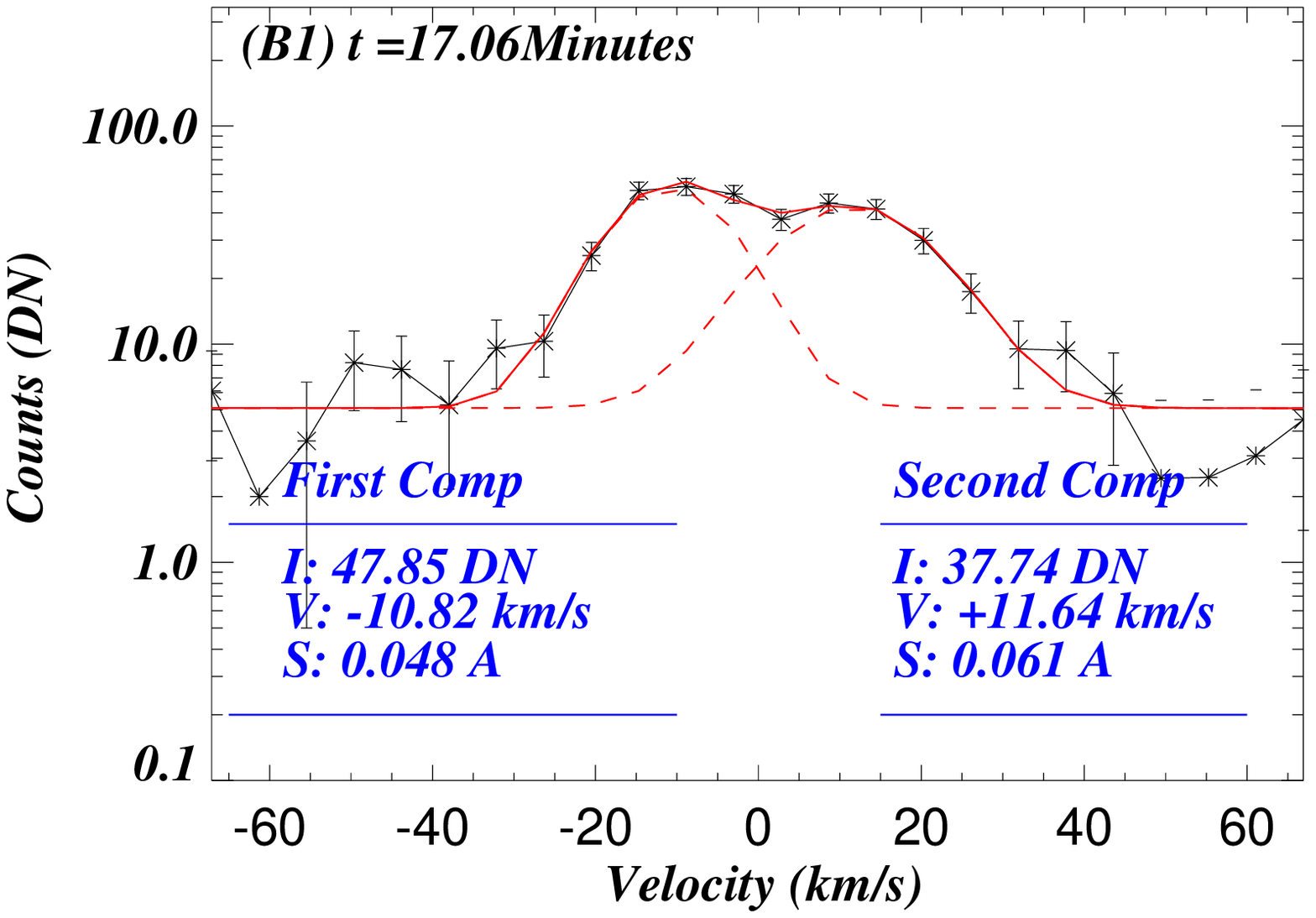}
\includegraphics[trim=3.0cm 0.5cm 1.5cm 0.0cm,scale=0.37]{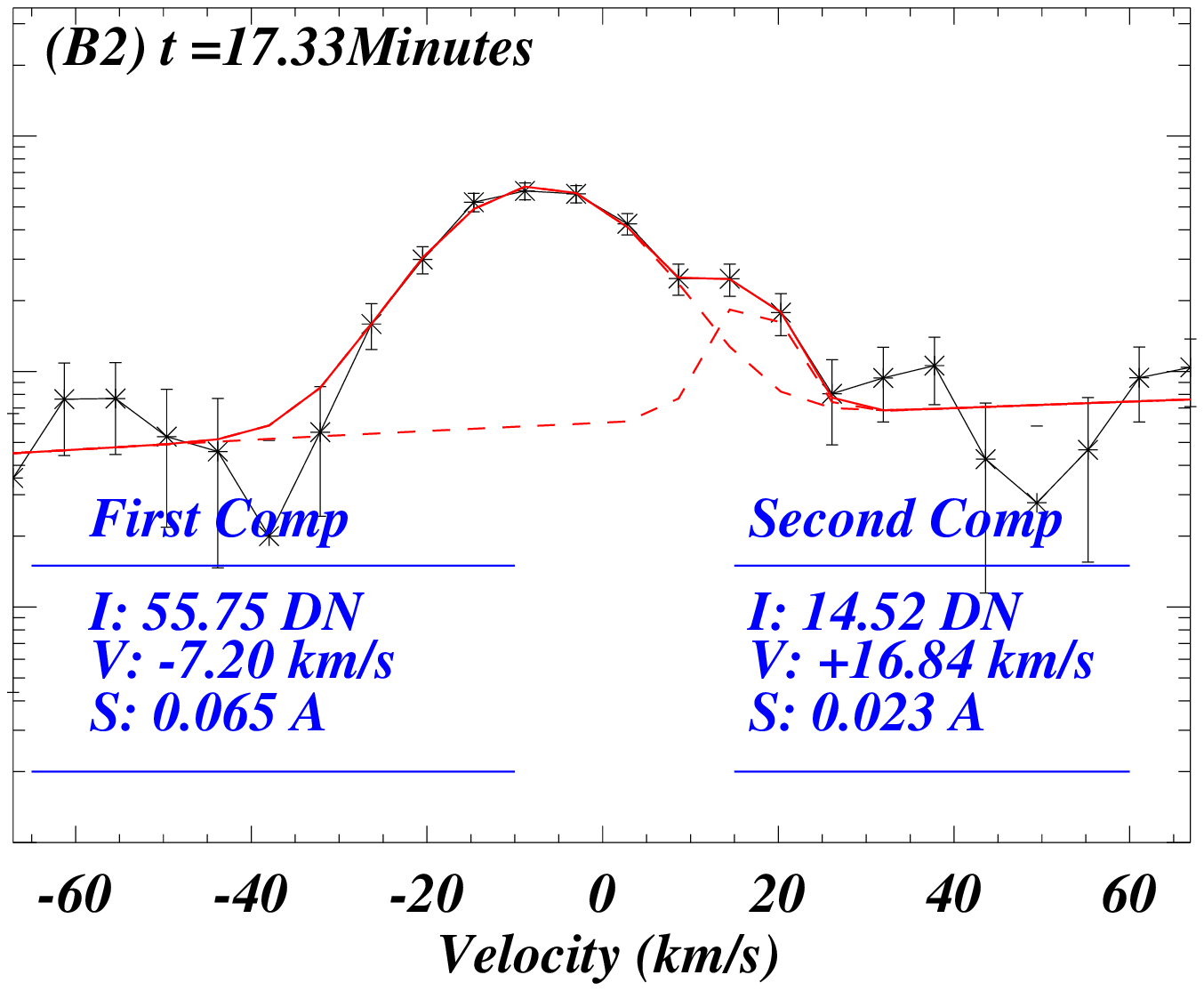}
\includegraphics[trim=3.0cm 0.5cm 1.5cm 0.0cm,scale=0.37]{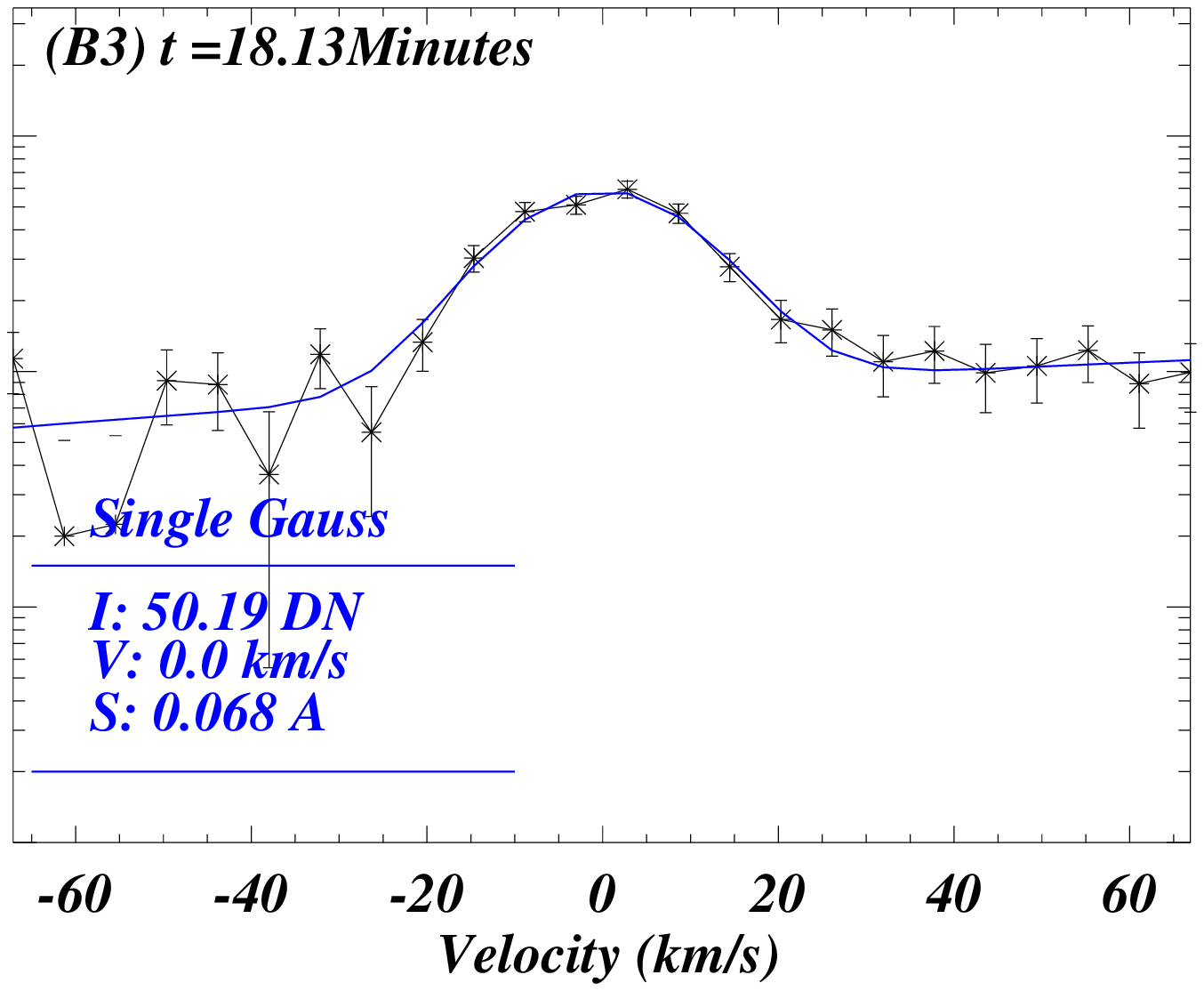}
}
\mbox{
\includegraphics[trim=0.0cm 0.5cm 1.5cm 0.0cm,scale=0.37]{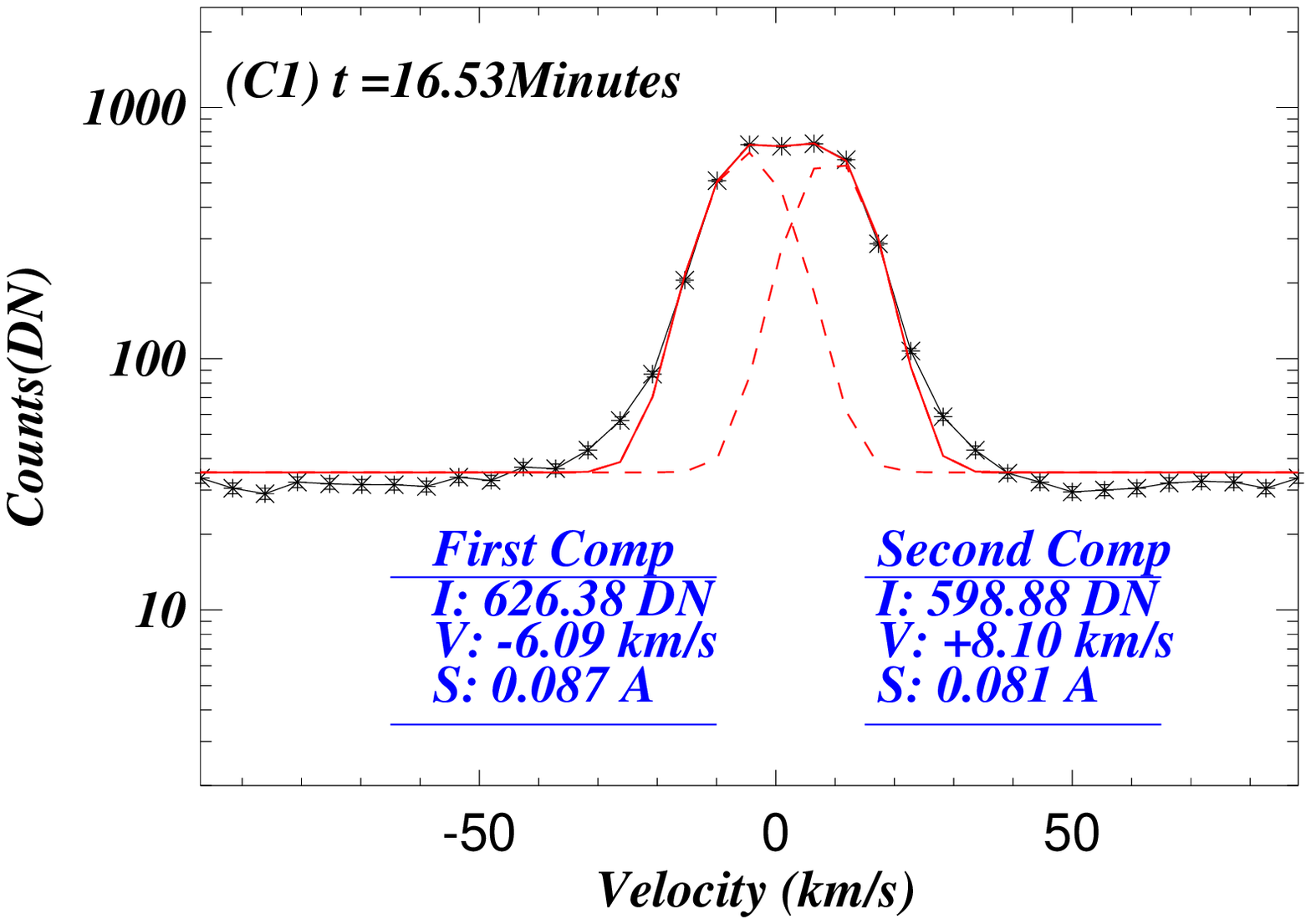}
\includegraphics[trim=3.0cm 0.5cm 1.5cm 0.0cm,scale=0.37]{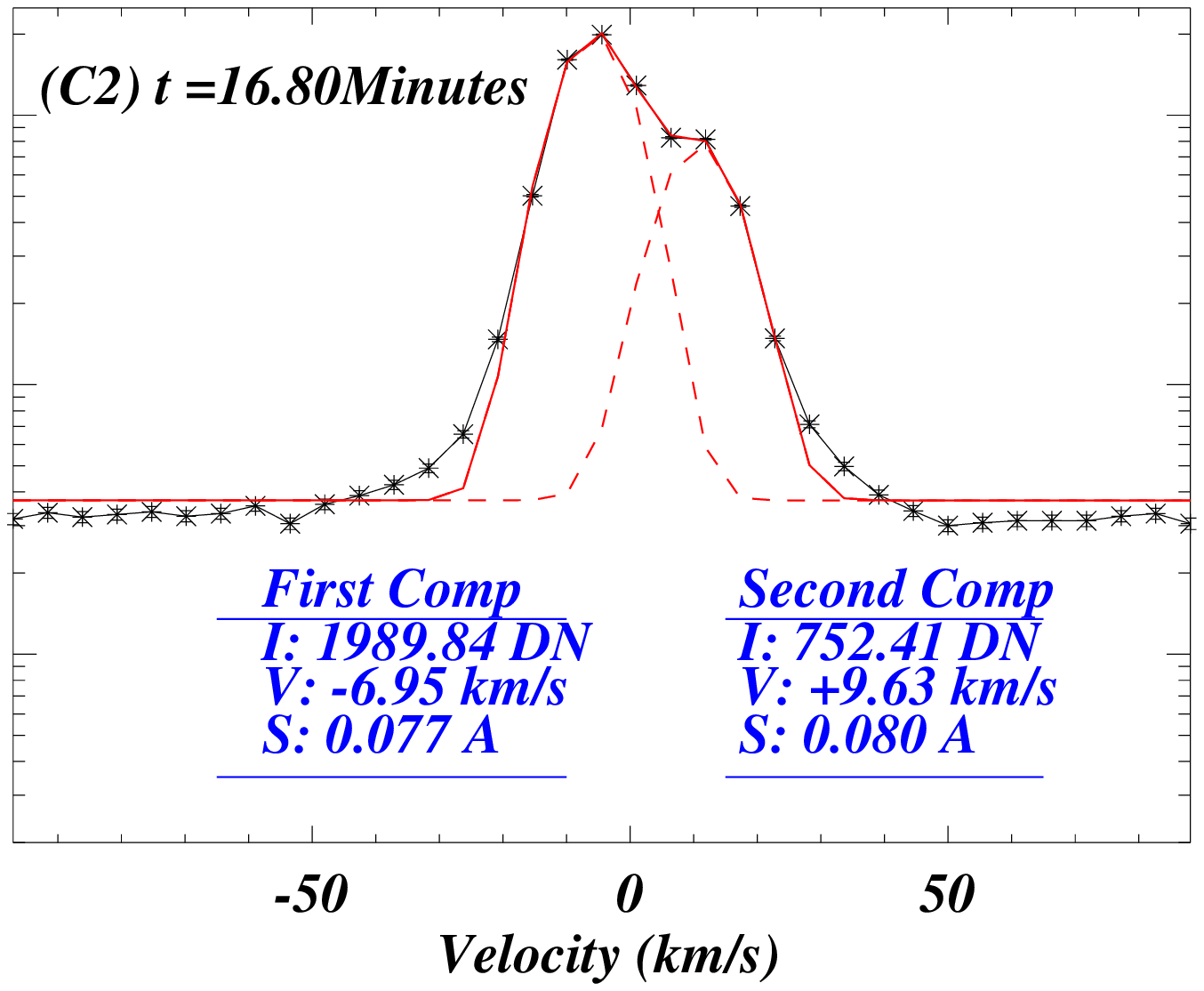}
\includegraphics[trim=3.0cm 0.5cm 1.5cm 0.0cm,scale=0.37]{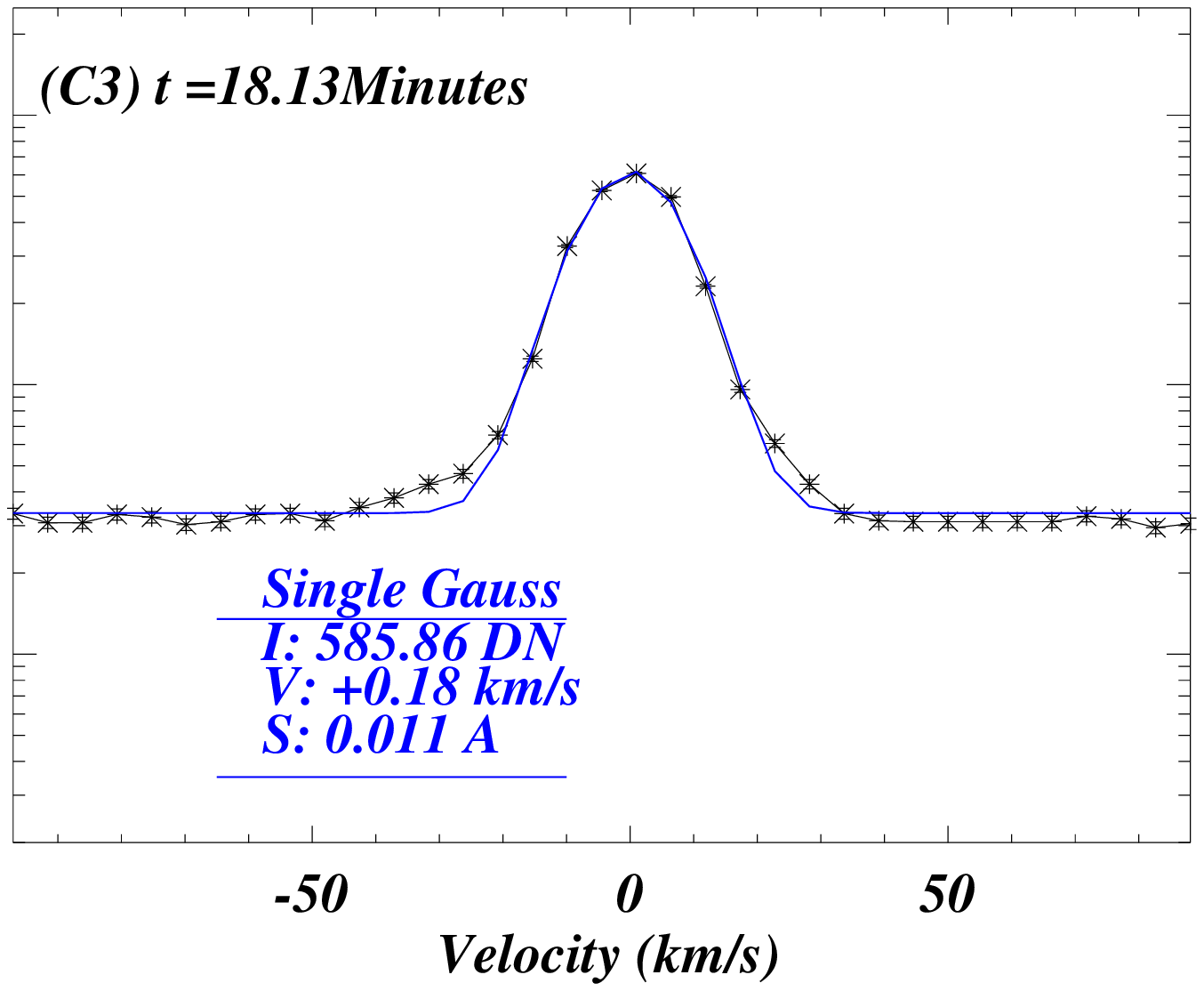}
}
\caption{Time evolution of the spectral line profile obtained at Y = -84.31$"$ for \ion{Si}{4}(top row), \ion{C}{2} (middle row), and Mg~{\sc ii} (bottom row). The black lines correspond to the original profiles, red (blue) curves correspond to double (single) Gaussian fits. The fit parameters for first and second components in case of double Gaussian fit and single component in case of single Gaussian fit is also labelled.}
\label{fig:all_profile}
\end{figure}

We find that the presence of a shock affects the profile of all three spectral lines. At the arrival of the shock, we observe clear double-peaked profiles for the \ion{C}{2} and \ion{Si}{4}. But the \ion{Mg}{2} line shows a flat-top profile that changes to double peak 
(see Figs.~\ref{fig:ref_fig} \&~\ref{fig:all_profile}). With passing time, the contribution from shock (i.e., blue-shifted component) dominates. In the end, all the line profiles change into single peak profiles. We characterized the double-peaked profiles by two Gaussians of different amplitude and width. However, the flat-top profiles of \ion{Mg}{2} was characterized by the two almost identical Gaussians. Thus, we may attribute the appearance of the flat-top profile to the equal contribution from upflows and downflows. Such flat-top profiles in \ion{Mg}{2} lines have also been reported by \cite{Carlsson2015} in plage regions.  \cite{Carlsson2015} have attributed this to the presence of hotter and denser plasma in plage regions compared to the quiet Sun.

Our observations show that the effect of the shocks in \ion{Mg}{2} lasts longer ($\sim$60{--}80.0~s) than in \ion{C}{2} and \ion{Si}{4} ($\sim$30{--}60~s). Moreover, the complete dynamics (i.e., shocks+down flows) lasts longer in \ion{Mg}{2} than \ion{C}{2} and \ion{Si}{4}. We note that the \ion{Mg}{2} line forms at the lower heights than \ion{C}{2} and \ion{Si}{4} and it experiences the shock first. This may be the cause of longer duration present in \ion{Mg}{2} line. It may also be related to the wider range of height of formation of \ion{Mg}{2} in comparison to \ion{C}{2} and \ion{Si}{4} lines \citep{Leen2013}.

\begin{figure}
\mbox{
\includegraphics[trim=1.0cm 0.5cm 1.0cm 0.5cm,scale=0.6]{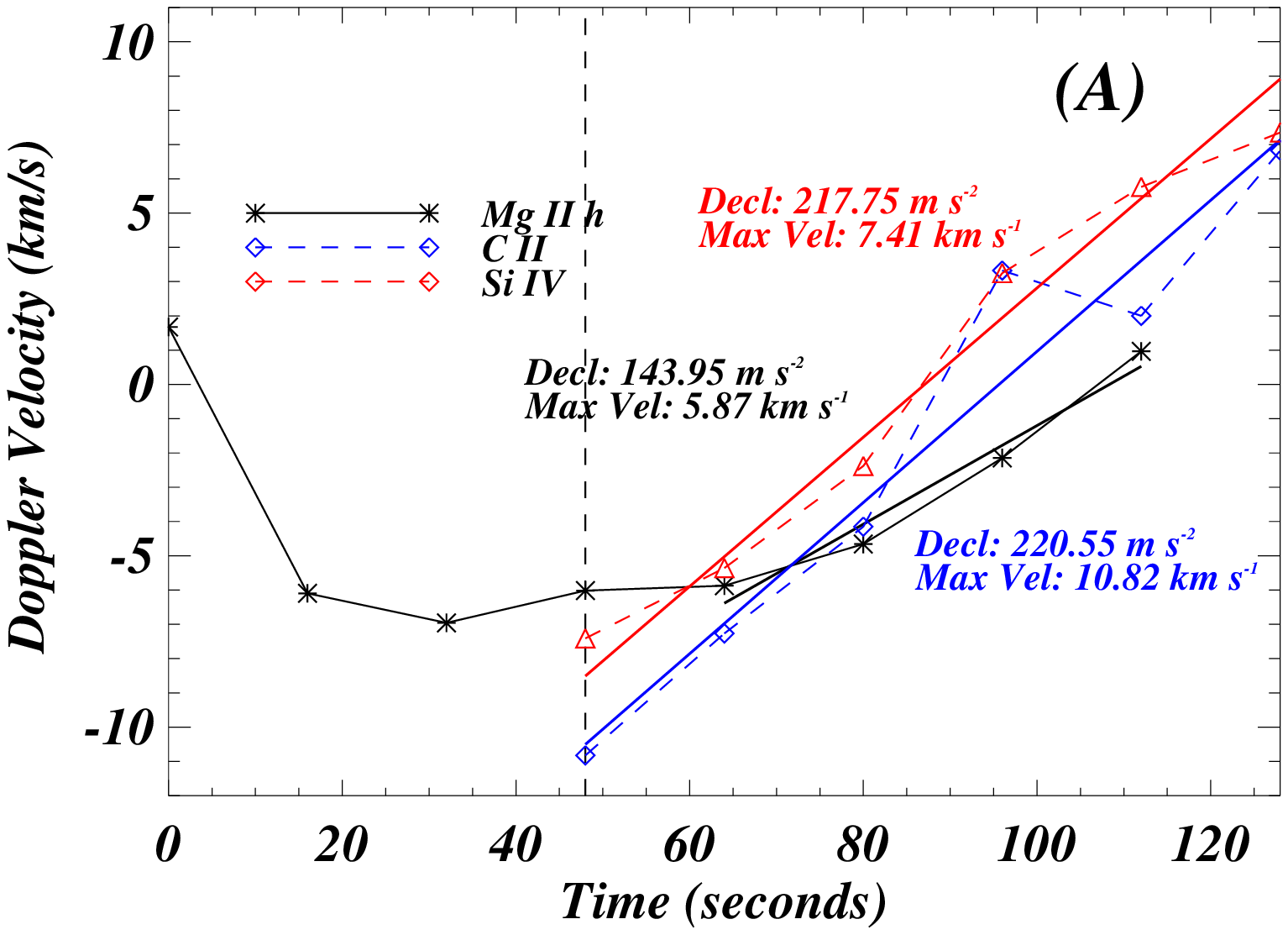}
\includegraphics[trim=0.5cm 0.0cm 1.0cm 1.0cm,scale=0.6]{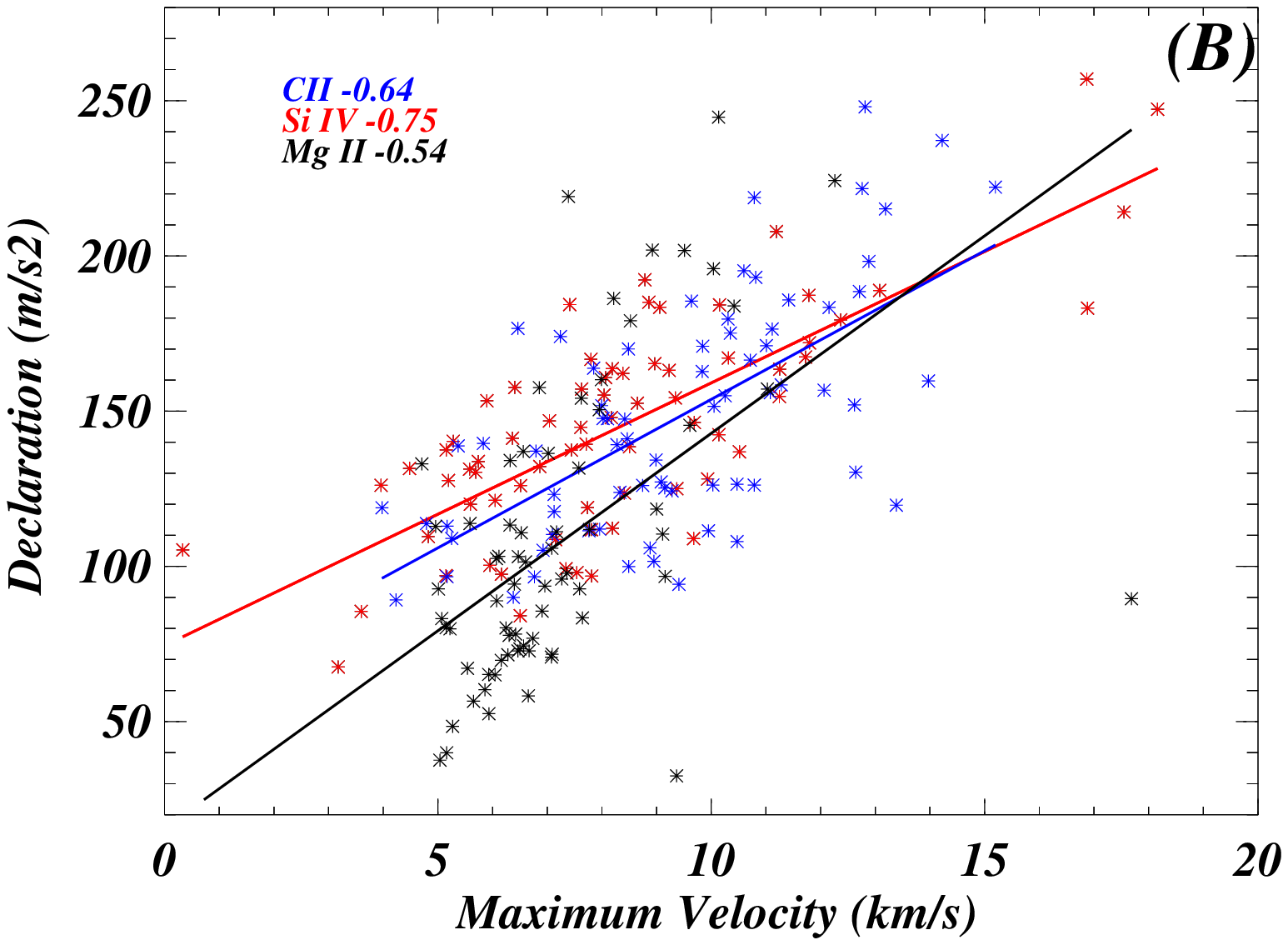}
}
\mbox{
\centering
\includegraphics[trim=-1.0cm 0.5cm -3.5cm 2.5cm,scale=1.0]{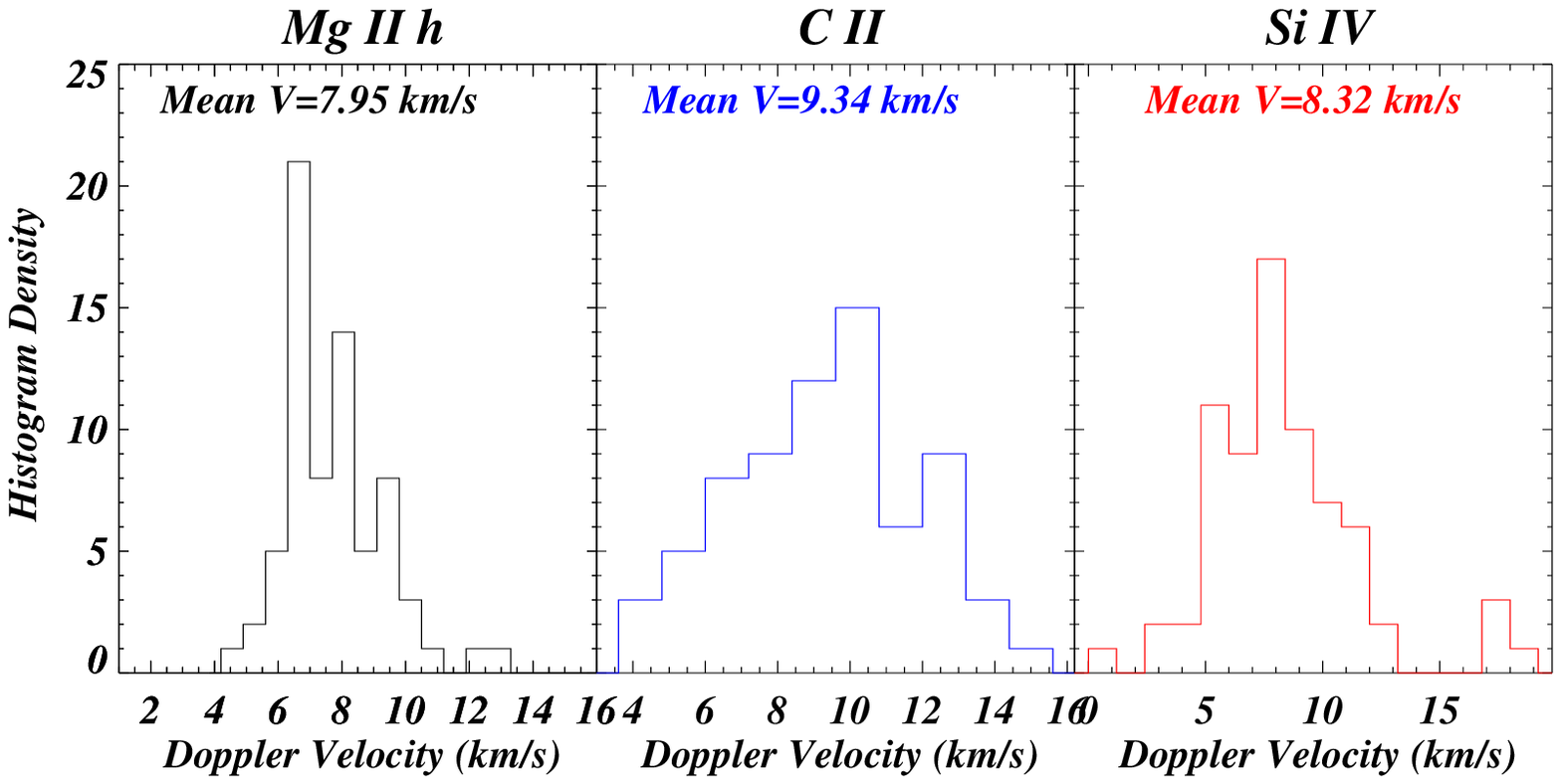}
}
\caption{Panel A: Evolution of Doppler velocity for the shock observed at $Y= -84.31\arcsec$ a function of time. The vertical dashed line marks the starts of shock signature in \ion{Si}{4} and \ion{C}{2}. The solid lines are fits to the data points in corresponding colors. Panel B: Scatter plot between deceleration and highest attained velocity for all the 72 shocks. Straight lines are fit to the corresponding data points. Bottom panels: Distribution of maximum velocities obtained by each shock. In all the panels, red corresponds to \ion{Si}{4}, blue corresponds to \ion{C}{2} and black corresponds to \ion{Mg}{2} lines. \label{fig:shock_acel_main}}
\end{figure}

The \ion{C}{2} and \ion{Si}{4} lines reveal the decrease in blueshifts with time that finally changes into redshifts (see Fig.~\ref{fig:shock_acel_main}.A). Such a pattern has also been reported by, e.g., \cite{voort2003, Centeno2006, Tian2014}. We perform a similar analysis to those by \cite{Tian2014} for \ion{Si}{4} line. We derive the deceleration of shocks in \ion{C}{2}, \ion{Mg}{2} and \ion{Si}{4} using the Doppler velocity-time curve. We observe that the shock decelerate in all the three spectral lines. However, before the deceleration, the \ion{Mg}{2} line also exhibits an acceleration followed by a constant speed preceding. This is an important observation made for the first time. However, the physics of line formation in optically thick plasmas is a complex process. Thus further work is required to make a firm conclusion on the kinematics of these shocks.

We find a positive correlation between the shock's highest velocity and declaration for all the three spectral lines see Fig.~\ref{fig:shock_acel_main}.B). This result is similar to those by \cite{Tian2014} but for only \ion{Si}{4}. The declaration of the shocks depends on the initial velocity. This finding is consistent with those by \citep[see e.g.,][]{voort2003, DePon2009, Tian2014}). 

\begin{figure}
\mbox{
\includegraphics[trim=1.0cm 0.5cm 2.0cm 0.5cm,scale=0.55]{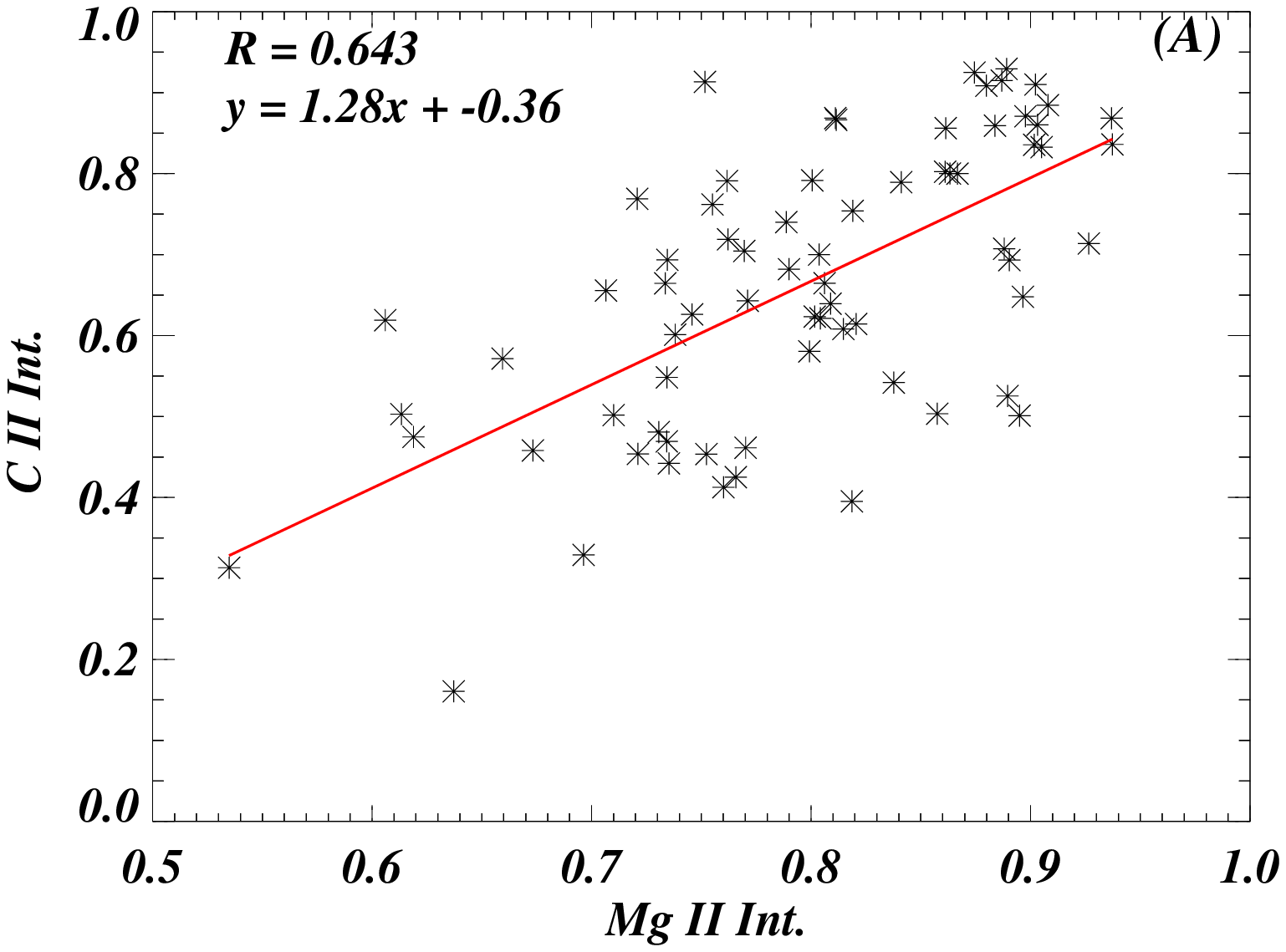}
\includegraphics[trim=0.0cm 0.5cm 1.5cm 0.5cm,scale=0.55]{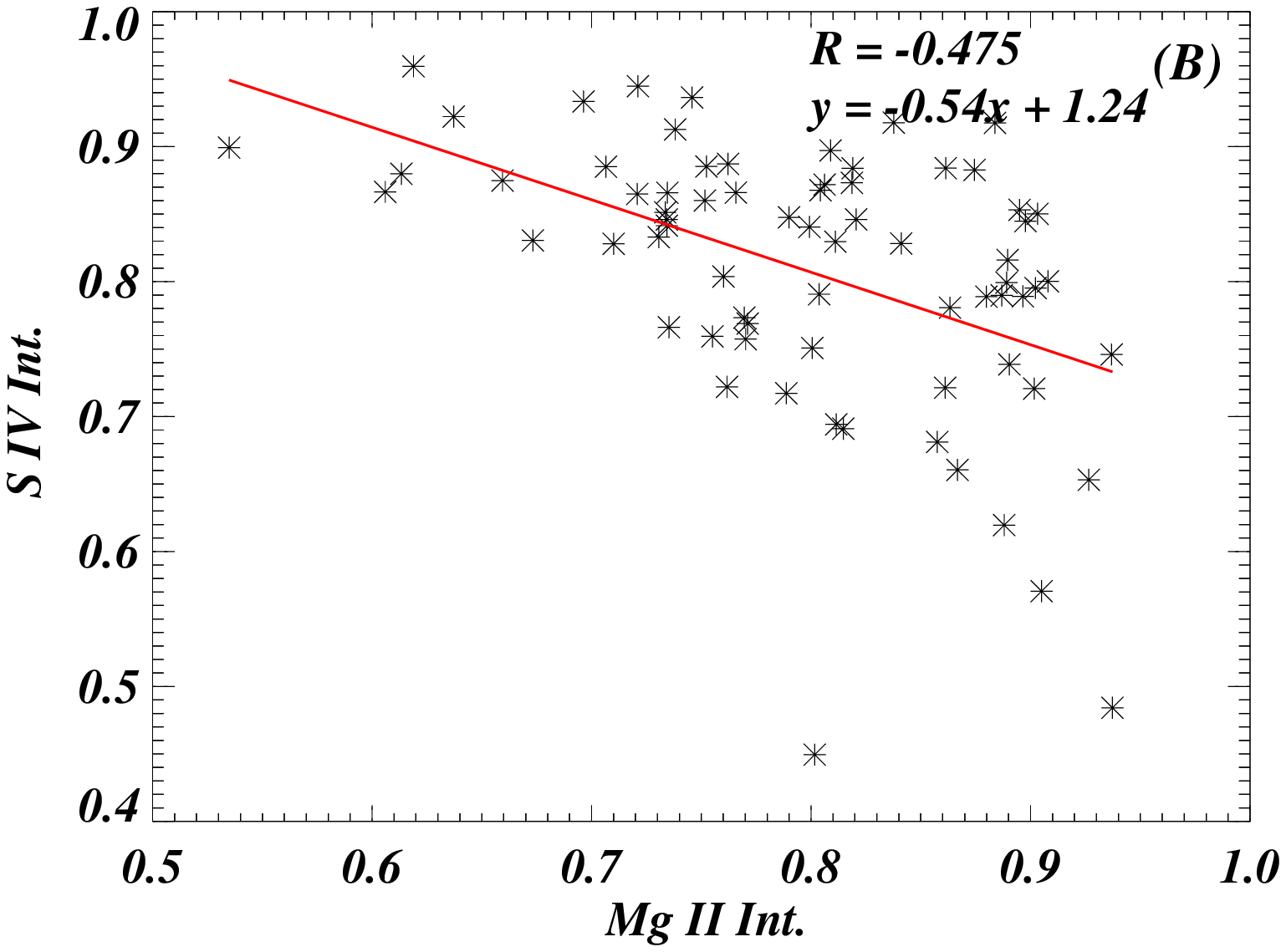}
}
\caption{Panel A: Contribution in \ion{C}{2} line intensities due to shock as a function of that in \ion{Mg}{2}. The solid line is a straight line fit to the data. Panel B: Same as in Panel A but for \ion{Si}{4} and \ion{Mg}{2}.}
\label{fig:shock_correl}
\end{figure}

Intensities contributed due to shock in \ion{Mg}{2} and \ion{C}{2} show positive correlation. But that in \ion{Mg}{2} and \ion{Si}{4} show negative correlation (see Fig.~\ref{fig:shock_correl}). The positive correlation suggests the amplification of shock within the broad chromosphere. This is also supported by the observed highest velocity in \ion{C}{2} (see Fig.~\ref{fig:shock_acel_main}.C). But, the negative correlation implies a lower amplitude of shock in the transition region. This may point towards observational evidence of the dissipation of shocks in the transition region. But, we note that the above findings involve optically thick lines of \ion{Mg}{2} and \ion{C}{2}. Thus, further work is required to establish this result. 

\begin{figure}
\centering
\mbox{
\includegraphics[trim=3.0cm 3.0cm 2.0cm 3.0cm,scale=1.0]{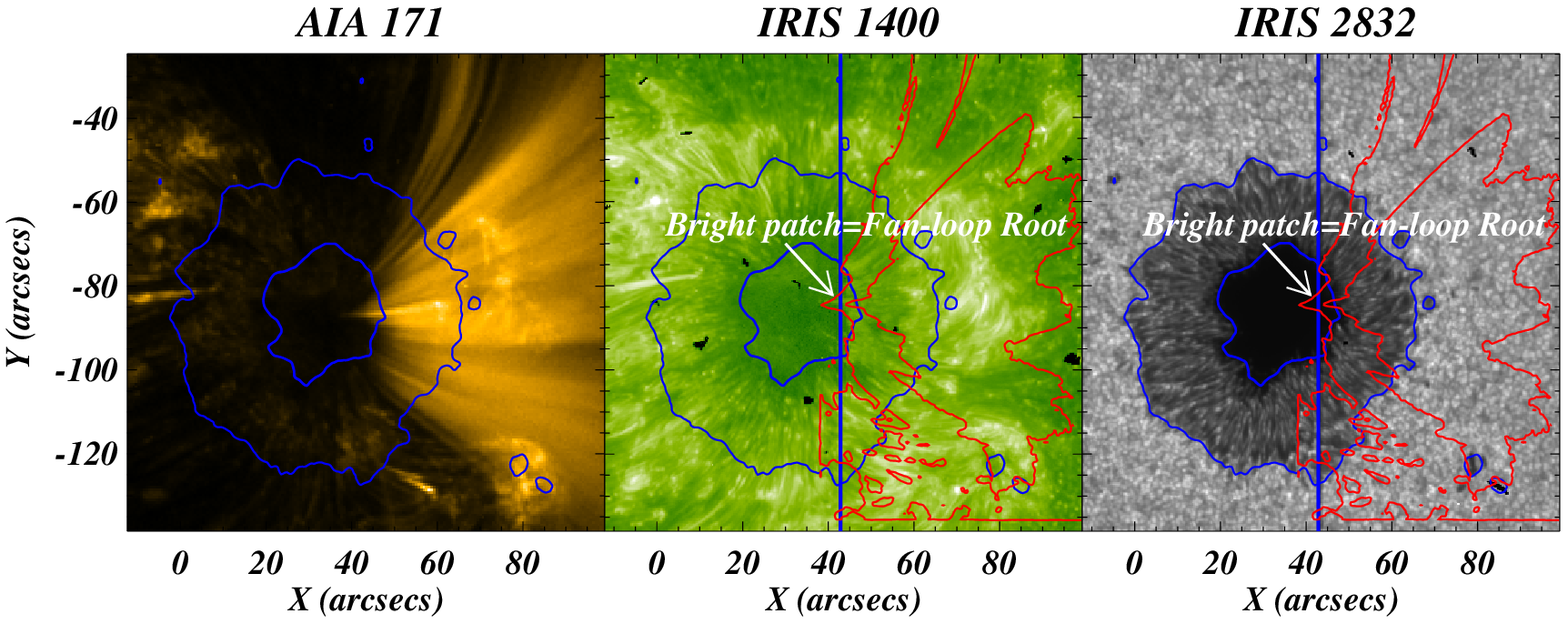}
}
\mbox{
\includegraphics[trim=3.0cm 4.0cm 2.0cm 3.0cm,scale=1.5]{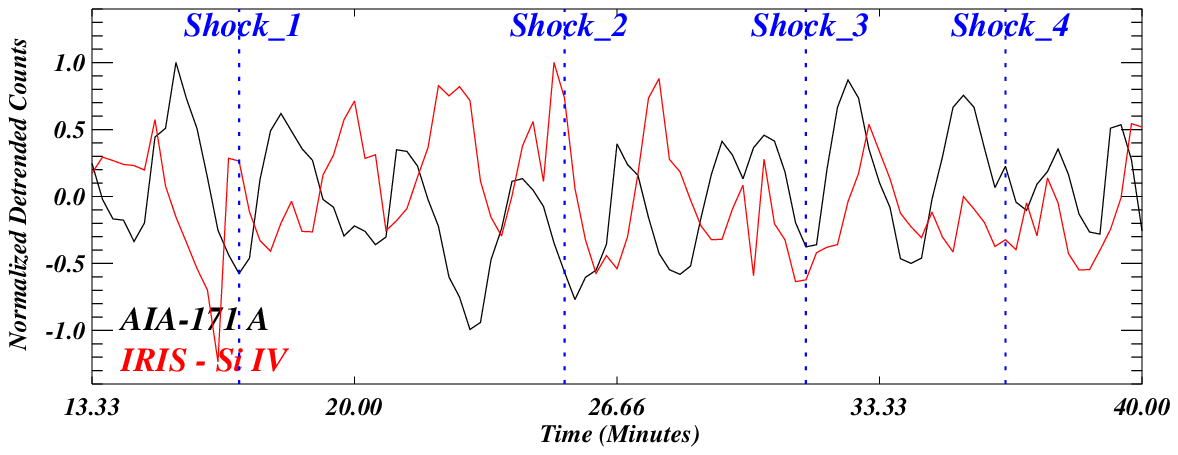}
}
\caption{The sunspot region observed with IRIS 2832~{\AA} (left) and IRIS 1400~{\AA} (middle) and the corresponding AIA 171~{\AA} image (right panel) showing fan loops. The blue (white) contours in left and middle (right) panel show sunspot umbra and penumbra. The red contous are AIA 171~{\AA} intensity showing fan loops.}
\label{fig:fan_loop}
\end{figure}

Finally, we observed that several fan loops are rooted within the bright-region in the umbra \citep[cf., Fig.~\ref{fig:fan_loop}; see also the results from][]{Sharma2017, Ghosh2017}. A qualitative comparison between AIA~171~{\AA} and \ion{Si}{4} light curves show a strong correlation. Such correlations suggest these shocks propagate further up into the lower corona and may play an essential role in the dynamics of fan loops.

%%---------------------------------------------------
\begin{acknowledgements} 
We thank the referee for constructive comments that helped improve the manuscript. PK thanks University of South Bohemia, \v{C}esk\'e Bud\v{e}jovice, Czech Republic for supporting this research. This work is partly supported by the Max-Planck Partner Group of MPS on Coupling and Dynamics of the Solar Atmosphere. IRIS is a NASA small explorer mission developed and operated by LMSAL with mission operations executed at NASA Ames Research Center and major contributions to downlink communications funded by ESA and the Norwegian Space Center. SDO observations are courtesy of NASA/SDO and the AIA, EVE, and HMI science teams. CHIANTI is a collaborative project involving George Mason University, the University of Michigan (USA), University of Cambridge (UK) and NASA Goddard Space Flight Center (USA).
\end{acknowledgements}
%%---------------------------------------------------
%%---------------------------------------------------

%%---------------------------------------------------
%%---------------------------------------------------
%----------------------------------------------------
%----------------------------------------------------
%----------------------------------------------------
%----------------------------------------------------
%\includegraphics[trim=0.5cm 0.0cm 1.5cm 1.0cm,scale=0.5]{shock_vel_decal.eps}
%%-------------------------------------------------
%%-------------------------------------------------
%%-------------------------------------------------
%%-------------------------------------------------
%%-------------------------------------------------
%%-------------------------------------------------
\appendix \label{append}
\renewcommand\thefigure{\thesection.\arabic{figure}}    
\section{Spectral Profiles Examples{--} Shock's Dynamics}\label{append:profile_evol}
The effects of spectral lines from one shock are displayed in Fig~\ref{fig:ref_fig}. Now, here we have shown the effects on spectral lines from two more shocks which are shown in Fig.~\ref{fig:append_one} and Fig.~\ref{fig:append_two}. Overall, we can say that the temporal behavior of spectral lines is very similar as we described earlier. In Fig.~\ref{fig:shock_acel}, we have Doppler velocity-time profiles from all three lines as we showed earlier in the main text. The behavior qualitatively matches the description above provided.   
%%-------------------------------------------------
\setcounter{figure}{0} 
%%-------------------------------------------------
\begin{figure}
\centering
\mbox{
\includegraphics[trim=6.0cm 0.5cm 1.5cm 0.0cm,scale=0.5]{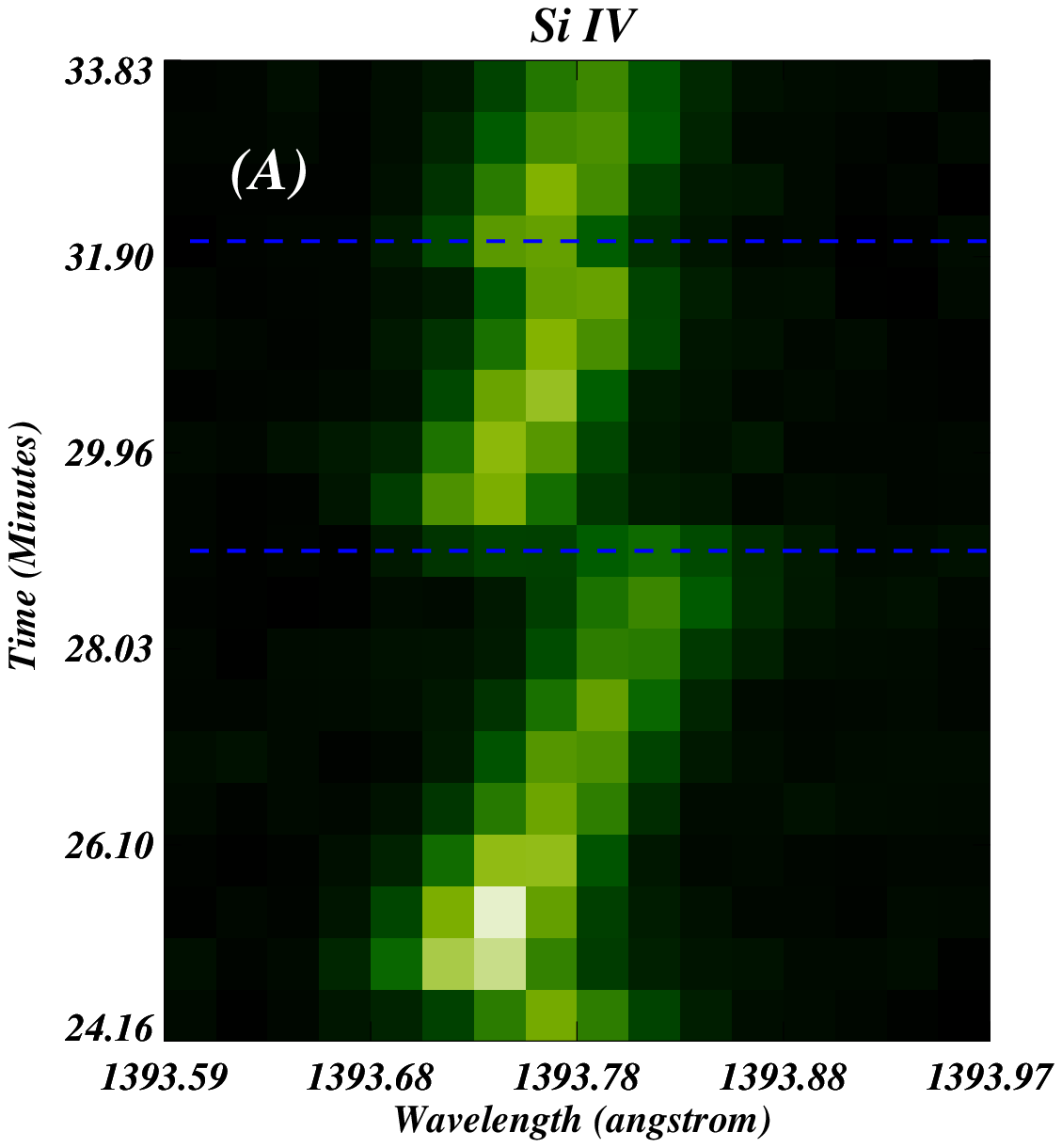}
\includegraphics[trim=6.0cm 0.5cm 1.5cm 0.0cm,scale=0.5]{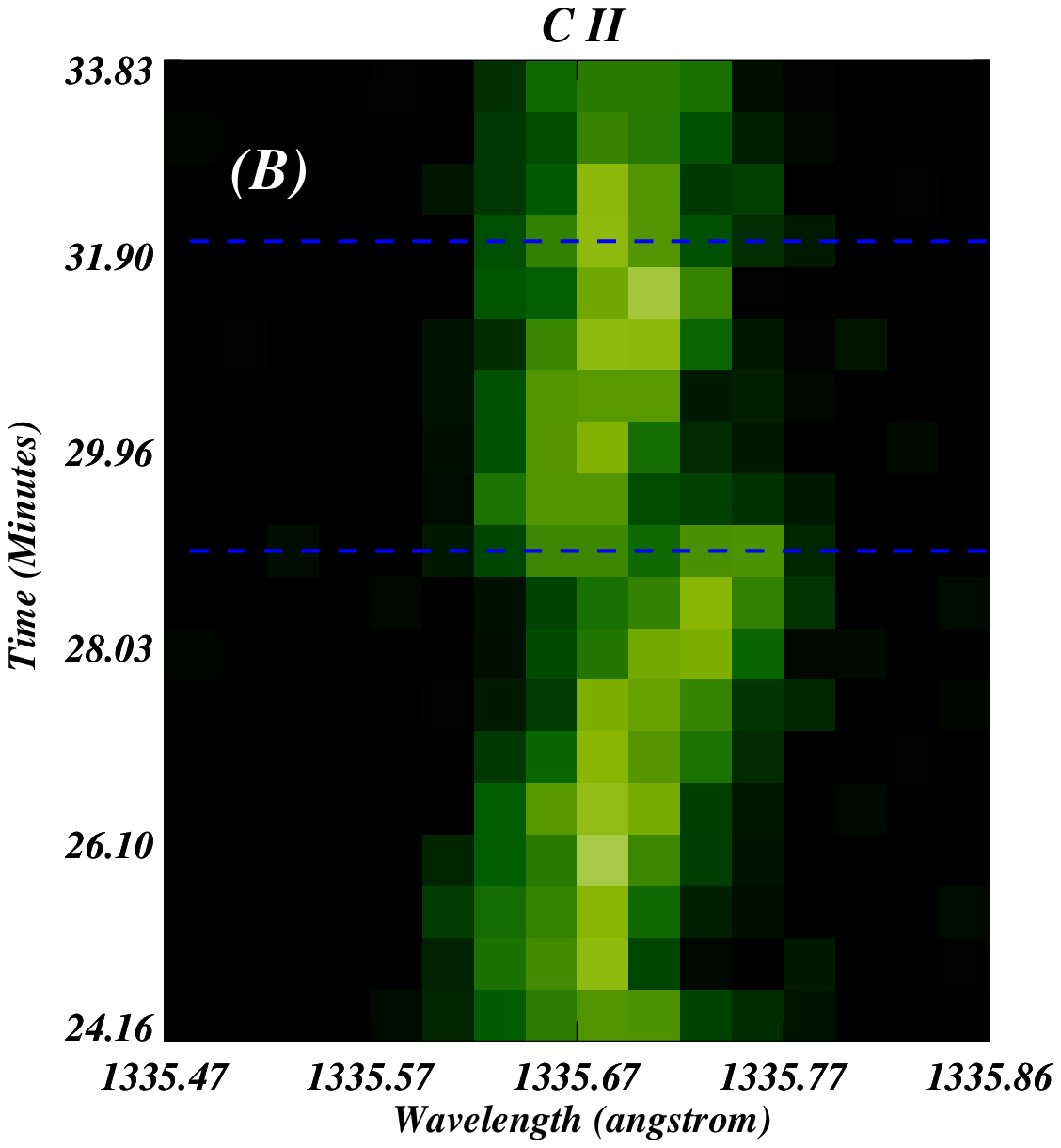}
\includegraphics[trim=6.0cm 0.5cm 1.5cm 0.0cm,scale=0.5]{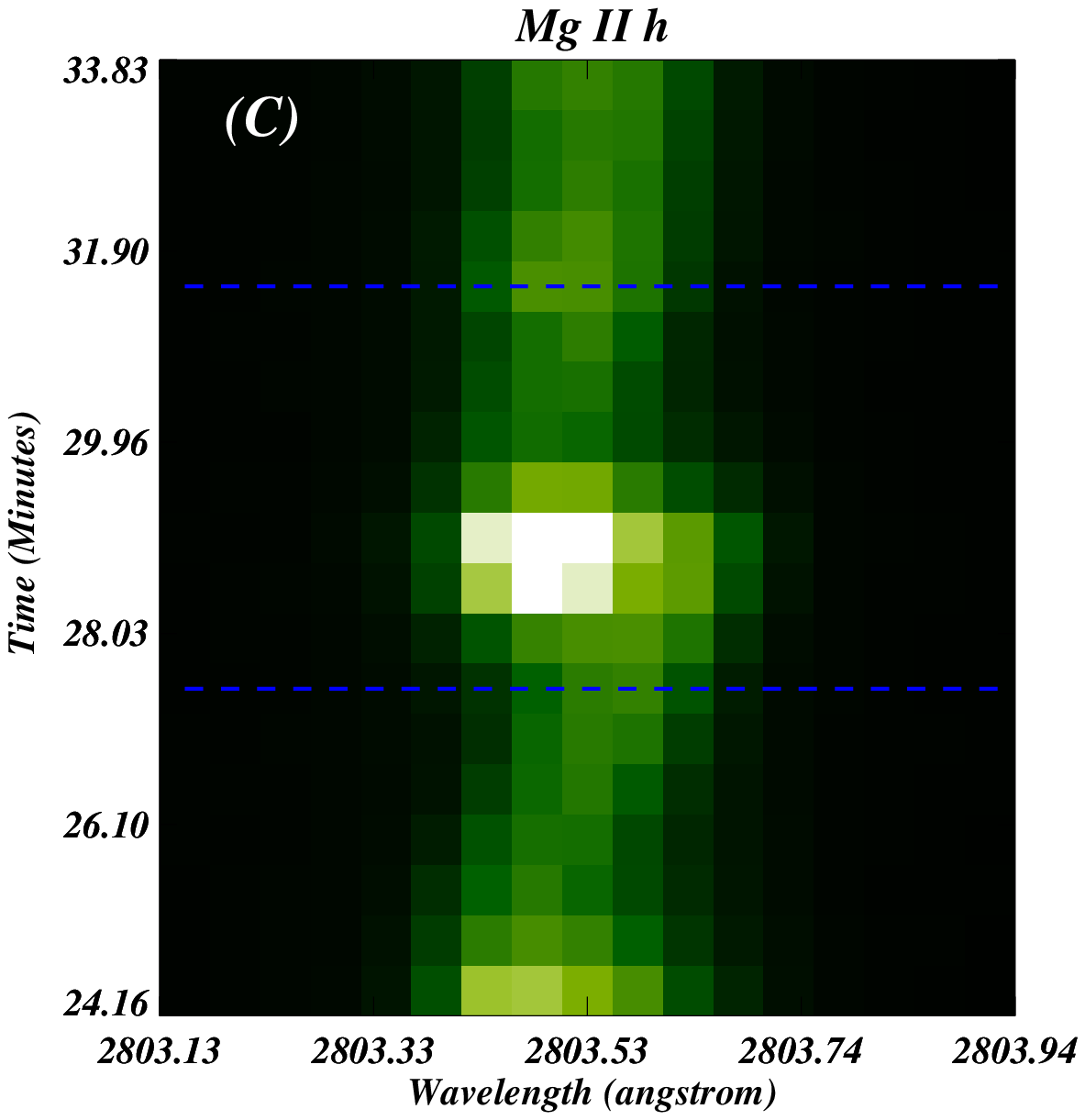}
}
%%-------------------------------------------------
\mbox{
\includegraphics[trim=5.0cm 0.5cm 1.5cm 0.0cm,scale=0.37]{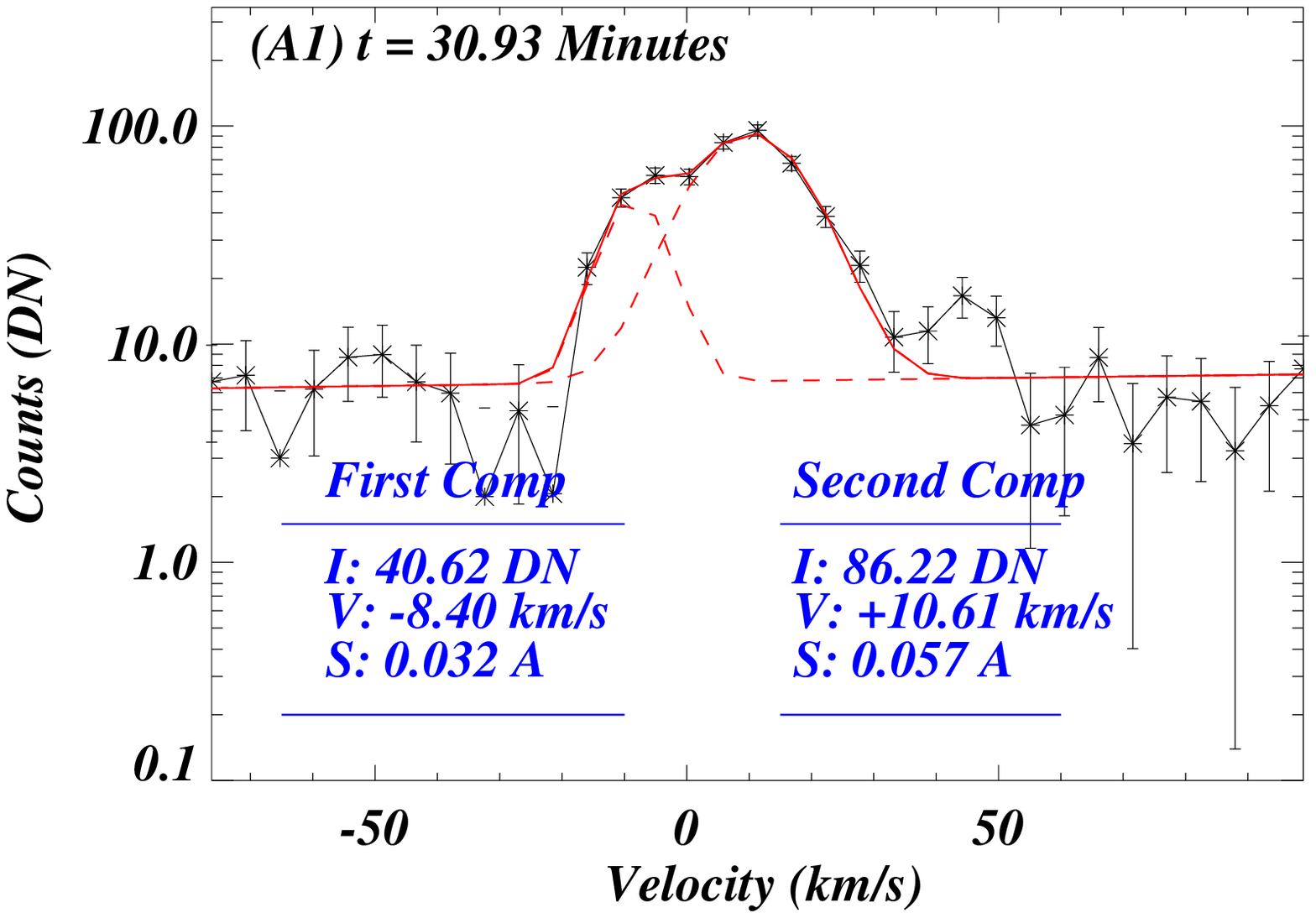}
\includegraphics[trim=3.0cm 0.5cm 1.5cm 0.0cm,scale=0.37]{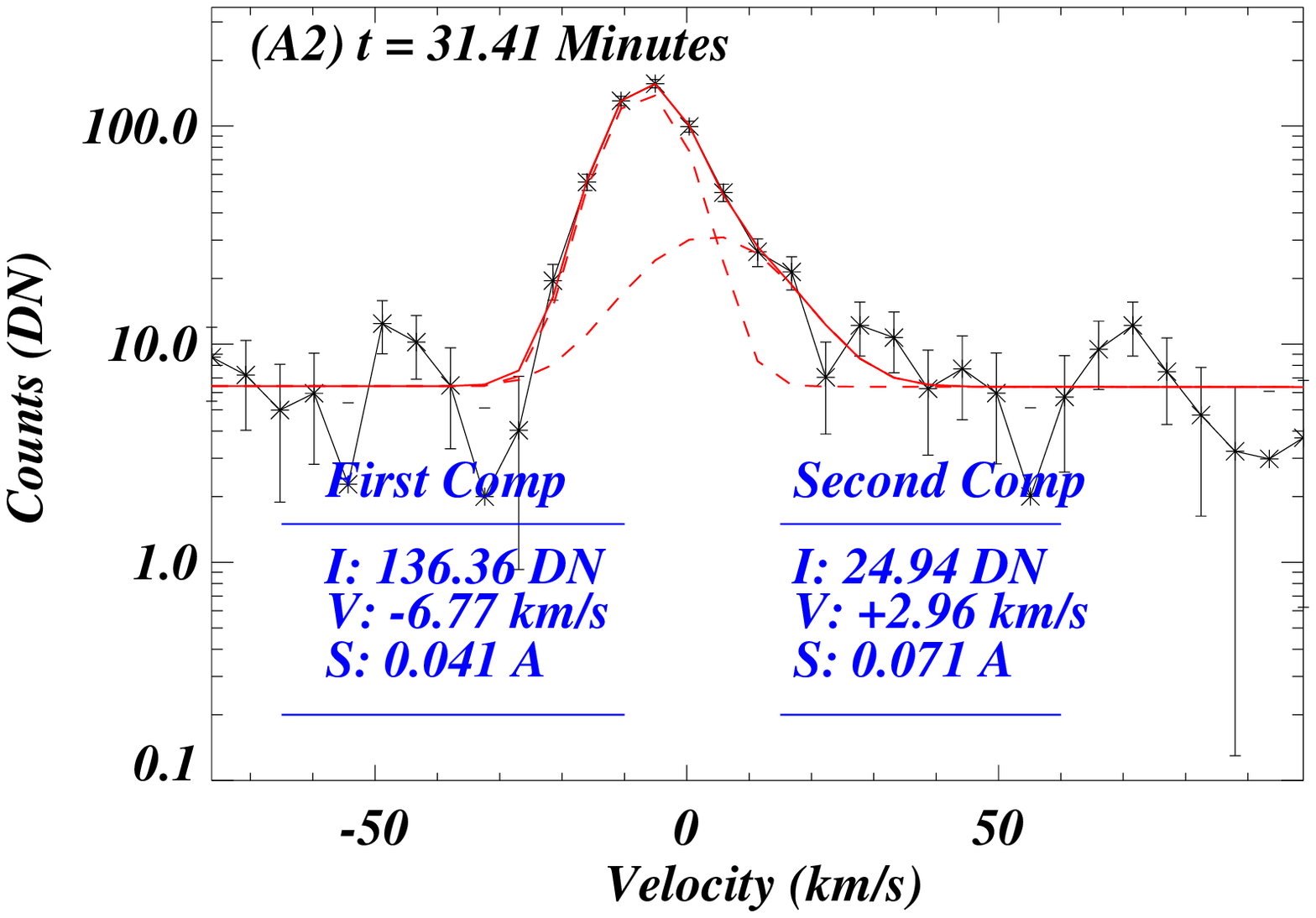}
\includegraphics[trim=3.0cm 0.5cm 1.5cm 0.0cm,scale=0.37]{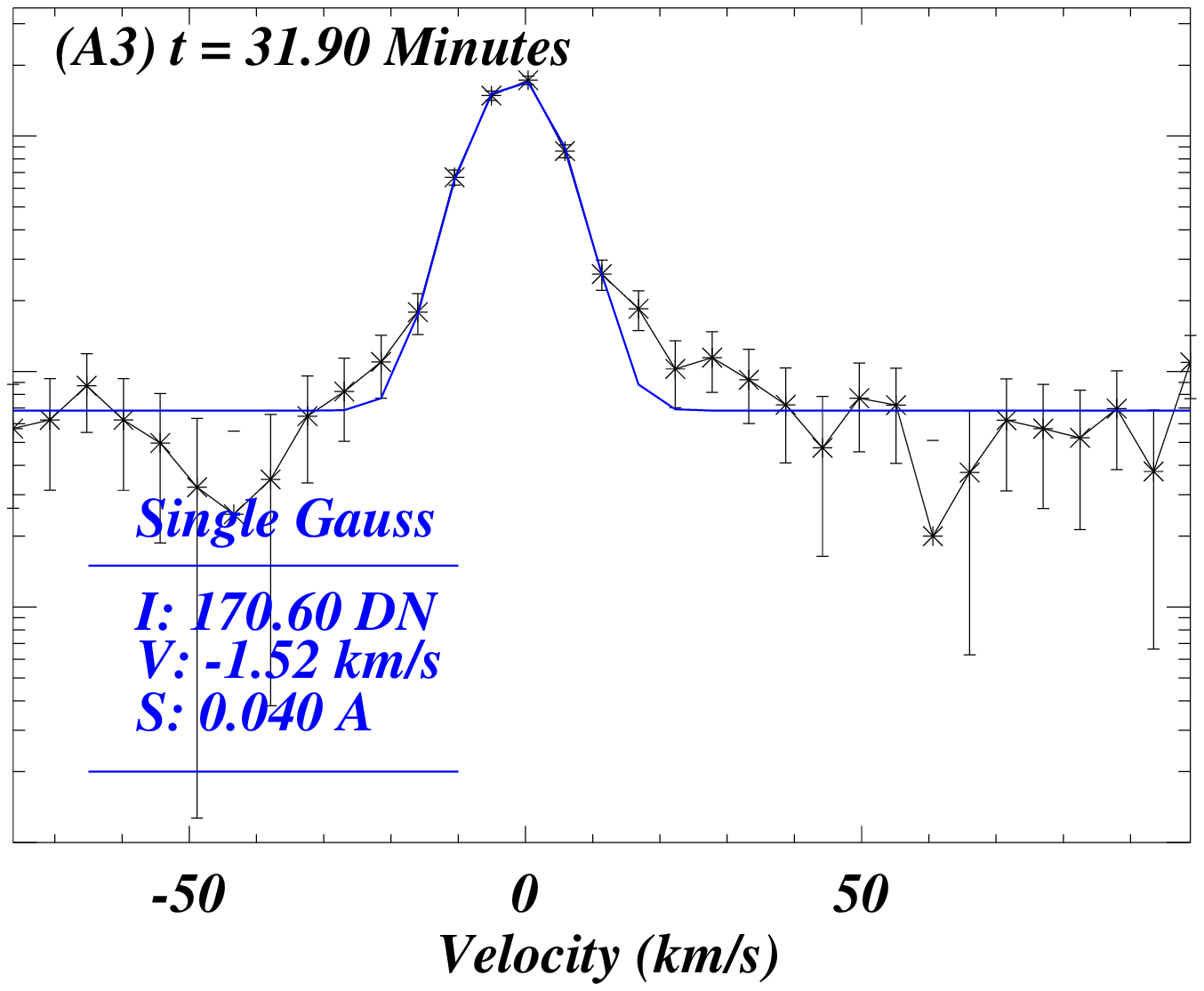}
\includegraphics[trim=3.0cm 0.5cm 1.5cm 0.0cm,scale=0.37]{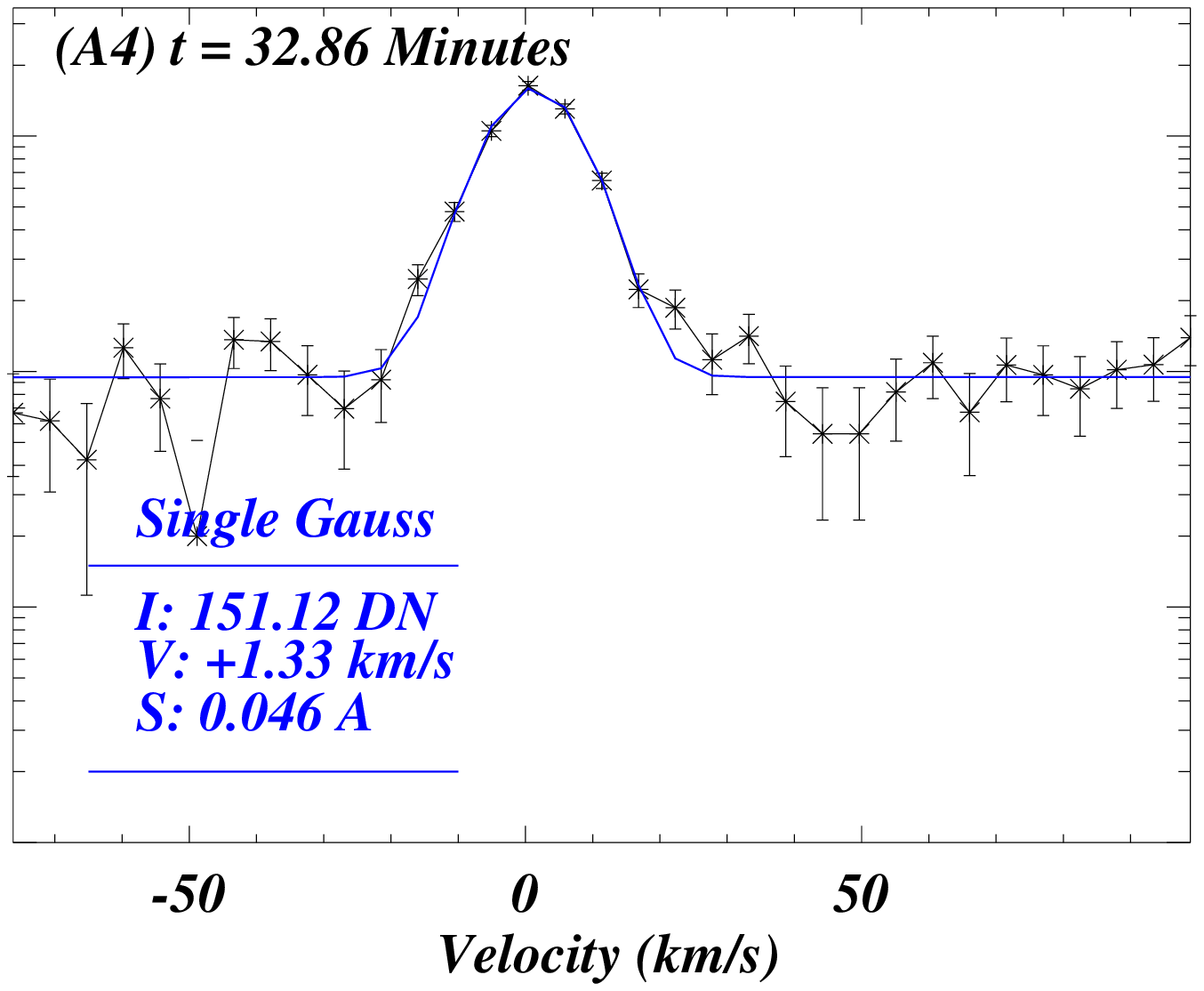}
}
%%-------------------------------------------------
\mbox{
\includegraphics[trim=5.0cm 0.5cm 1.5cm 0.0cm,scale=0.37]{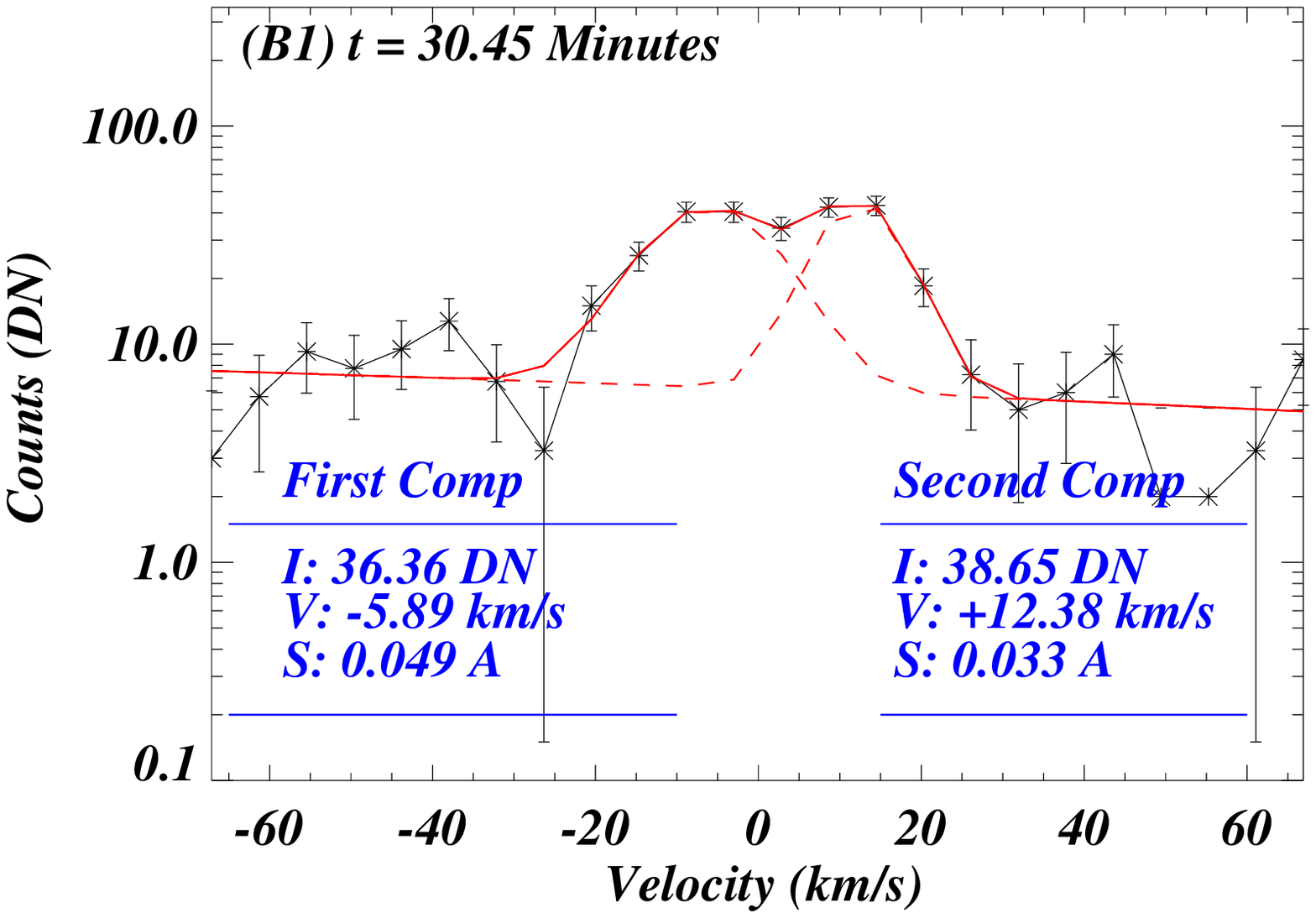}
\includegraphics[trim=3.0cm 0.5cm 1.5cm 0.0cm,scale=0.37]{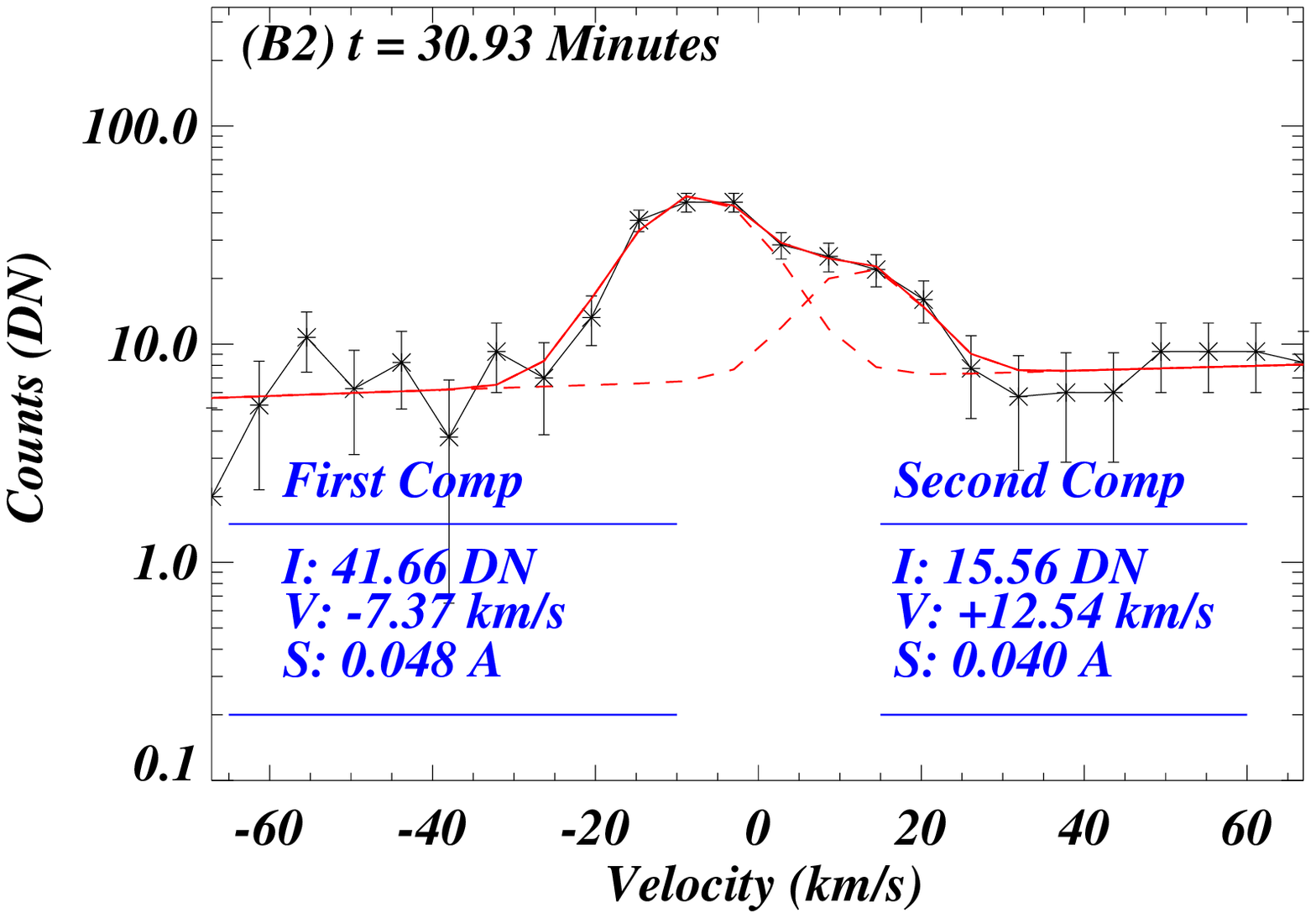}
\includegraphics[trim=3.0cm 0.5cm 1.5cm 0.0cm,scale=0.37]{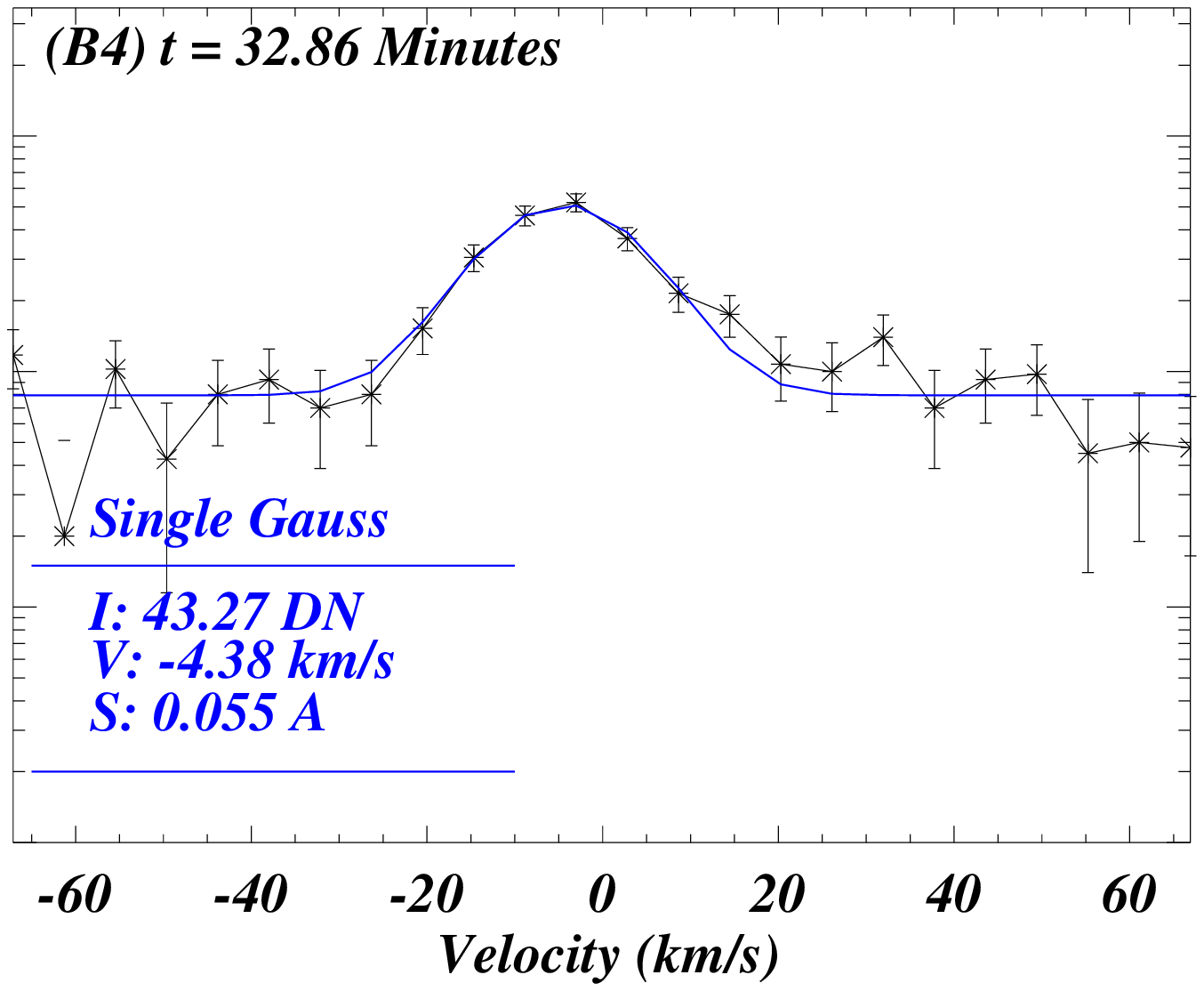}
\includegraphics[trim=3.0cm 0.5cm 1.5cm 0.0cm,scale=0.37]{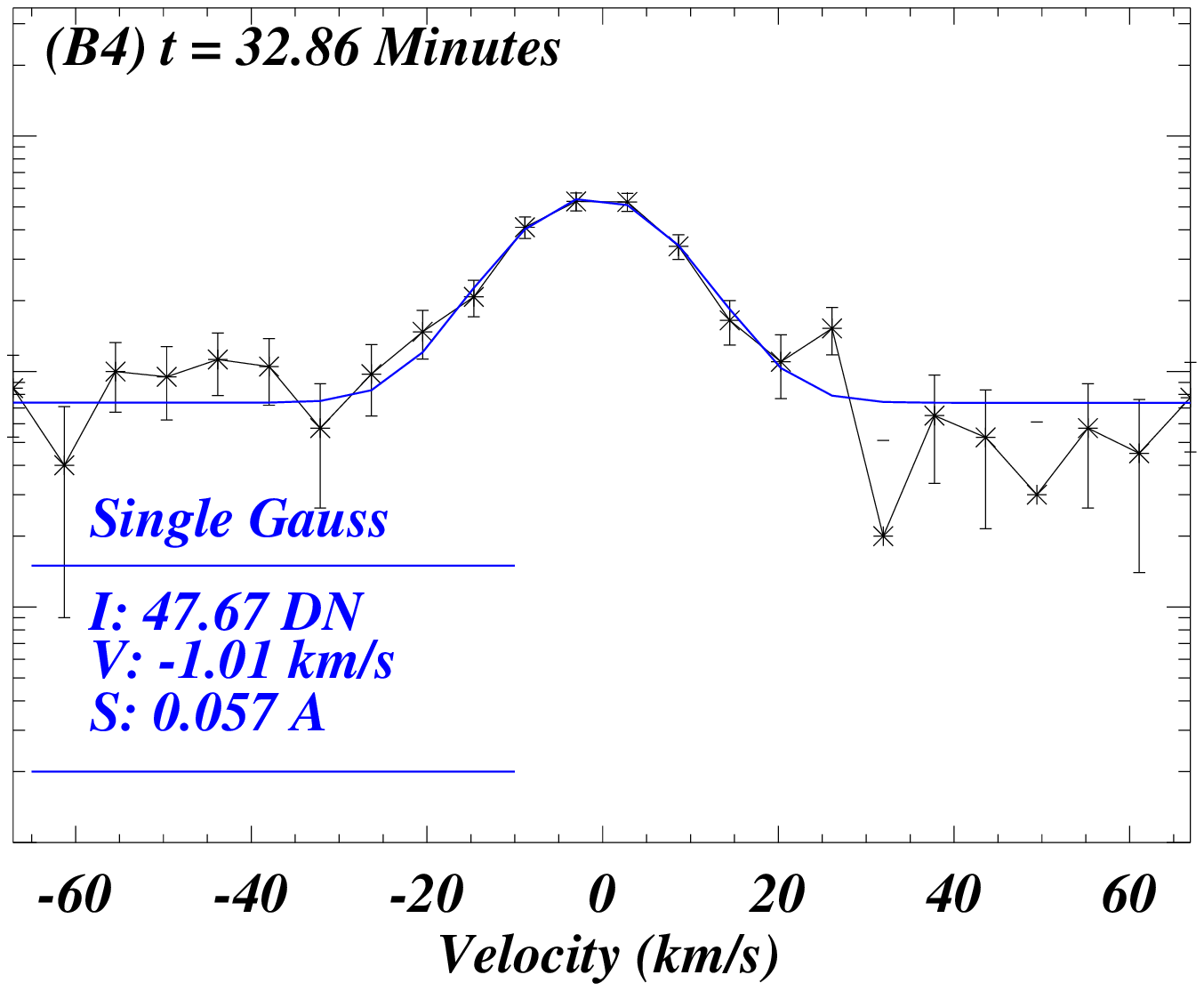}
}
%%-------------------------------------------------
\mbox{
\includegraphics[trim=5.0cm 0.5cm 1.5cm 0.0cm,scale=0.37]{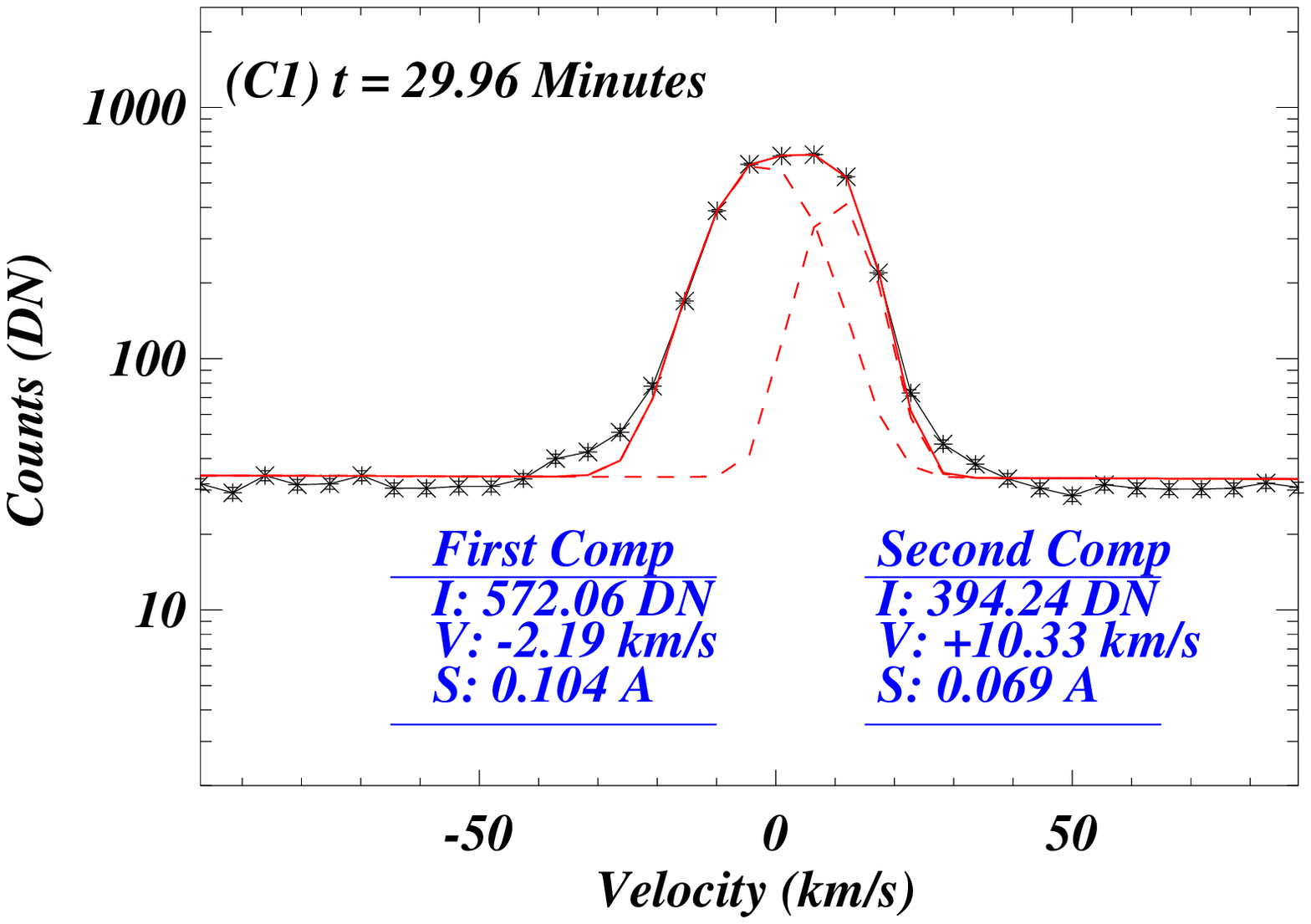}
\includegraphics[trim=3.0cm 0.5cm 1.5cm 0.0cm,scale=0.37]{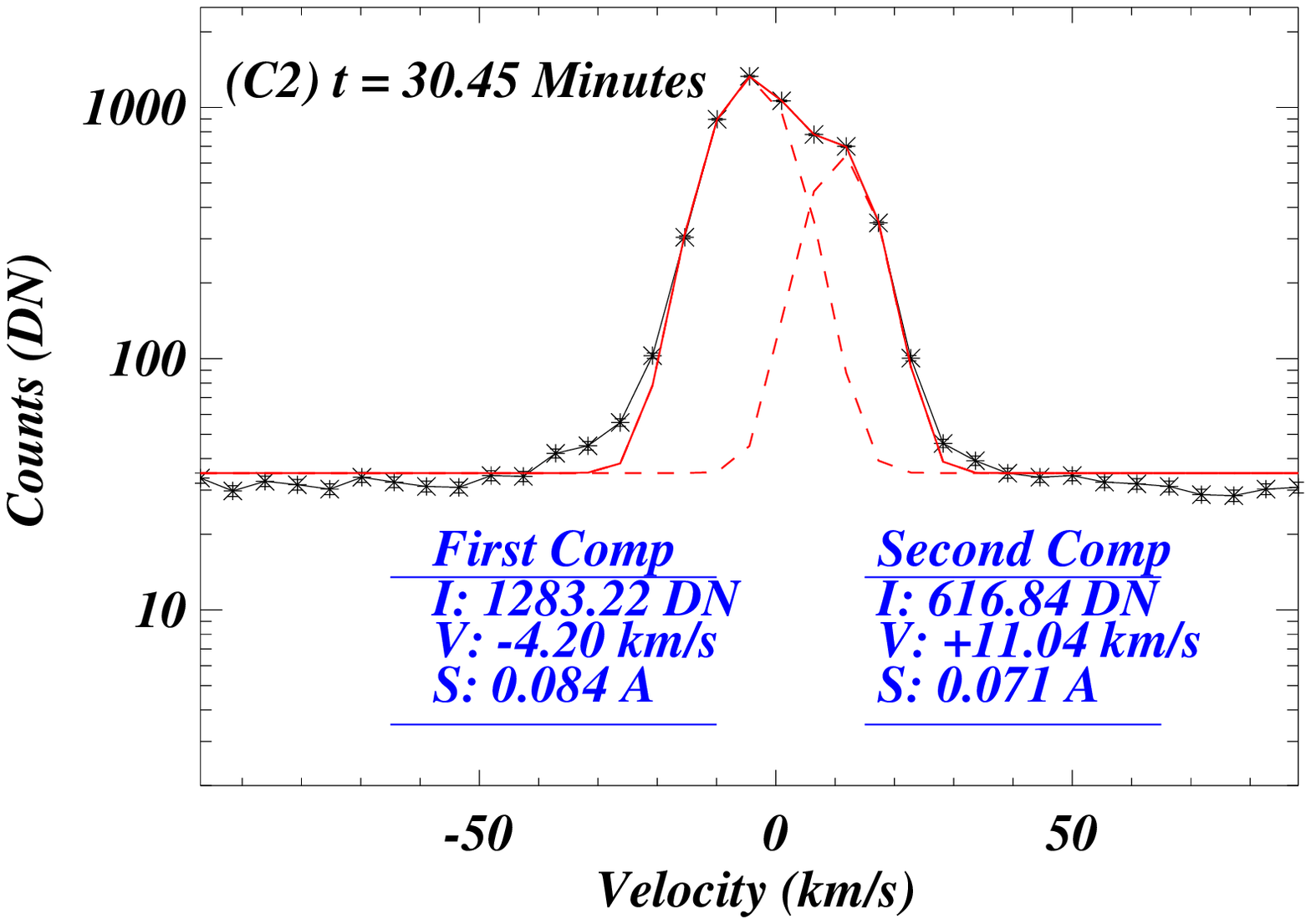}
\includegraphics[trim=3.0cm 0.5cm 1.5cm 0.0cm,scale=0.37]{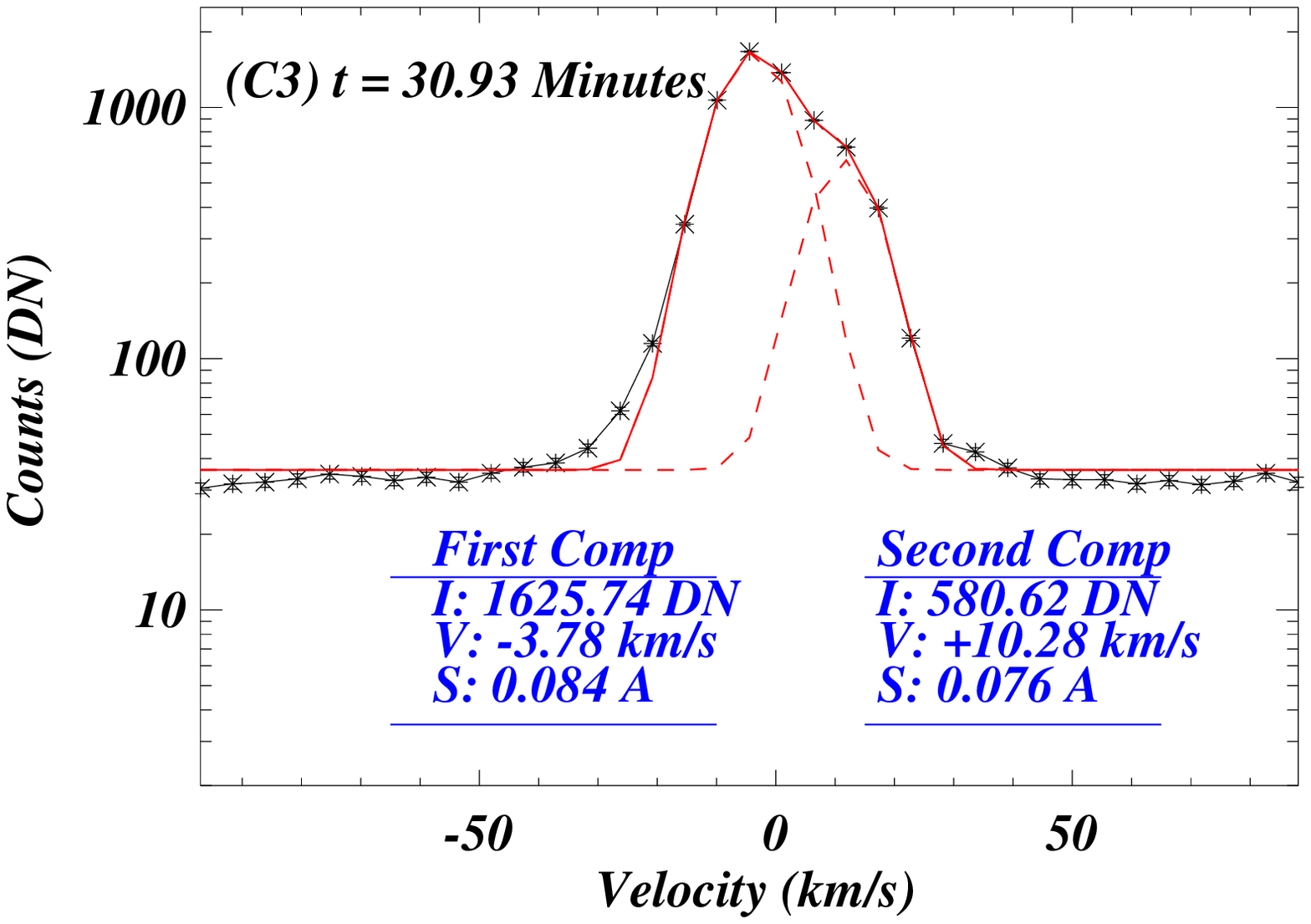}
\includegraphics[trim=3.0cm 0.5cm 1.5cm 0.0cm,scale=0.37]{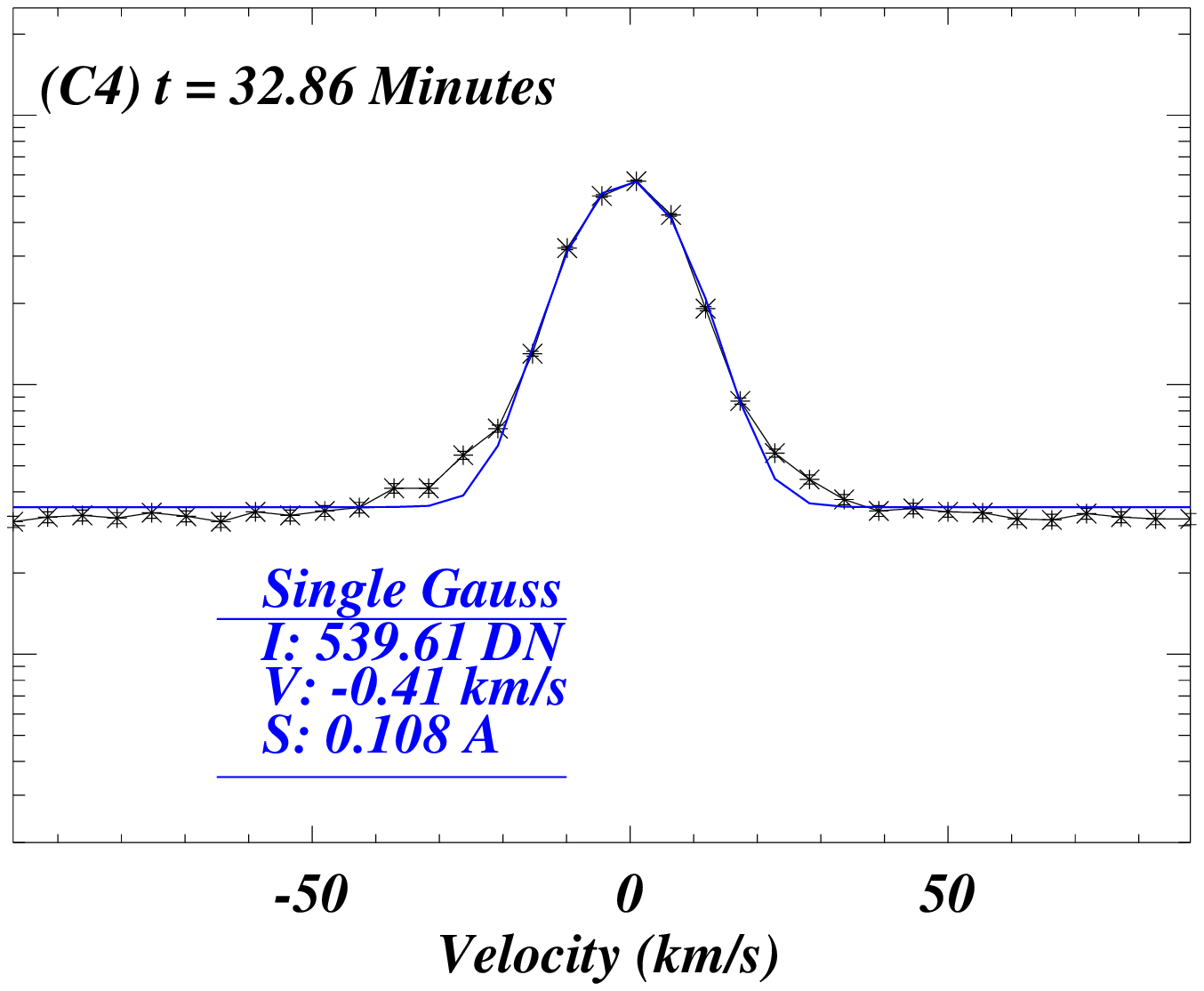}
}
\caption{Same as the middle panel in Fig.~\ref{fig:ref_fig} and Fig.~\ref{fig:all_profile} but for location y = 83.98$"$}
\label{fig:append_one}
\end{figure}
%%-------------------------------------------------
\begin{figure}
\centering
\mbox{
\includegraphics[trim=6.0cm 0.5cm 1.5cm 0.0cm,scale=0.5]{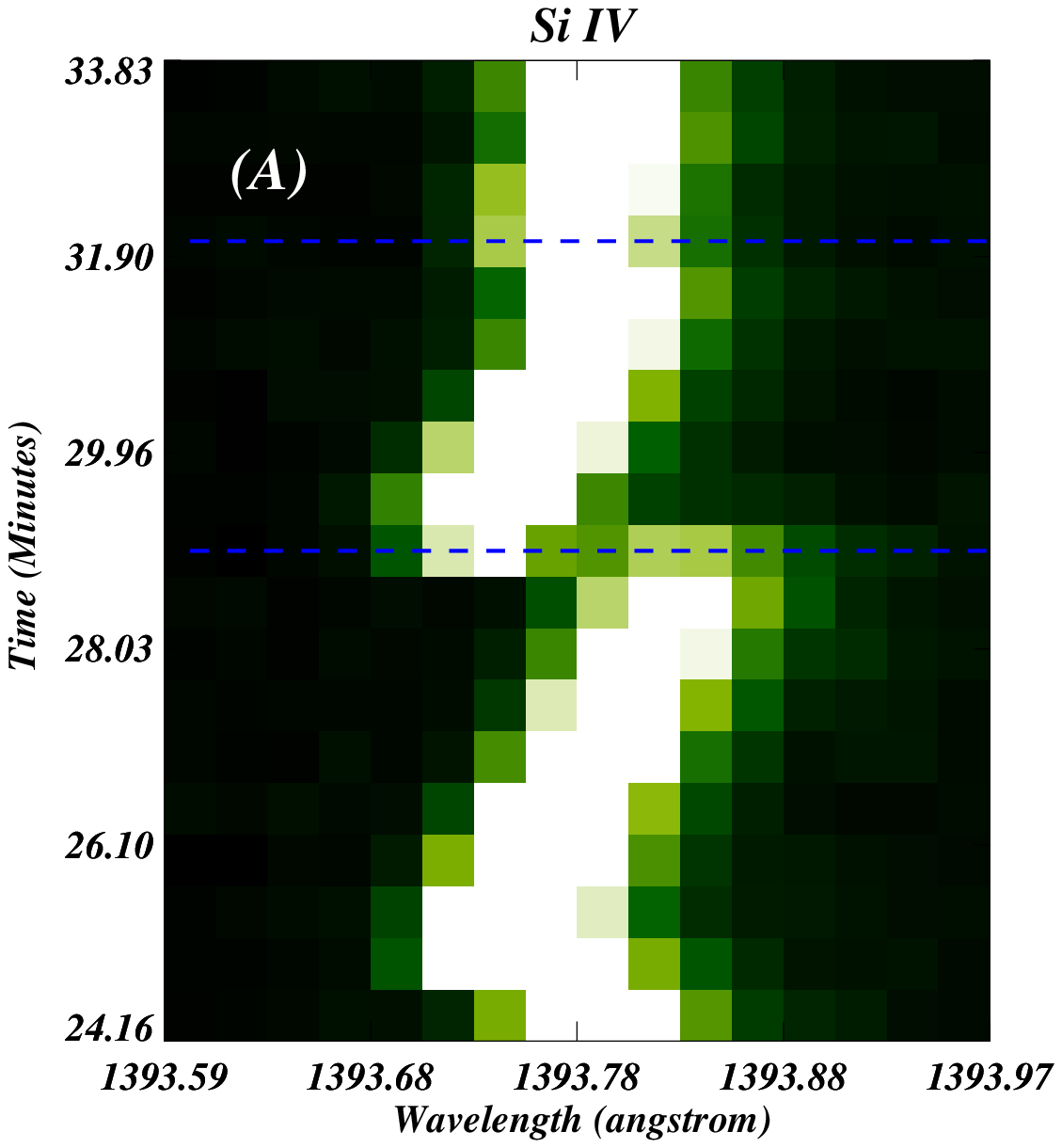}
\includegraphics[trim=6.0cm 0.5cm 1.5cm 0.0cm,scale=0.5]{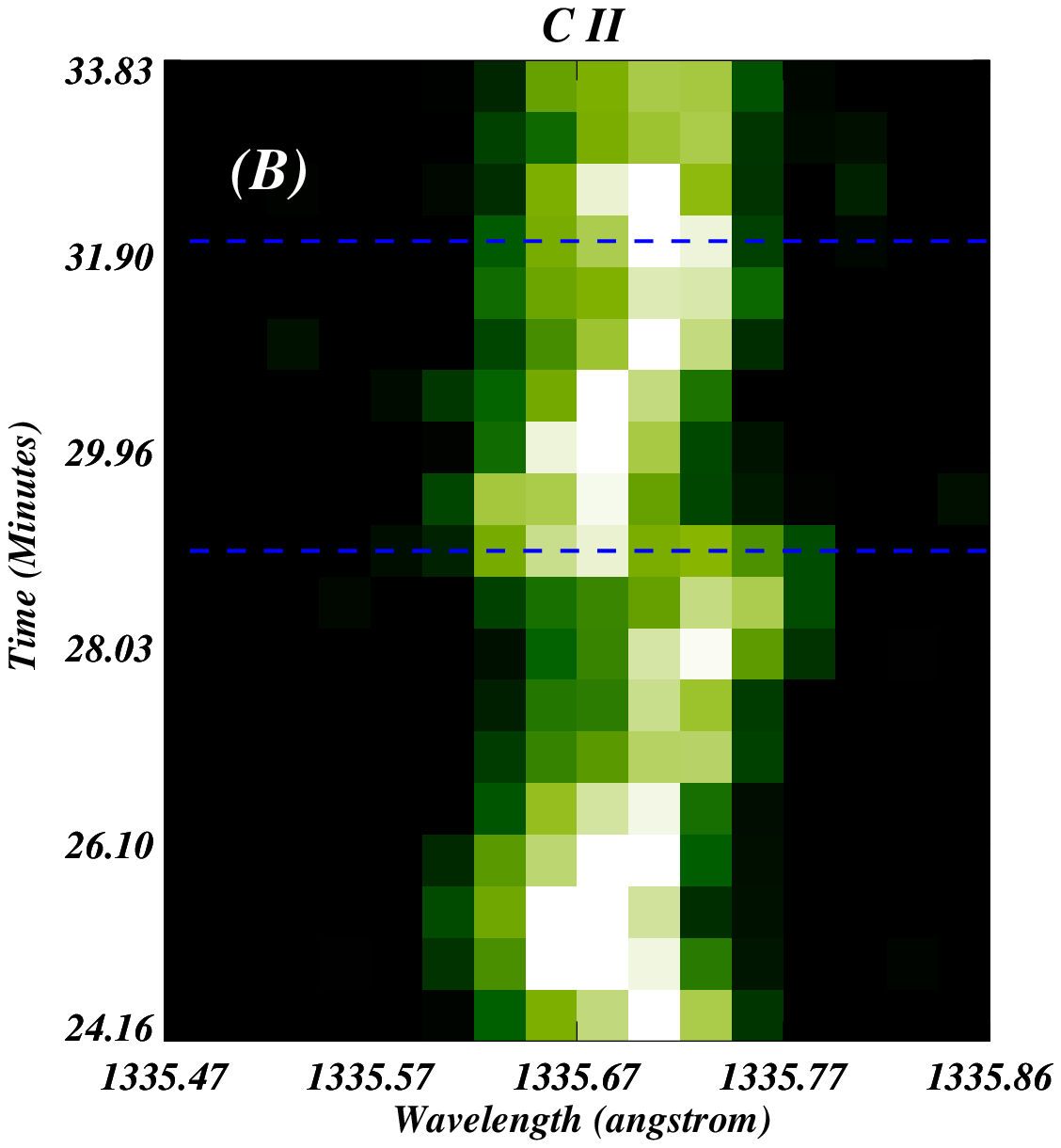}
\includegraphics[trim=6.0cm 0.5cm 1.5cm 0.0cm,scale=0.5]{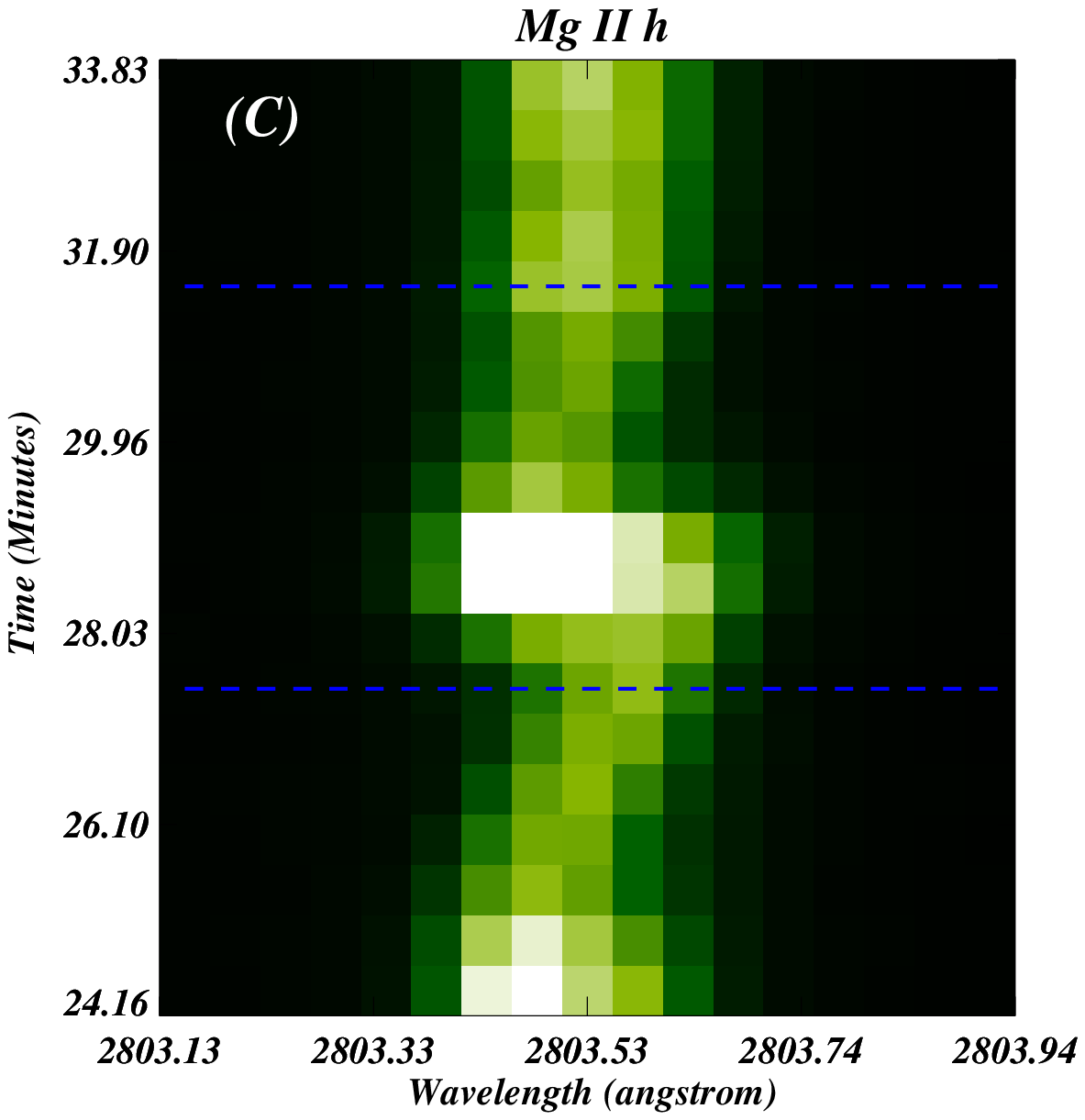}
}
%%-------------------------------------------------
\mbox{
\includegraphics[trim=5.0cm 0.5cm 1.5cm 0.0cm,scale=0.37]{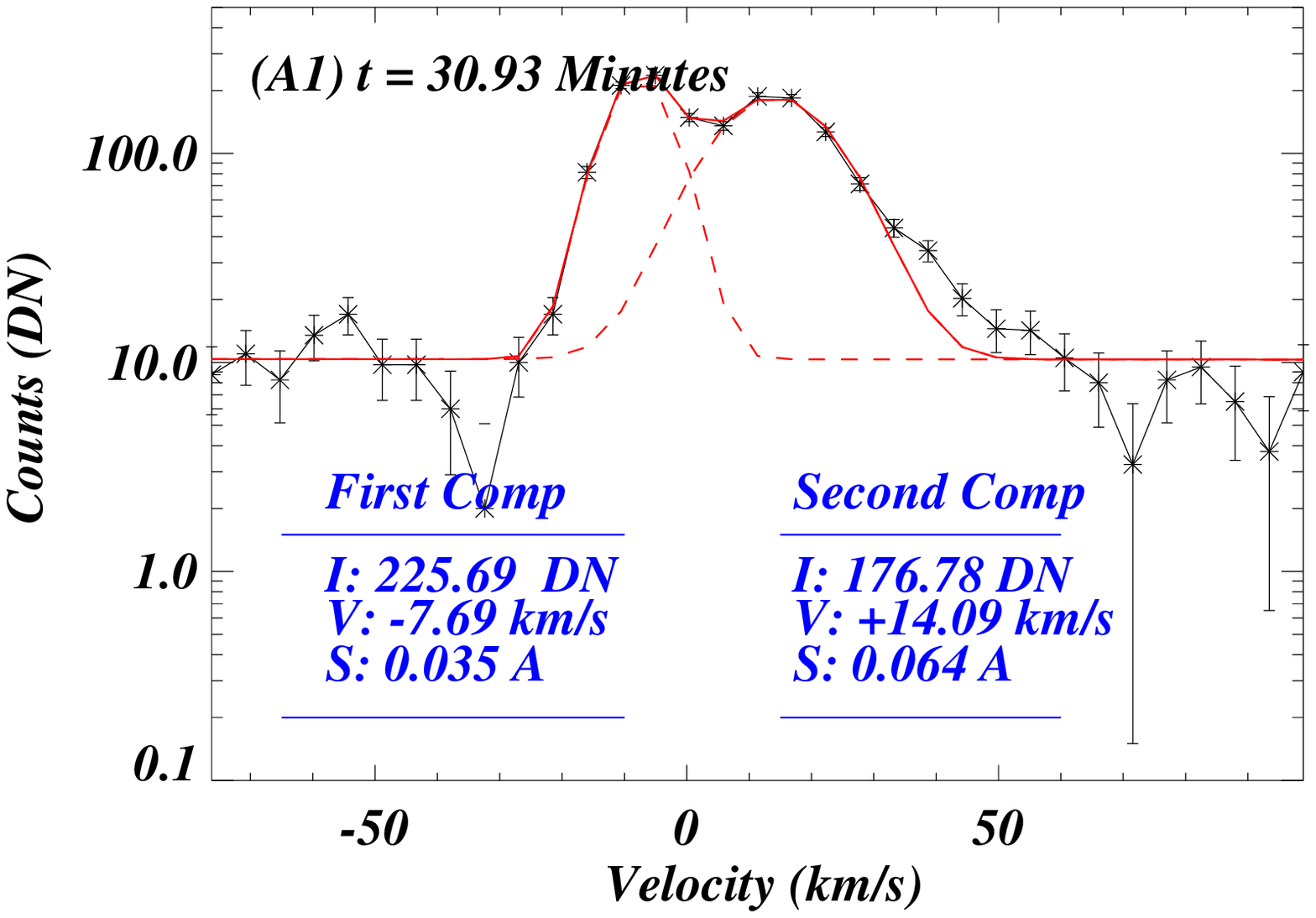}
\includegraphics[trim=3.0cm 0.5cm 1.5cm 0.0cm,scale=0.37]{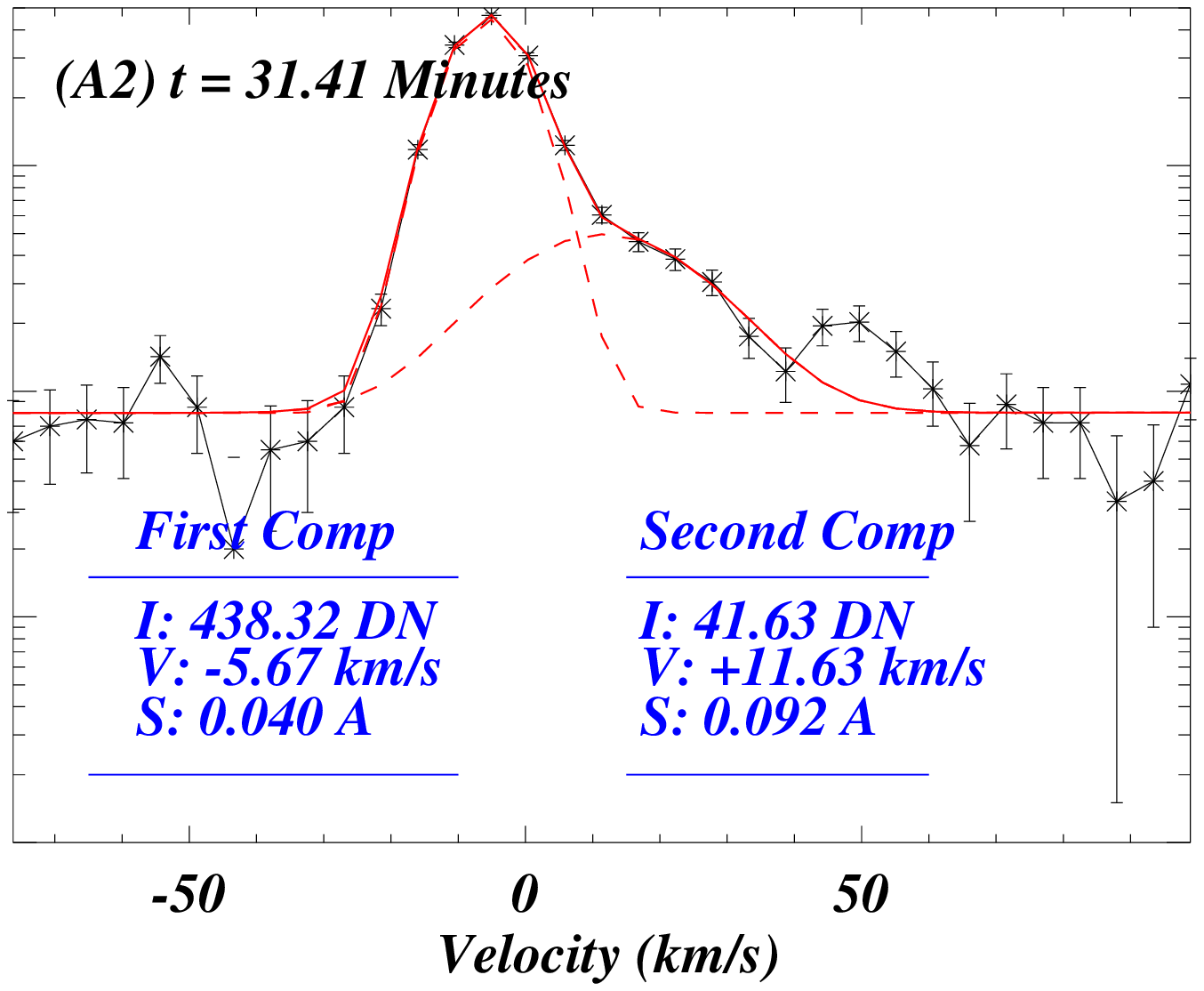}
\includegraphics[trim=3.0cm 0.5cm 1.5cm 0.0cm,scale=0.37]{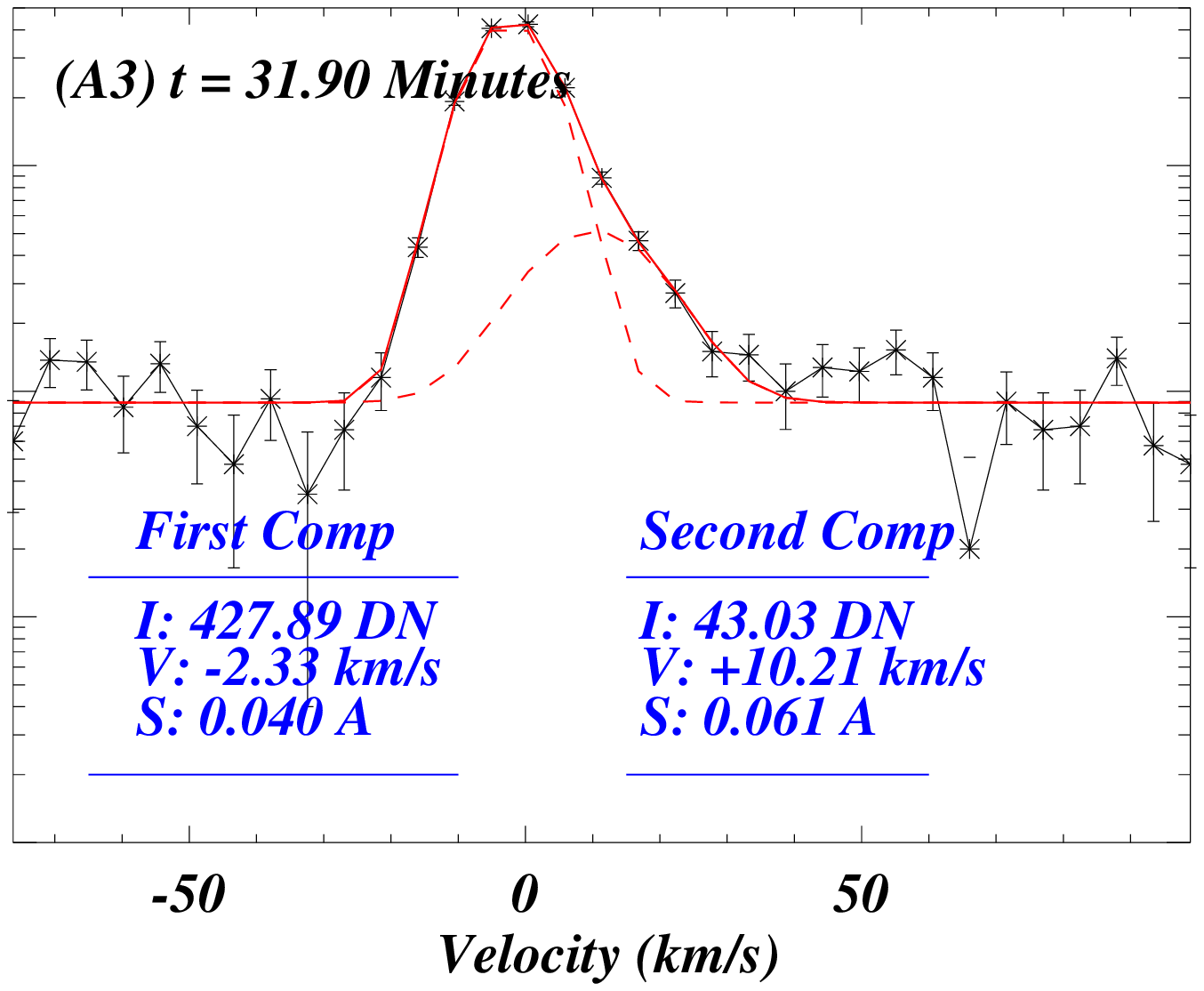}
\includegraphics[trim=3.0cm 0.5cm 1.5cm 0.0cm,scale=0.37]{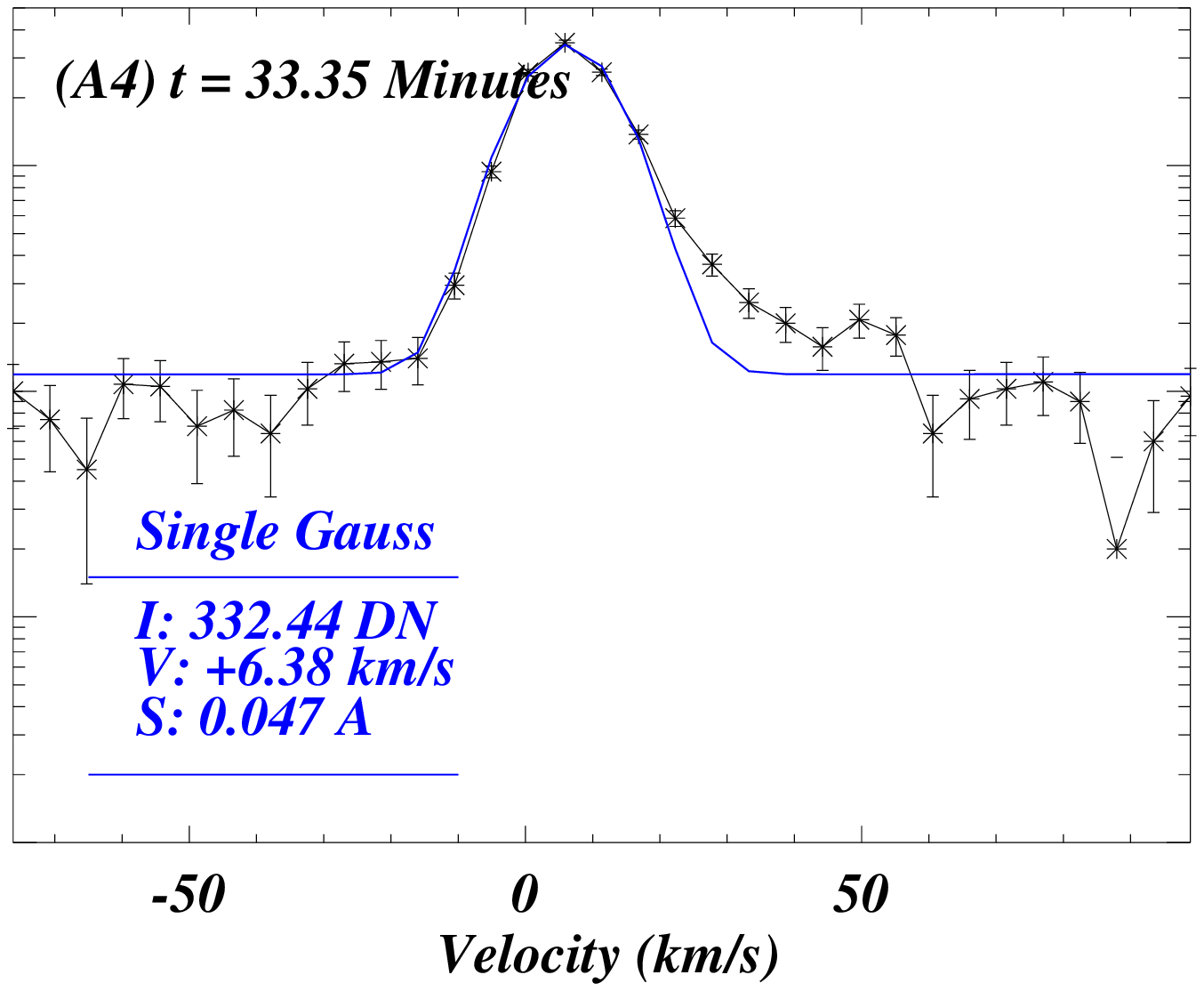}
}
%%-------------------------------------------------
\mbox{
\includegraphics[trim=5.0cm 0.5cm 1.5cm 0.0cm,scale=0.37]{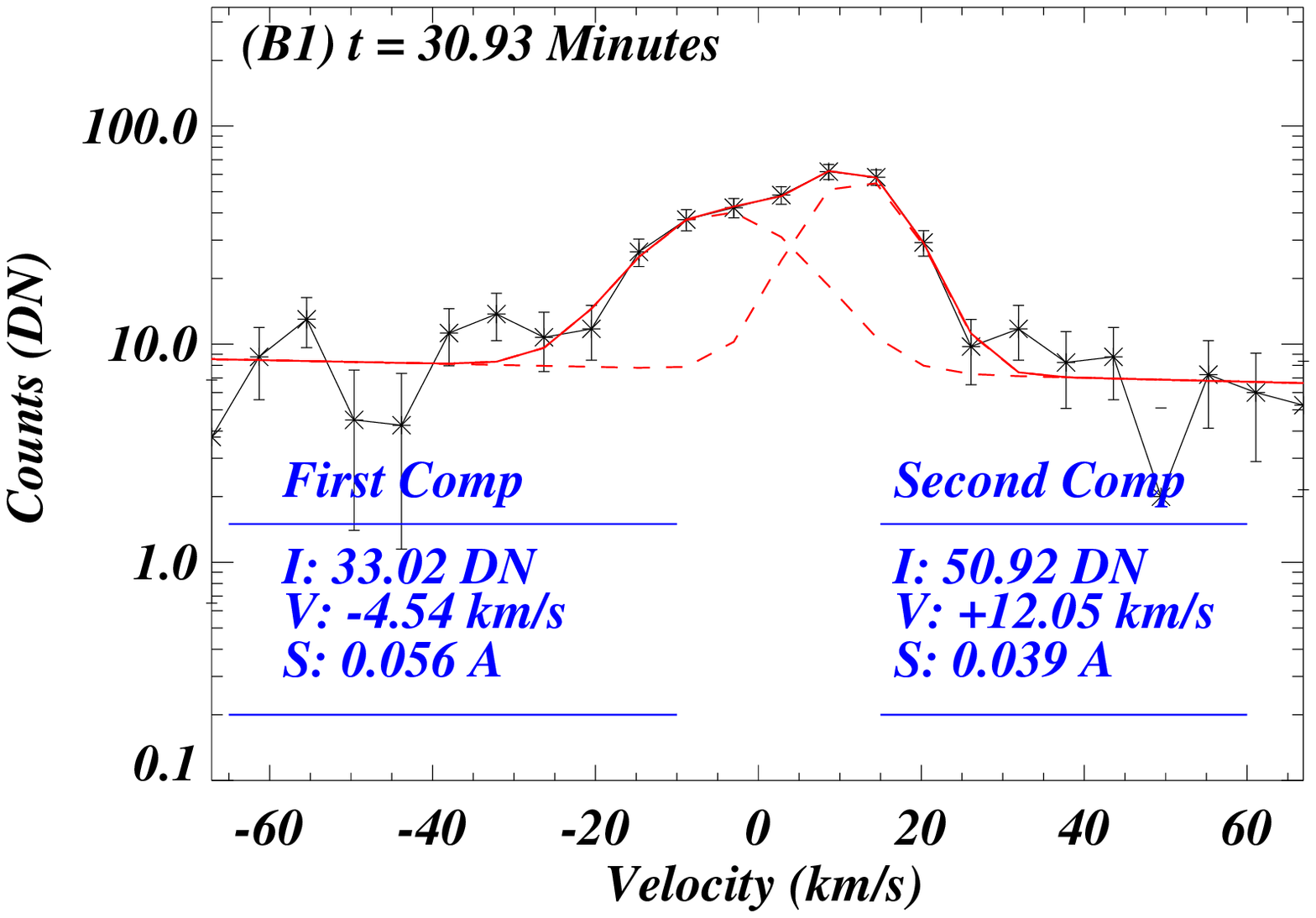}
\includegraphics[trim=3.0cm 0.5cm 1.5cm 0.0cm,scale=0.37]{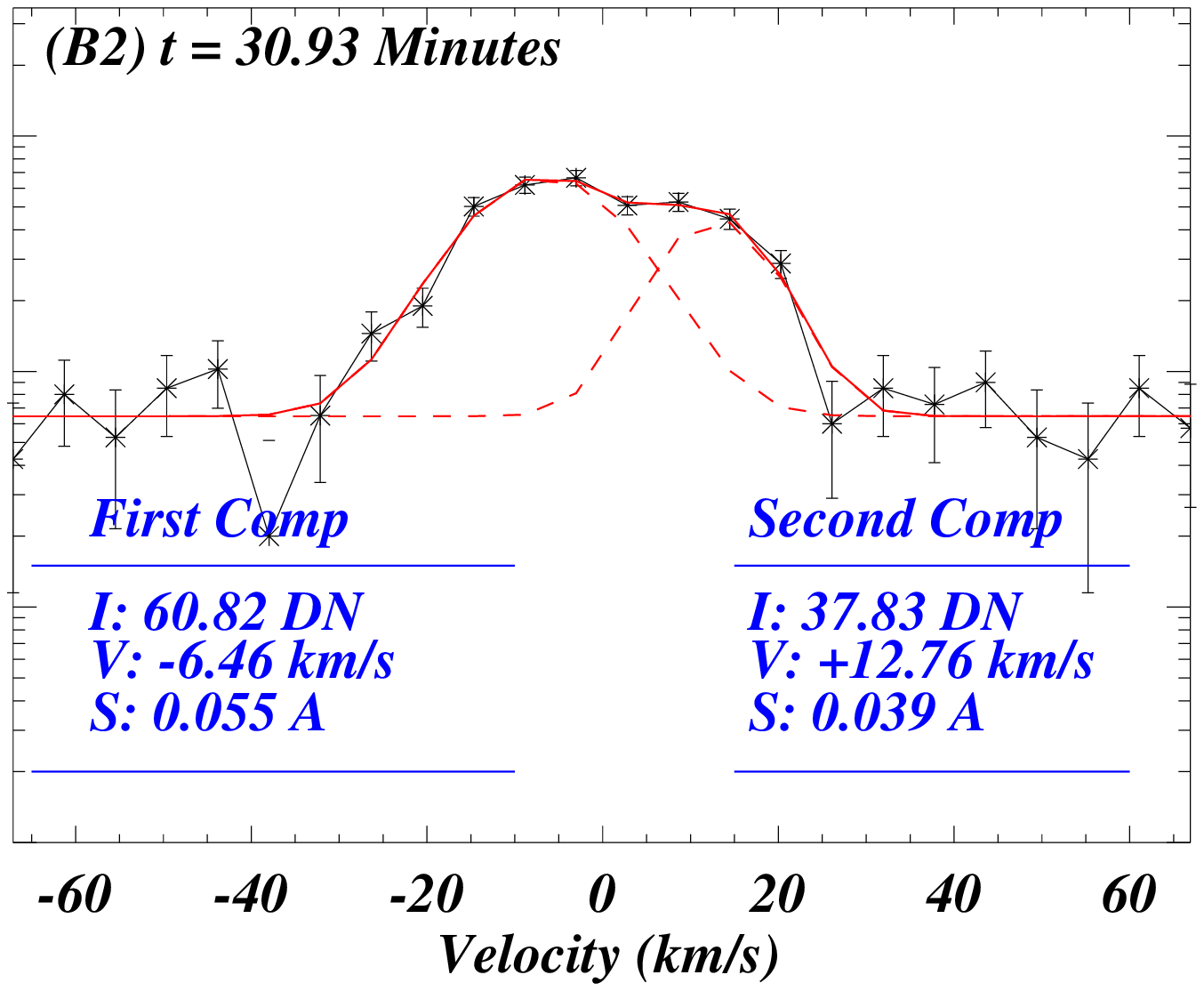}
\includegraphics[trim=3.0cm 0.5cm 1.5cm 0.0cm,scale=0.37]{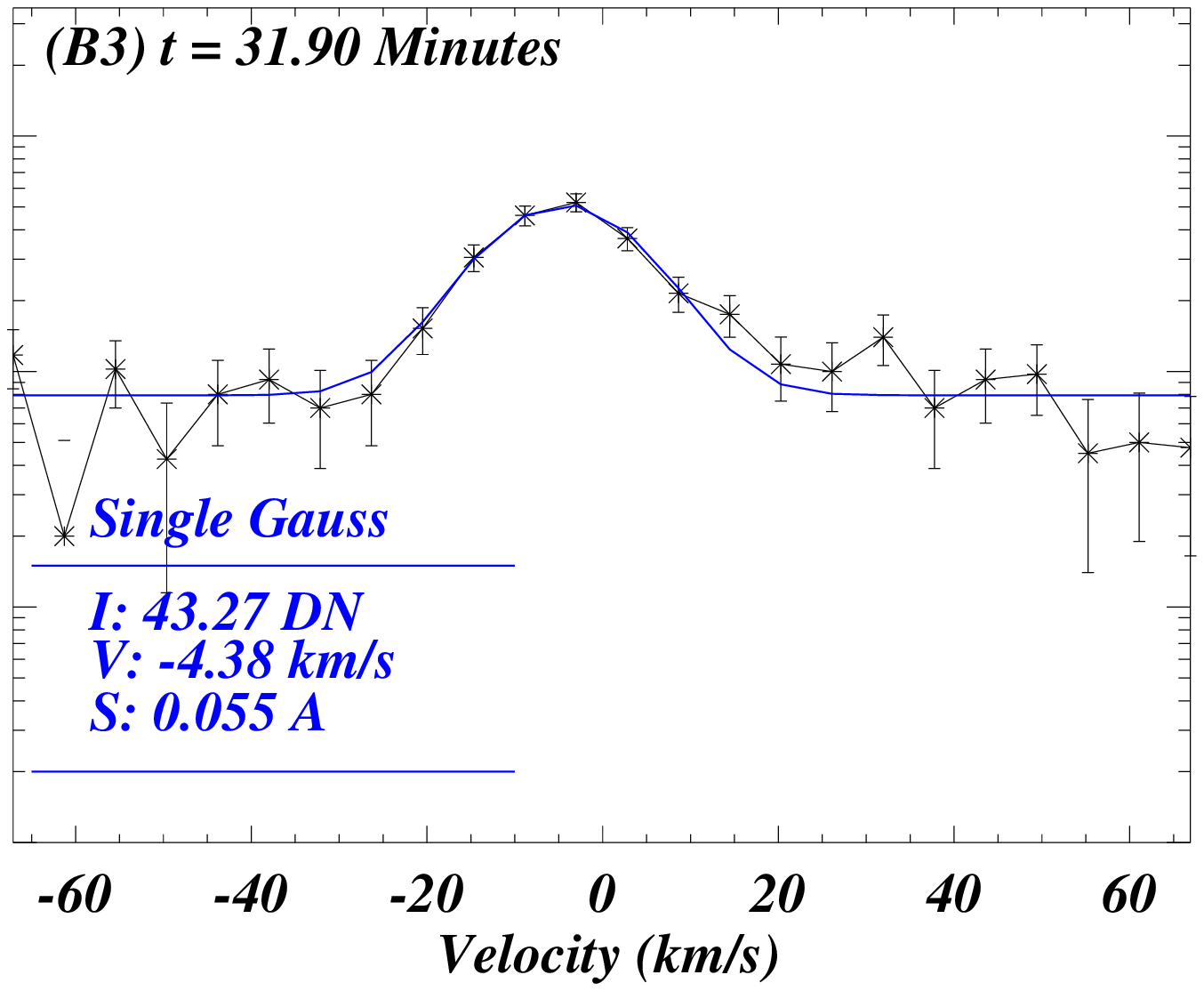}
\includegraphics[trim=3.0cm 0.5cm 1.5cm 0.0cm,scale=0.37]{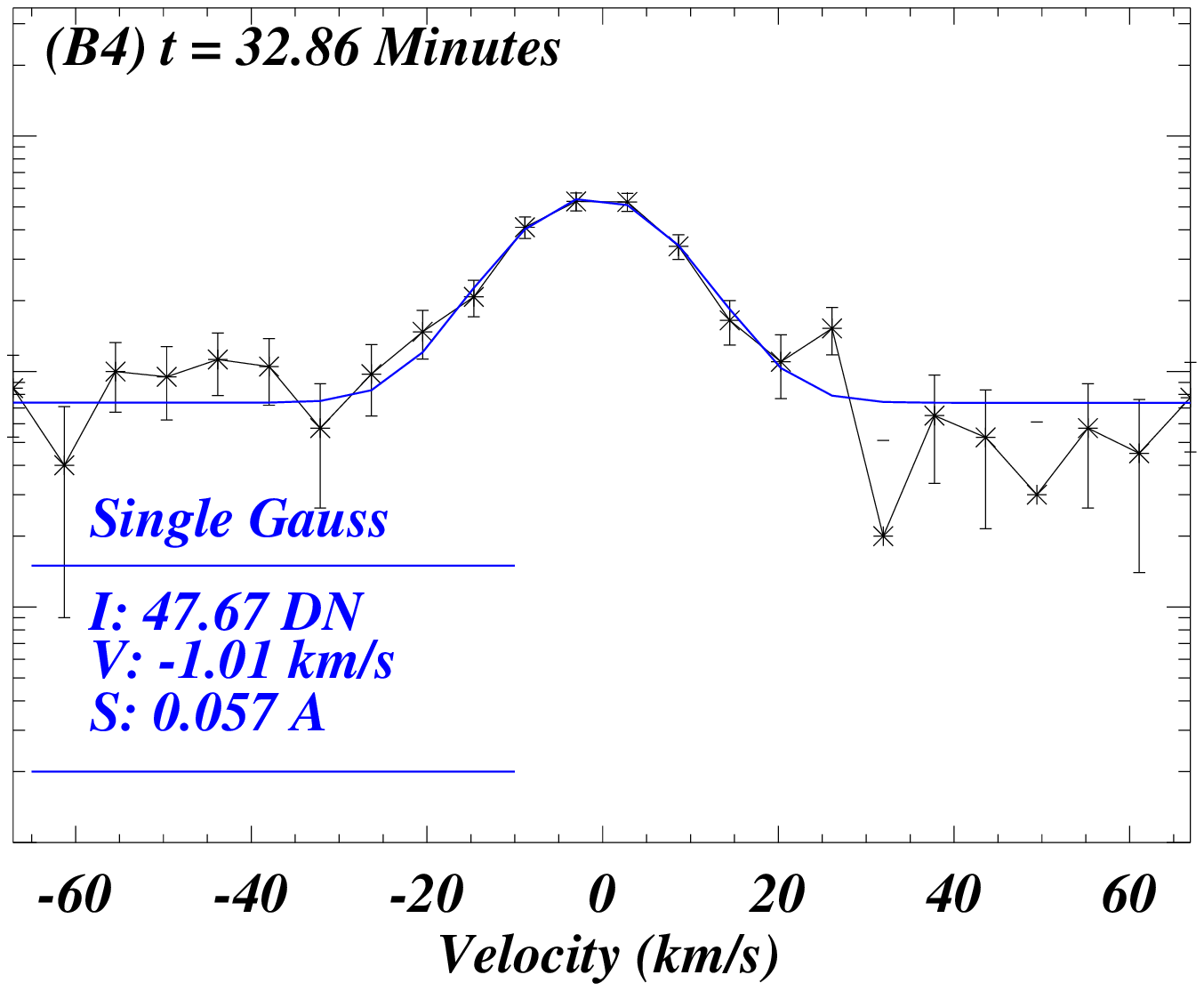}
}
%%%%%%Third Shock: Need to be Replaced%%%%%%%%%%%%%%%%%%%%%%%%%%%%%%%%%%%%%%%%%
\mbox{
\includegraphics[trim=5.0cm 0.5cm 1.5cm 0.0cm,scale=0.37]{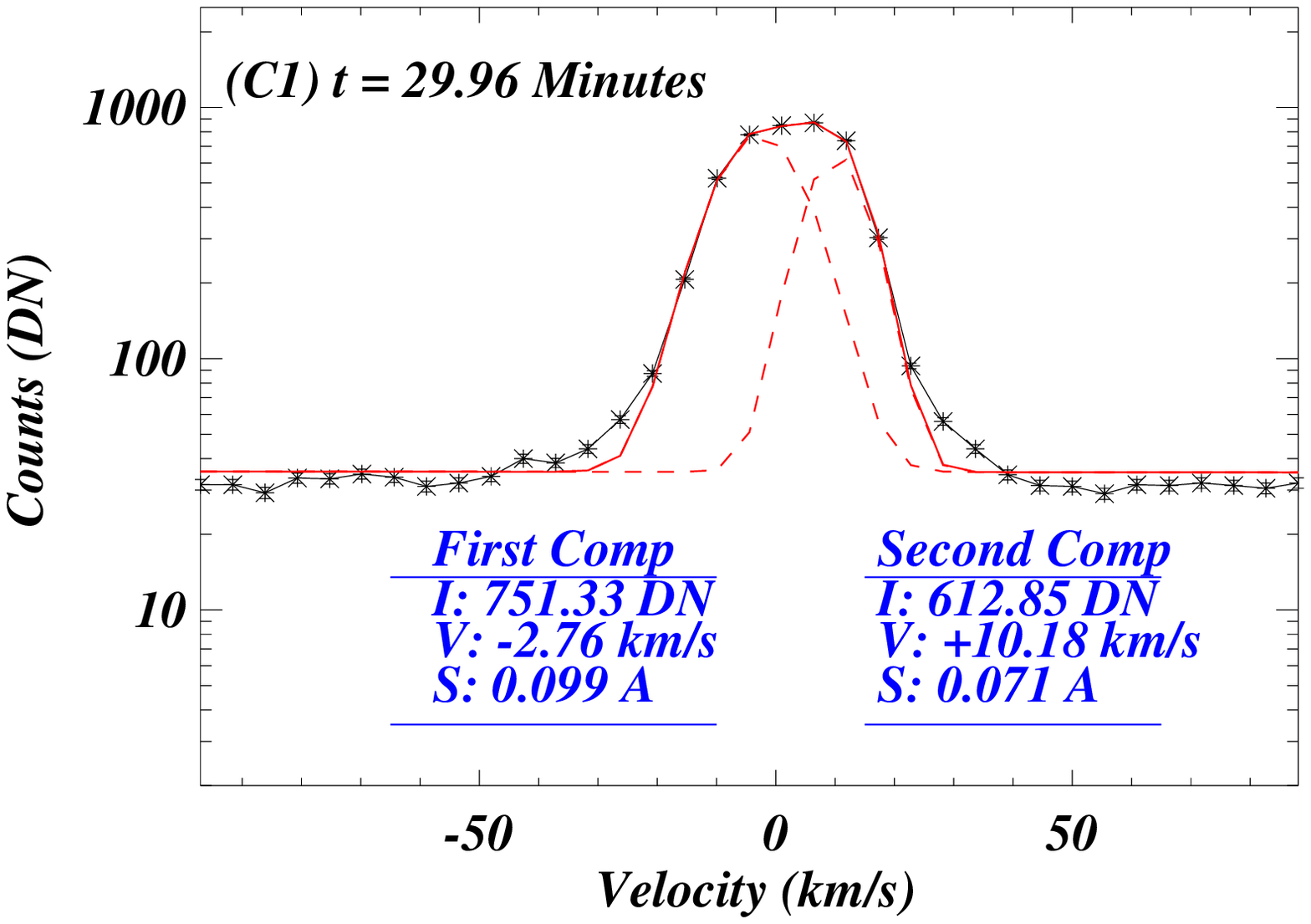}
\includegraphics[trim=3.0cm 0.5cm 1.5cm 0.0cm,scale=0.37]{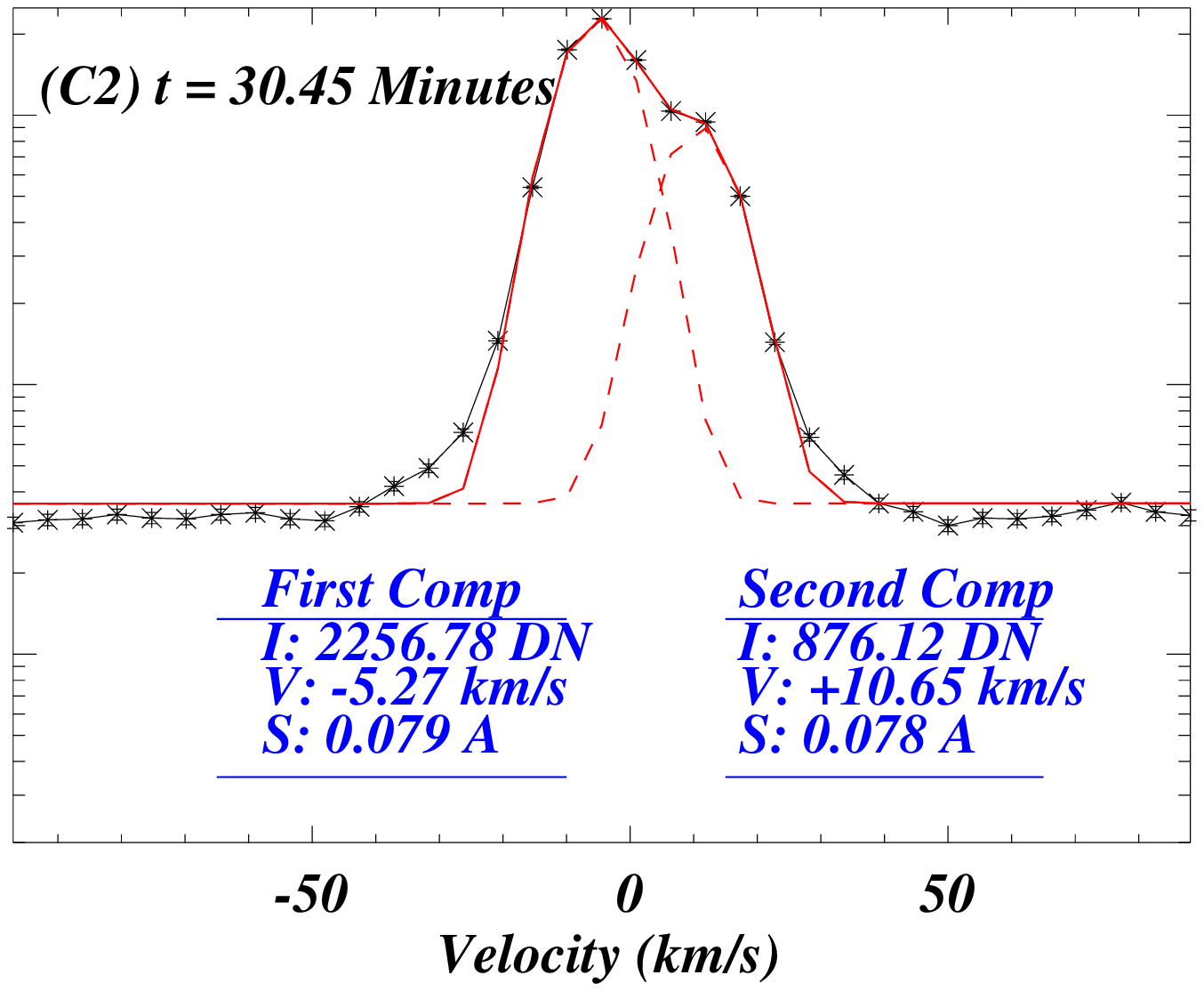}
\includegraphics[trim=3.0cm 0.5cm 1.5cm 0.0cm,scale=0.37]{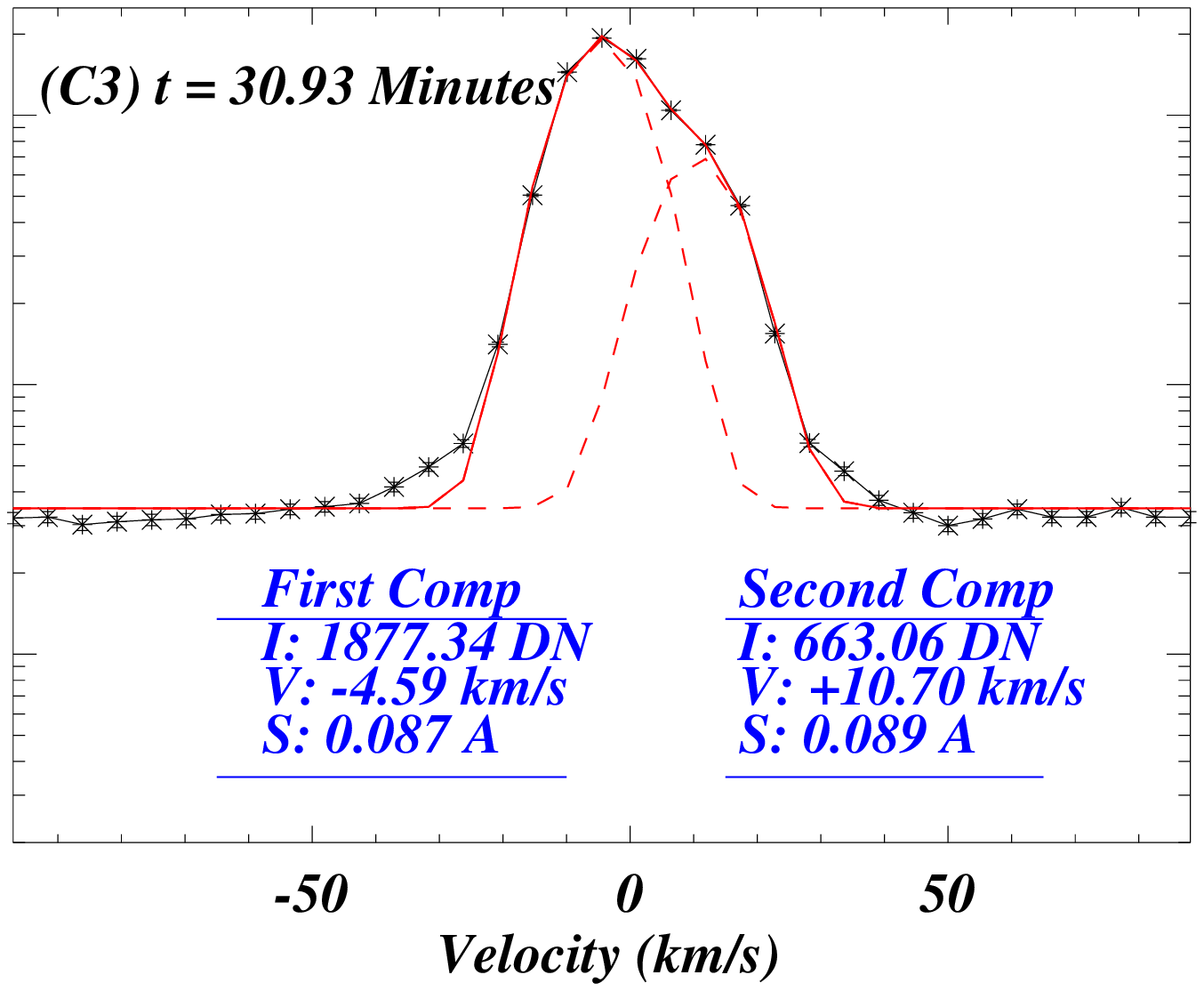}
\includegraphics[trim=3.0cm 0.5cm 1.5cm 0.0cm,scale=0.37]{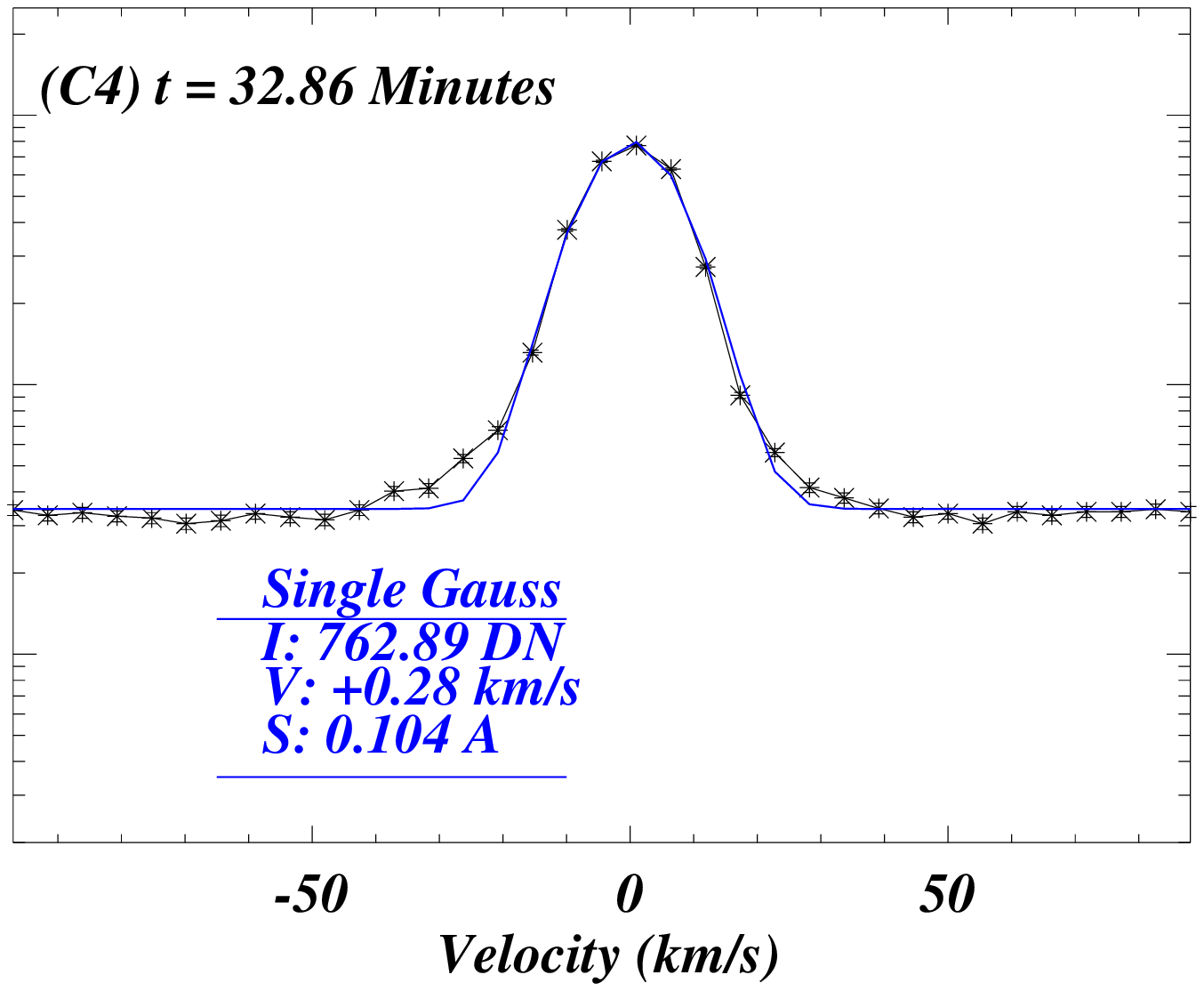}
}

\caption{Same as the middle panel in Fig.~\ref{fig:ref_fig} and Fig.~\ref{fig:all_profile} but for the location y = 84.65$"$}
\label{fig:append_two}
\end{figure}
%%-------------------------------------------------
\begin{figure}
\mbox{
\includegraphics[trim=1.0cm 0.5cm 1.0cm 0.5cm,scale=0.5]{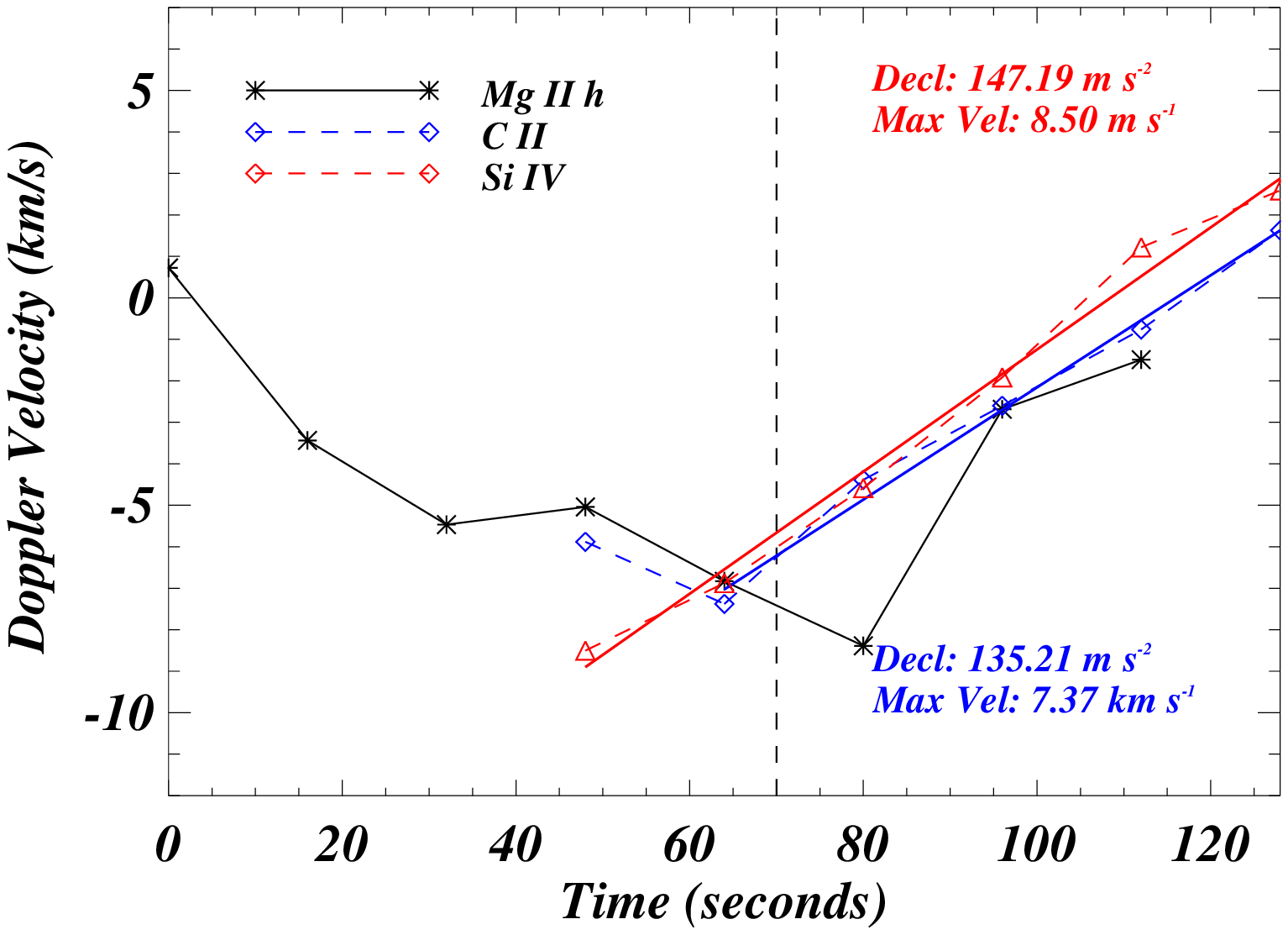}
\includegraphics[trim=0.5cm 0.0cm 1.5cm 1.0cm,scale=0.5]{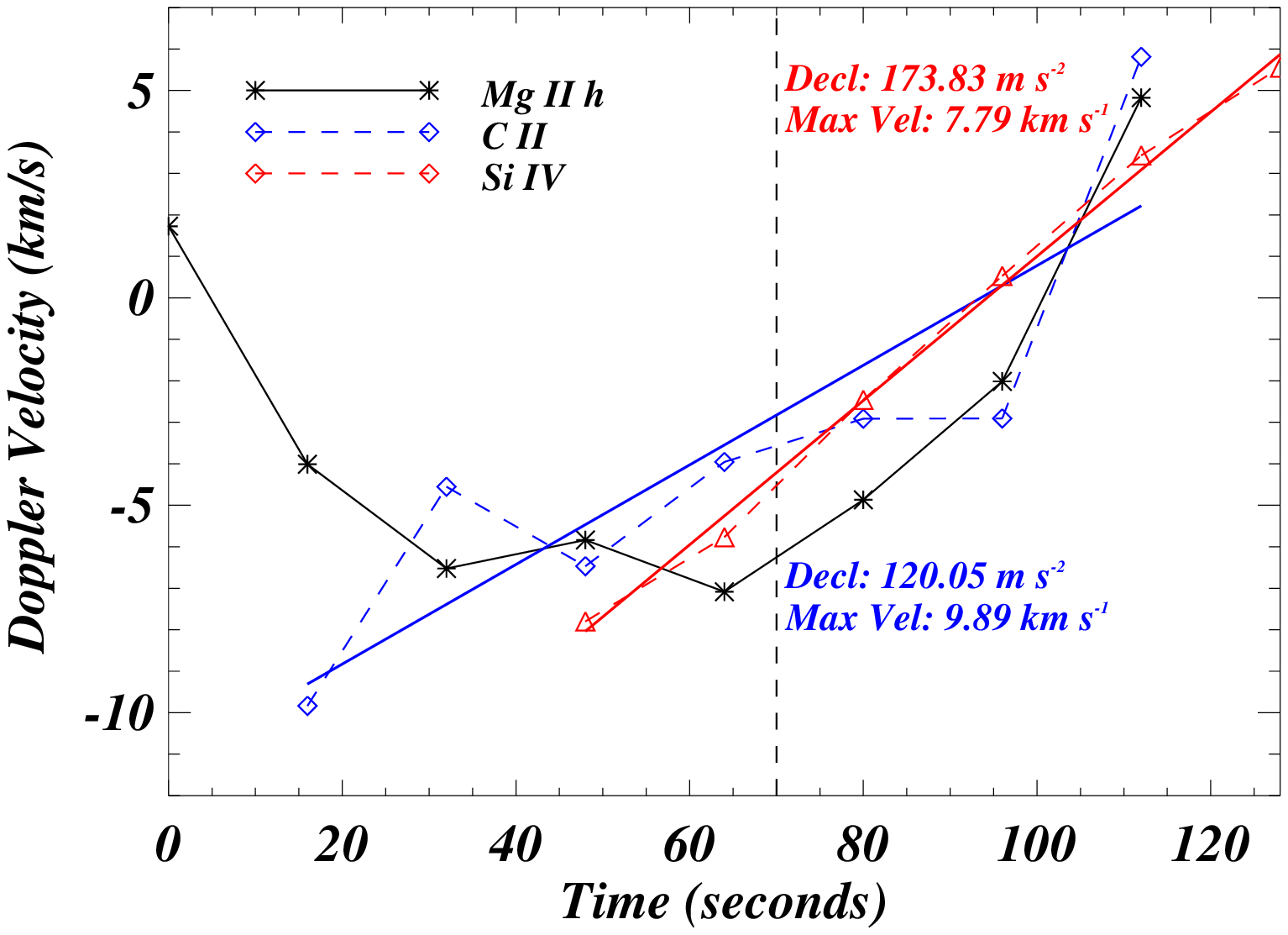}
}
\caption{Same as top-right panel of Figure~\ref{fig:shock_acel_main} but for the shocks displayed in the Fig.~\ref{fig:append_one} (left-panel) and Fig.~\ref{fig:append_two} (right-panel).}
\label{fig:shock_acel}
\end{figure}

%%-------------------------------------------------
%\setcounter{0}
%\setcounter{section}{0}
%\appendix \label{append_iris_aia}
%\renewcommand\thefigure{\thesection.\arabic{figure}} 
\renewcommand\appendixname{appendix}  \label{append_iris_aia}  
\section{Relation between \ion{Si}{4} and AIA~171~{\AA} {--} Shock Signature in AIA 171~{\AA}}\label{append:iris_aia_correl}

In order to further establish the association between the shocks observed in \ion{Si}{4} and AIA~{171}, in the right-panel of Fig.~\ref{fig:iris_aia}, we plot IRIS \ion{Si}{4} time-series (left column, top-panel) along with 6$^{th}$ order polynomial fit (red curve). The 6$^{th}$ order polynomial was used to remove the large scale trend from the light curve. The trend removed light curve is shown in the 2$^{nd}$ panel.  Similar procedure is applied to AIA~171~{\AA} time-series, and the results are shown in the 3$^{rd}$ and 4$^{th}$ panels (from top, left column) of Fig.~\ref{fig:iris_aia}. 

Finally, after processing the time-series, we have compared IRIS \ion{Si}{4} (red) and AIA~171~{\AA} (black) light curves obtained at $Y = -84.65${\arcsec} (top right panel) and at $Y = -84.31${\arcsec} (bottom right panel). We can see that both the light curves are well correlated with each other with some apparent time-lag. We have also marked the times at which the shock were identified by the vertical dashed blue line. 
 
%%-------------------------------------------------
\setcounter{figure}{0} 
\begin{figure}
\mbox{
\includegraphics[trim=3.0cm 1.5cm 3.0cm 2.5cm,scale=0.8]{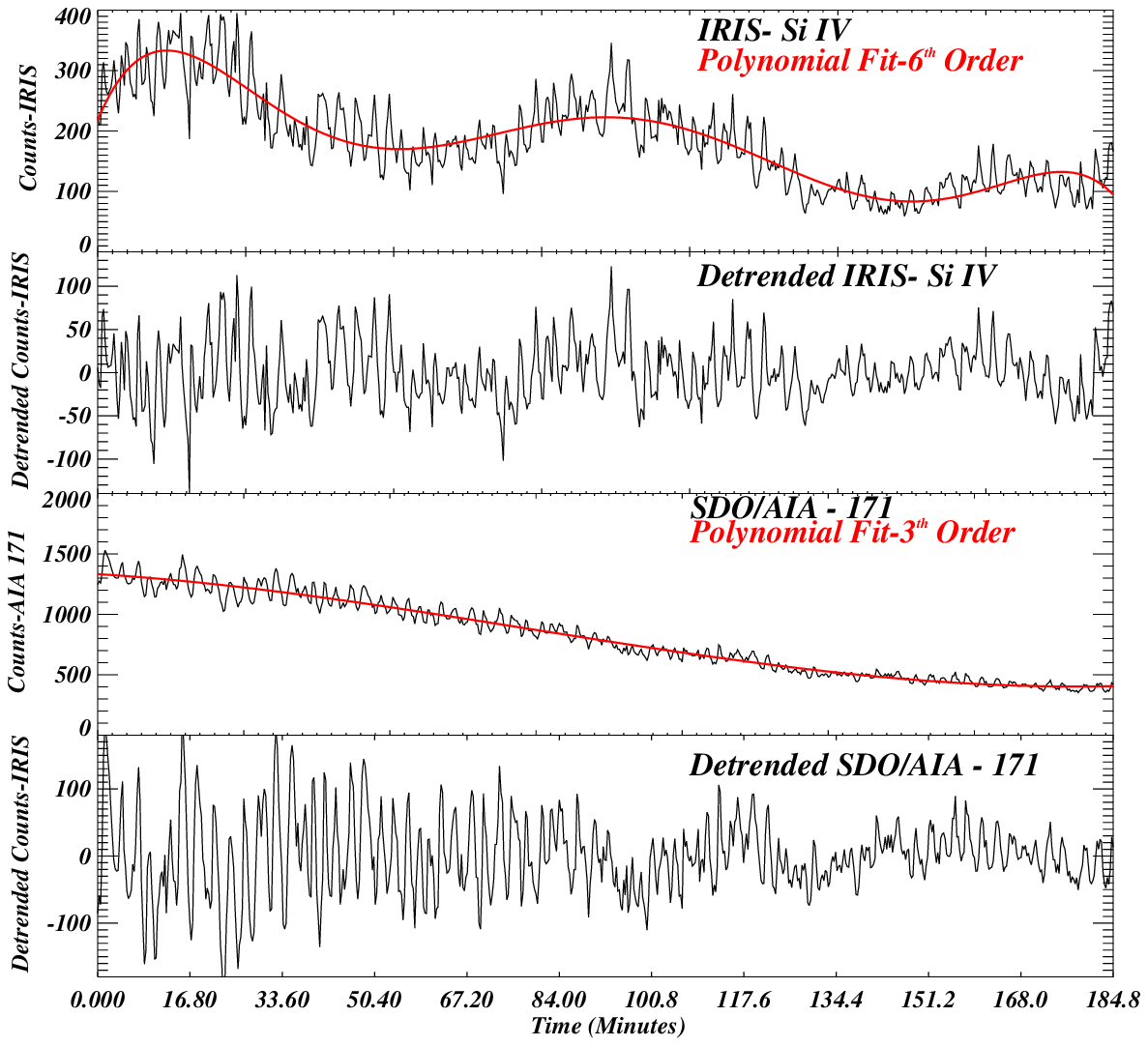}
\includegraphics[trim=3.0cm 0.0cm 3.5cm 3.0cm,scale=0.8]{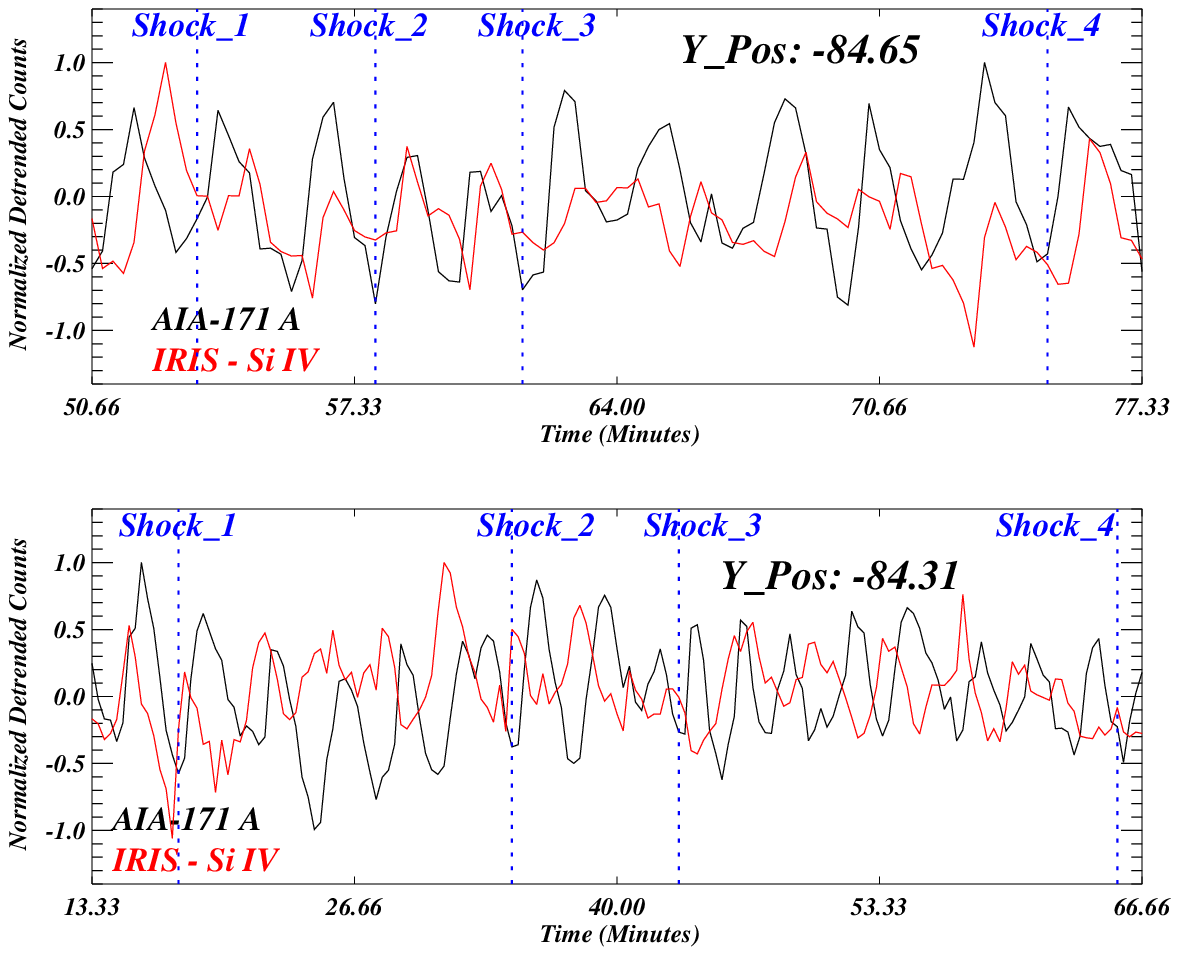}
}
\caption{Left panel: \ion{Si}{4} (top) and AIA~171~{\AA}) (third)  along with polynomial fit in red colors. The second and bottom-panel shows detrended light curves. Right panel: comparison between \ion{Si}{4} and AIA~171~{\AA} detrended light curves obtained at two different locations for the time range 50.66 - 77.33 minutes and from 13.33 to 66.66 minutes.}\label{fig:iris_aia}
\end{figure}
%%-------------------------------------------------
\end{document}